\documentclass[11pt,usenatbib]{article}
\usepackage{authblk}
\usepackage
[a4paper,
 left=0.6in,
 right=0.6in,
 top=0.65in, 
 bottom=0.8in,
] 
{geometry}
\usepackage[utf8]{inputenc}
\usepackage{color}
\usepackage{graphicx}
\usepackage[numbers,sort&compress,square]{natbib}
\usepackage{amsmath}
\usepackage{amssymb}
\usepackage[colorlinks=true, citecolor=blue]{hyperref}
\usepackage[flushleft]{threeparttable}
\usepackage{comment}
\usepackage{wrapfig}
\usepackage{enumitem}
\usepackage{pdfpages}
\usepackage{comment}

\usepackage{tabulary}
\usepackage{multirow}
\usepackage{upgreek}

\newcommand{\jcap}{J. Cosmol. Astropart. Phys.}
\newcommand{\prl}{Phys. Rev. Letters}
\newcommand{\apj}{ApJ}
\newcommand{\mnras}{MNRAS}
\newcommand{\apjl}{ApJL}

\newcommand{\nat}{Nature}
\newcommand{\aap}{A \&  A}

\newcommand{\aj}{AJ}

\newcommand{\ssr}{Space Sci. Rev. }
\newcommand{\araa}{ARA\&  A}

\newcommand{\prd}{Phys. Rev. D}

\newcommand{\baas}{Bull. Am. Astron. Soc.}
\newcommand{\aapr}{A\&A Rev.}
\newcommand{\physrep}{Phys. Reports}
\newcommand{\nar}{New Astr. Rev.}

\newcommand{\msun}{{\rm M}_{\odot}}
\newcommand{\procspie}{Proc. of the SPIE}

\begin{document}
\includepdf[page=1]{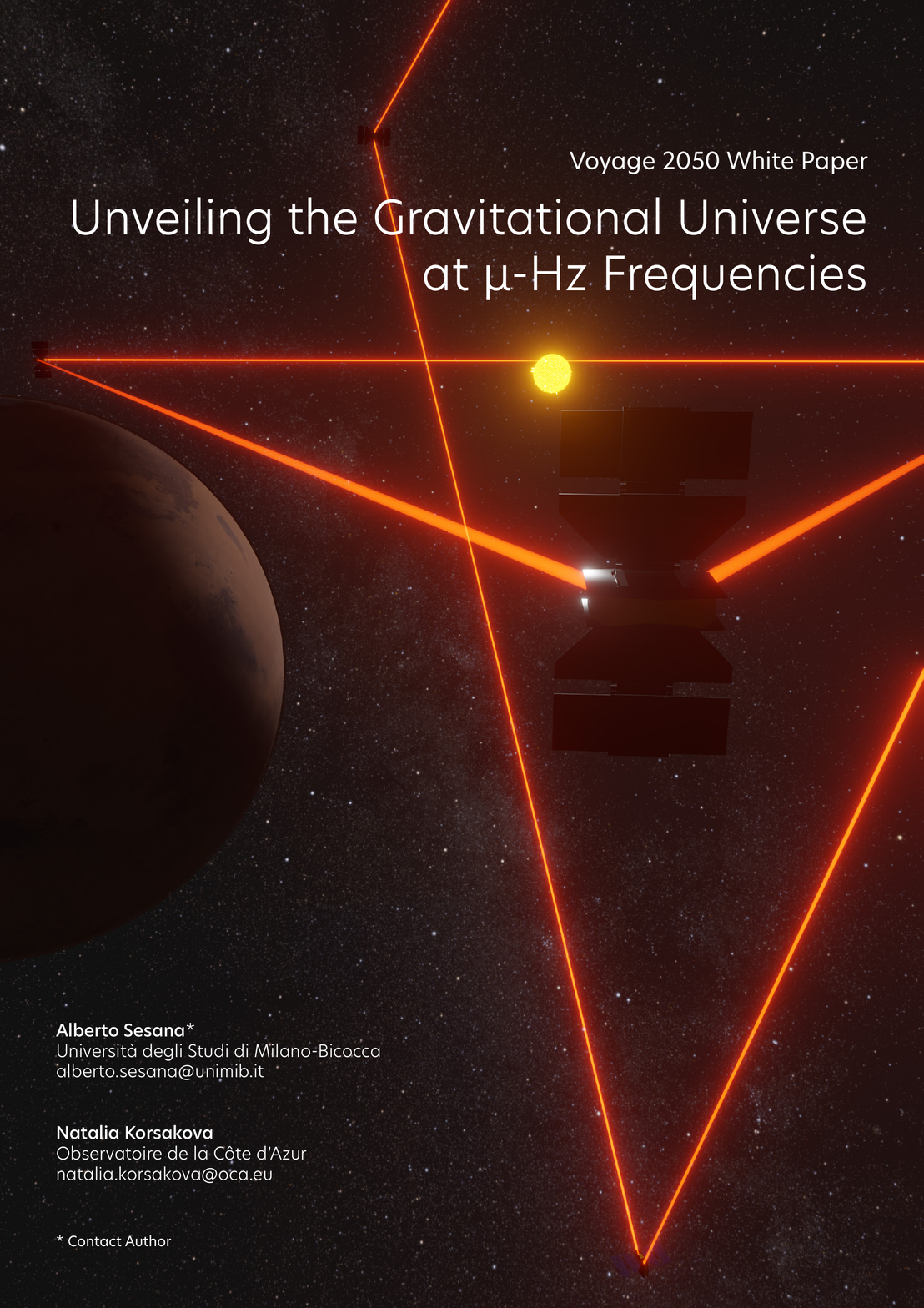}
%
%
%
 \thispagestyle{empty}
 \Large
\noindent\textbf{Members of the proposing team:}
\normalsize 
\\
\\
\begin{small}
\begin{description}
\item
[Manuel Arca Sedda]
Universit\"{a}t Heidelberg, {\it Heidelberg, Germany}
\item
[Vishal Baibhav]
Johns Hopkins University, {\it Baltimore (MD), USA}
\item
[Enrico Barausse]
Scuola Internazionale Superiore di Studi Avanzati, {\it Trieste, Italy}
\item
[Simon Barke]
University of Florida, {\it Gainesville (FL), USA}
\item
[Emanuele Berti]
Johns Hopkins University, {\it Baltimore (MD), USA}
\item
[Matteo Bonetti]
Universit\`{a} degli Studi di Milano-Bicocca, {\it Milan, Italy}
\item
[Pedro~R. Capelo]
University of Zurich,  {\it Zurich, Switzerland}
\item
[Chiara Caprini]
Laboratoire Astroparticule et Cosmologie, {\it Paris, France}
\item
[Juan Garcia-Bellido]
Universidad Aut\'{o}noma de Madrid, {\it Madrid, Spain}
\item
[Zoltan Haiman]
Columbia University, {\it New York (NY), USA}
\item
[Karan Jani]
Georgia Institute of Technology, {\it Atlanta (GA), USA}
\item
[Oliver Jennrich]
European Space Agency, Noordwijk, Netherlands
\item
[Peter Johansson]
University of Helsinki, {\it Helsinki, Finland}
\item
[Fazeel Mahmood Khan]
Institute of Space Technology, {\it Islamabad, Pakistan}
\item
[Valeriya Korol]
Leiden Observatory, {\it Leiden, Netherlands}
\item
[Natalia Korsakova]
Observatoire de la C\^{o}te d'Azur, {\it Nice, France}
\item
[Astrid Lamberts]
Observatoire de la C\^{o}te d'Azur, {\it Nice, France}
\item
[Alessandro Lupi]
Scuola Normale Superiore, {\it Pisa, Italy}
\item
[Alberto Mangiagli]
Universit\`{a} degli Studi di Milano-Bicocca, {\it Milan, Italy}
\item
[Lucio Mayer]
University of Zurich, {\it Zurich, Switzerland}
\item
[Germano Nardini]
University of Stavanger, {\it Stavanger, Norway}
\item
[Fabio Pacucci]
Kapteyn Astronomical Institute, {\it Groningen, Netherlands}
\item
[Antoine Petiteau]
Laboratoire Astroparticule et Cosmologie, {\it Paris, France}
\item
[Alvise Raccanelli]
CERN, {\it Geneva, Switzerland}
\item
[Surjeet Rajendran]
Johns Hopkins University, {\it Baltimore (MD), USA}
\item
[John Regan]
Dublin City University, {\it Dublin, Ireland}
\item
[Alberto Sesana]
Universit\`{a} degli Studi di Milano-Bicocca, {\it Milan, Italy}
\item
[Lijing Shao]
Kavli Institute for Astronomy and Astrophysics, Peking Univeristy, {\it Beijing, China}
\item
[Alessandro Spallicci]
Universit\'{e} d'Orl\'{e}ans, {\it Orl\'{e}ans, France}
\item
[Nicola Tamanini]
Albert Einstein Intitute, {\it Golm, Germany}
\item
[Marta Volonteri]
Institut d'Astrophysique de Paris, {\it Paris, France}
\item
[Niels Warburton]
University College Dublin, {\it Dublin, Ireland}
\item
[Kaze Wong]
Johns Hopkins University, {\it Baltimore (MD), USA }
\item
[Miguel Zumalacarregui]
University of California at Berkeley, {\it Berkeley (CA), USA}
\end{description}
\end{small}

\pagebreak 
\setcounter{page}{1}


\begin{abstract}
  We propose a space-based interferometer surveying the gravitational wave (GW) sky in the milli-Hz to $\upmu$-Hz frequency range. By the 2040s', the $\upmu$-Hz frequency band, bracketed in between the Laser Interferometer Space Antenna (LISA) and pulsar timing arrays, will constitute the largest gap in the coverage of the astrophysically relevant GW spectrum. Yet many outstanding questions related to astrophysics and cosmology are best answered by GW observations in this band. We show that a $\upmu$-Hz GW detector will be a truly overarching observatory for the scientific community at large, greatly extending the potential of LISA. Conceived to detect massive black hole binaries from their early inspiral with high signal-to-noise ratio, and low-frequency stellar binaries in the Galaxy, this instrument will be a cornerstone for multimessenger astronomy from the solar neighbourhood to the high-redshift Universe.  
\end{abstract}

\section{Introduction: the GW landscape post-LISA}

As we enter the era of gravitational wave (GW) astrophysics, the Universe unfolds by revealing the most extreme and energetic events abiding the laws of gravity. In the first and second observing runs, the Laser Interferometer Gravitational-wave Observatory (LIGO) and Virgo detected cosmic whispers from several colliding black hole binaries (BHBs, \cite{TheLIGOScientific:2016pea,LIGOScientific:2018mvr}) and from a neutron star (NS) binary (NSB, \cite{abb17a}), this latter accompanied by spectacular electromagnetic (EM) emission visible at all wavelengths \cite{abb17b}. We now know that BHBs form in great numbers and routinely merge along the cosmic history, and that their dynamics in the strong field is consistent (within measurement errors) with general relativity (GR) \cite{TheLIGOScientific:2016src}. We know that colliding NSs are the engines of short gamma-ray bursts, they give rise to radioactive decay powered kilonovae, and they pollute the interstellar medium with heavy elements (e.g. \cite{2017Natur.551...75S,2017ApJ...848L..27T}). In short, GWs broke onto the stage, bringing the promise of revolutionizing our understanding of astrophysics, cosmology and fundamental physics \cite{blspwh11}.

This revolution will be completed in the next two decades, when observatories on the ground and in space will survey the GW Universe across the frequency spectrum, from the kilo-Hz down to the nano-Hz. In the $0.3$--$10^4$~Hz window, third-generation (3G) ground-based detectors, such as the Einstein Telescope (ET, \cite{2010CQGra..27s4002P}) and Cosmic Explorer (CE, \cite{2019arXiv190704833R}), will detect millions of stellar-origin CO binaries (BHBs, NSBs, and NS-BH binaries) out to $z>10$. At $10^{-4}<f<0.1$~Hz, the Laser Interferometer Space Antenna (LISA, \cite{2017arXiv170200786A}) will: observe the coalescence of massive black hole (MBH) binaries (MBHBs) in the mass range $3\times 10^3<M<10^7$~$\msun$ everywhere in the Universe; probe the population of double white dwarfs (DWDs) in the Galaxy; capture COs slowly inspiralling onto MBHs, mapping out their geometric structure; pierce through the cosmic microwave background (CMB) to probe the physics of the early Universe \cite{2012CQGra..29l4016A,2016PhRvD..93b4003K,Caprini:2015zlo,2016JCAP...04..002T,2016JCAP...12..026B,2017PhRvD..95j3012B,kor17}. At even lower frequencies, pulsar timing arrays (PTAs, \cite{2016MNRAS.458.1267V}) and the square kilometre array (SKA, \cite{2009IEEEP..97.1482D}) will probe the $10^{-9}<f<10^{-7}$~Hz window, unveiling the adiabatic inspiral of the most massive BHs in the Universe, inhabiting the cores of the most massive galaxies at $z<1$ \cite{2015aska.confE..37J,2018ApJ...864..113R}. The panorama will be completed by advanced polarization experiments such as the Probe of Inflation and Cosmic Origins (PICO, \cite{2019arXiv190210541H}) and the Cosmic Origins Explorer (COrE, \cite{2011arXiv1102.2181T}), attempting to probe the B-mode polarization imprinted by a relic stochastic GW background (SGWB) of cosmological origin onto the CMB.

In parallel, new advanced facilities capturing EM radiation in all bands will be operational on the ground and in space. Thirty-meter class optical telescopes, such as the Extremely Large Telescope (ELT, \cite{2007Msngr.127...11G}) and Thirty Meter Telescope (TMT, \cite{2013JApA...34...81S}), will reveal the assembly of the first galaxies out to $z>10$; the next generation of X-ray missions, like the ESA L2 Advanced Telescope for High Energy Astrophysics (Athena, \cite{2013arXiv1306.2307N}) and the proposed NASA Lynx \cite{2017SPIE10397E..0SG}, will probe the emergence of the first quasars; the James Webb Space Telescope (JWST, \cite{2006SSRv..123..485G}) and the Wide Field Infrared Survey Telescope (WFIRST, \cite{2015arXiv150303757S}) will pierce through the first billion years of galaxy formation, and will potentially unveil the nature of dark energy; the SKA itself will probe the ionization history of the Universe with 21cm tomography. Light and gravity will work together to unveil the intimate nature of ultradense dark matter (DM) in NSB mergers, multimessenger observations of NSBs will offer a new way to measure the geometry of the Universe, perhaps we will catch EM signals from merging MBHBs out to moderate redshifts, gaining new insights into the galaxy and MBH assembly processes as well as the interplay between dynamic gravity and relativistic plasma.  

Yet, the foreseeable achievements of the coming two decades of exploration of the Cosmos will inevitably leave unanswered questions and bring new challenges. In the context of the Voyage 2050 program, we explore in this White Paper, the possibility of a $\upmu$-Hz space based GW mission, bridging the gap between the milli-Hz and nano-Hz frequency windows surveyed by LISA and PTAs, respectively. By the 2040s', this will likely be the largest gap in the coverage of the astrophysically relevant GW spectrum, and open questions related to a variety of topics, ranging from the emergence of the $z>7$ quasars to the physics of over-contact binaries, may well find their answer in there. In Section~\ref{sec:sources}, we sketch the sensitivity of a possible detector extending the range of space-based interferometry to $\approx 10^{-7}$~Hz and highlight its most relevant observational capabilities. We refer to those to construct a detailed science case in Section~\ref{sec:science} and present a more detailed strawman mission concept in Section~\ref{sec:strawman}.   

\section{The observational potential of a $\upmu$-Hz space-based detector}
\label{sec:sources}

To support the case for a $\upmu$-Hz detector, we anchor our discussion to a specific heliocentric constellation of satellites, that will be described in more detail in Section \ref{sec:strawman}. The main idea is to place three spacecraft in a nearly Martian orbit, forming an equilateral triangle, thus allowing the construction of two independent Michelson via time delay interferometry (TDI, \cite{2005LRR.....8....4T}), analogue to LISA. Ideally, the instrument will feature a second identical constellation, at an angle with respect to the ecliptic. The rotation of the two constellations in different planes with a 1.8~year period, will allow sky localization via Doppler modulation at a LISA level, if not better. From now on, we will refer to this detector as $\upmu$Ares. The observational potential of $\upmu$Ares described below is based on a 10~year mission duration.

\begin{figure}
\centering
 \includegraphics[scale=0.85]{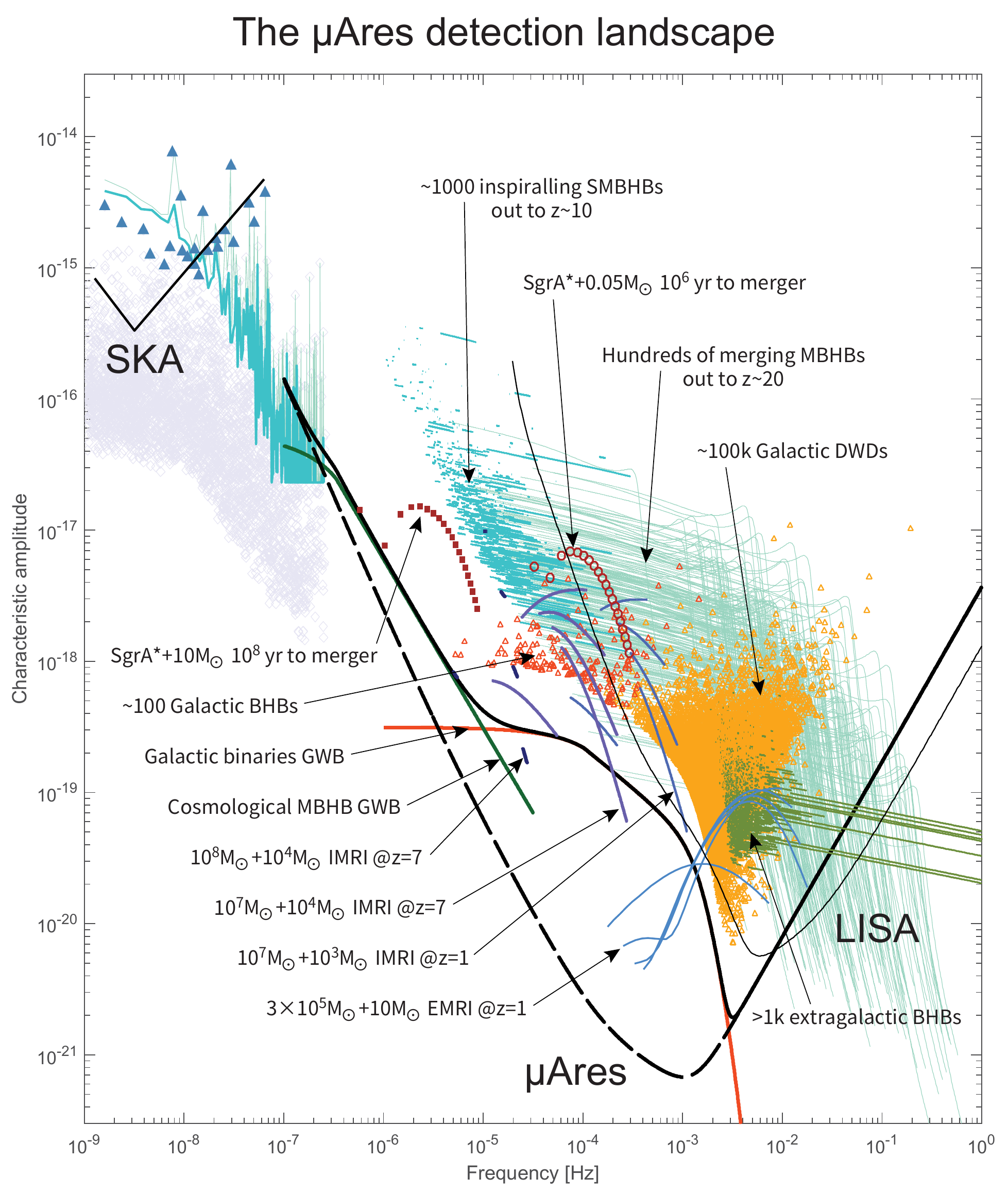} 
\caption{\footnotesize{$\upmu$Ares sky-averaged sensitivity curve (thick black curve; dashed: instrument only; solid: including astrophysical foregrounds), compared to LISA (thin solid black curve) and SKA (solid black line at the top left). Sources in the SKA portion of the figure include individual signals from a population of MBHBs (pale violet), resulting in an unresolved GWB (jagged blue line) on top of which the loudest sources can be individually resolved (dark blue triangles). The vast diversity of $\upmu$Ares sources is described by the labels in the figure. For all Galactic sources (including DWDs, BHBs, and objects orbiting SgrA$^*$), the frequency drift during the observing time has been assumed to be negligible. We thus plot $h\sqrt{n}$, where $n$ is the number of cycles completed over the mission lifetime, assumed to be 10 years. In this case, the signal-to-noise ratio (SNR) of the source is given by the height of its marker over the sensitivity curve. Extragalactic sources (including BHBs, MBHBs, EMRIs, and IMRIs) generally drift in frequency over the observation time. We thus plot the standard $h_{\rm c}=h(f^2/\dot{f})$. In this case, the SNR of the source is given by the area enclosed in between the source track and the sensitivity curve. In both cases, when multiple harmonics are present, SNR summation in quadrature applies.}}
\label{Fig1}
\end{figure}

Figure \ref{Fig1} shows the $\upmu$Ares sensitivity, together with a detailed overview of expected and potential sources. With nearly 400M~km armlength, $\upmu$Ares can achieve a characteristic strain sensitivity of $h_{\rm c}\approx 10^{-18}$ at $f=10^{-6}$~Hz, with more than two orders of magnitude frequency gain compared to LISA. Note that, due to the enhanced laser power (see Section~\ref{sec:strawman}), its sensitivity is only $\approx 3$ times worse than LISA at $f>0.01\,$~Hz, and $\approx 3$ times better around $0.003\,$~Hz, being limited by Galactic DWDs at lower frequencies. Although the milli-Hz range is not the specific focus of this proposal, we note that $\upmu$Ares might in fact outperform LISA `in the bucket', offering as a potential bonus a deeper view onto extreme mass ratio inspirals (EMRIs) and stellar-origin BHBs.

We now enumerate the observational potential of this design, separating sources in Galactic, and extragalactic, proceeding in order of increasing `cosmological distance' to the observer. When population models for specific sources are available, we list expected detection numbers, whereas for more speculative sources we highlight the reach of the detector. 

\vspace{0.3cm}

\par\noindent\rule{\textwidth}{0.4pt}
\noindent {\bf The Milky Way as a GW factory:}
\begin{itemize}[topsep=0pt,itemsep=-1pt,partopsep=0pt, parsep=0pt]
\item ${\cal O}(10^5)$ resolved Galactic DWDs, down to $f\approx 10^{-4}$~Hz (model from \cite{lam19});
\item ${\cal O}(100)$ resolved Galactic BHBs, down to $f\approx 10^{-5}$~Hz (model from \cite{2018MNRAS.480.2704L});
\item thousands mixed (CO + main sequence star), contact, and over-contact binaries, mostly in the unexplored $10^{-5}$--$10^{-4}$~Hz frequency range;
\item COs with masses down to $\approx10^{-3}$~$\msun$ orbiting SgrA$^*$ out to distances of several AU, including up to ${\cal O}(100)$ stellar BHs and many more brown dwarfs.
    \end{itemize}  
\vspace{0.2cm}
    {\bf Astrophysical sources, from the local Universe to the dawn of galaxy formation:}
    \begin{itemize}[topsep=0pt,itemsep=-1pt,partopsep=0pt, parsep=0pt]
\item ${\cal O}(10^3)$ extragalactic BHBs, improving by an order of magnitude over LISA;
\item EMRIs around MBHs in the mass range $10^5$--$10^6$~$\msun$ out to $z>3$;
\item intermediate mass ratio inspirals (IMRIs) involving $10^3$--$10^4$~$\msun$ intermediate mass black holes (IMBHs) inspiralling onto $10^7$--$10^8$~$\msun$ MBHs out to $z\gtrsim 6$;
\item $\approx 50$ inspiralling MBHBs with $M_1<10^{6}\msun$ in dwarf galaxies at $z<0.5$ (model from \cite{2019MNRAS.486.4044B});
\item ${\cal O}(100)$ merging MBHBs along the cosmic history with high SNR (model from \cite{2019MNRAS.486.4044B});
\item ${\cal O}(10^3)$ inspiralling MBHBs at $z<10$, some of them caught thousands of years from coalescence (model form \cite{2019MNRAS.486.4044B});
\item bursts from $M>10^6$~$\msun$ direct seed formation out to high redshift.  
    \end{itemize}  
\vspace{0.2cm}
   {\bf Stochastic signals; astrophysical foregrounds and cosmological backgrounds:}
   \begin{itemize}[topsep=0pt,itemsep=-1pt,partopsep=0pt, parsep=0pt]
\item characterization of the DWDs confusion noise over two decades of frequencies, with SNR$>1000$ and detection of confusion noise from contact binaries with periods ranging from few days to few hours in the $10^{-5}$--$10^{-4}$~Hz frequency range;
\item unresolved foreground arising from unresolved MBHBs at $10^{-7}<f<10^{-5}$~Hz;
\item unresolved foregrounds arising from extragalactic BNSs and BHBs (similar to LISA);
\item cosmological SGWBs down to  $h^2\Omega_{\rm gw}\sim 6 \times 10^{-17}$ at $2 \times 10^{-4}$~Hz (nearly four orders of magnitude more sensitive than LISA).  
   \end{itemize}
\vspace{0.2cm}
{\bf Early source localization.}   
Besides detection at high SNR, extending the sensitivity to low frequencies crucially implies that merging MBHBs will be detected early in their inspiral and localized in the sky well before merger. Assuming two triangular non-coplanar constellations, {\it $\upmu$Ares will localize merging SMBHs at $z<5$ to better than 10~deg$^2$ a year before final coalescence} (see Figure~\ref{fig:skyloc} and discussion in Section~\ref{sec:multimulti}), opening avenues in multimessenger astronomy which will still be impracticable with LISA.  
\par\noindent\rule{\textwidth}{0.4pt}

\section{Science goals}
\label{sec:science}
With a general overview of the detector sensitivity and of the observable sources, we now describe the core science enabled by $\upmu$Ares-like detector. We define five main themes:
    \begin{enumerate}[topsep=0pt,itemsep=-1pt,partopsep=0pt, parsep=0pt]
    \item the MBH Universe;
    \item multimessenger and multiband view;
    \item general relativity and beyond;
    \item cosmology and cosmography;
    \item the Milky Way.  
    \end{enumerate}  
The following subsections describe in detail individual specific themes, with the main science goals summarized in the boxes at the end of each subsection.

\subsection{The massive black hole Universe}

\subsubsection{The emergence of high redshift quasars}

Quasars (QSOs) are among the brightest sources of EM radiation in the Universe and they are currently observed as far back as $z=7.5$ (\cite{2018Natur.553..473B}, see also \cite{2006AJ....132..117F,2011Natur.474..616M,2015Natur.518..512W}). These sources are powered by supermassive black holes (SMBHs)\footnote{Massive and supermassive black holes (MBHs and SMBHs) are interchangeably used in the literature to refer to black holes with $M> 10^6$~$\msun$. Here we use SMBH only when referring to the $M>10^9$~$\msun$ systems powering quasars. We instead use MBH and MBHB when referring to those systems as GW sources for $\upmu$Ares.} on typical mass scales of $10^9$--$10^{10}$~M$_\odot$, characterized by luminosities up to $10^{47}$--$10^{48}$~erg~ s$^{-1}$, and are currently observed at cosmic epochs up to about 800~Myr after the Big Bang, while theoretical models (e.g. \cite{2001PhR...349..125B})  and observations (e.g. \cite{2018Natur.555...67B}) suggest that the first stars and black holes started forming 100--200~Myr after the Big Bang. It is thus very remarkable that the mass assembly of the first MBHs occurred very rapidly, in less than $\sim 600$~Myr. Although the direct detection of the first stars and black holes is not yet feasible, it is a primary goal for any future surveys. In fact, observing these primordial objects is a crucial step towards constraining the properties of the very early Universe (see, e.g. \cite{2019BAAS...51c.557H,2019BAAS...51c.117P}).

The MBHs powering these quasars should have grown from initial ``seeds'', formed at $z=10$--30, and their assembly histories over cosmic time is the end result of many different mechanisms, often acting at the same time: seeding, accretion, and mergers (see Section \ref{sec:seeds}). Hence, the accelerated assembly process needed for the high-redshift QSOs must rely on enhancing either the initial seed mass, the growth rate, or the merger rate. In either case, non-conventional physical mechanisms are required: either an accelerated growth of BHs, well above the Eddington limit (e.g. \cite{1979MNRAS.187..237B,2005ApJ...633..624V,2015MNRAS.448..104P,2016MNRAS.459.3738I,2016MNRAS.456.2993L}),  the emergence and eventual collapse of a supermassive star, formed as a result of unusually rapid accretion \cite{2012ApJ...756...93H,2018MNRAS.474.2757H,2018MNRAS.478.5037R}, an extremely massive black hole seed formed by relativistic ``dark collapse'' of a supermassive cloud by-passing the supermassive star stage \cite{2015ApJ...810...51M}, a major contribution from MBHB mergers despite ejections caused by anisotropic emission of GWs \cite{2004ApJ...613...36H}, or formation channels with much more massive seeds related to topological defects or large-curvature perturbations in the early Universe~\cite{2015JCAP...06..007B,2018MNRAS.478.3756C} (see also Section~\ref{subsec:multimultiMBHB}).

More generally, the complementarity of accretion and mergers extends beyond the exceptional, but rare, high-redshift quasars. All MBHs in the Universe are expected to experience both, and to grow through these processes. The crucial question is: what is the most relevant process as a function of cosmic time and environment? The contribution of accretion can be traced by EM observations, converting the emitted energy  into the mass that created such energy: this is So\l{}tan's argument \cite{1982MNRAS.200..115S}. GW observations detect mergers of black holes and thus measure their contribution. Only the powerful combination of these two probes can provide a complete census of how MBHs obtained their mass. Theoretical models suggest that the growth by MBHB mergers becomes more important at low redshifts and for masses higher than $10^8$~M$_\odot$, as cold gas dwindles in the most massive galaxies \citep{2007MNRAS.382.1394M,2014MNRAS.440.1590D,2015ApJ...799..178K}.

\begin{figure*}
 \begin{minipage}[b]{0.45\textwidth}
   \centering
   \includegraphics[scale=0.12]{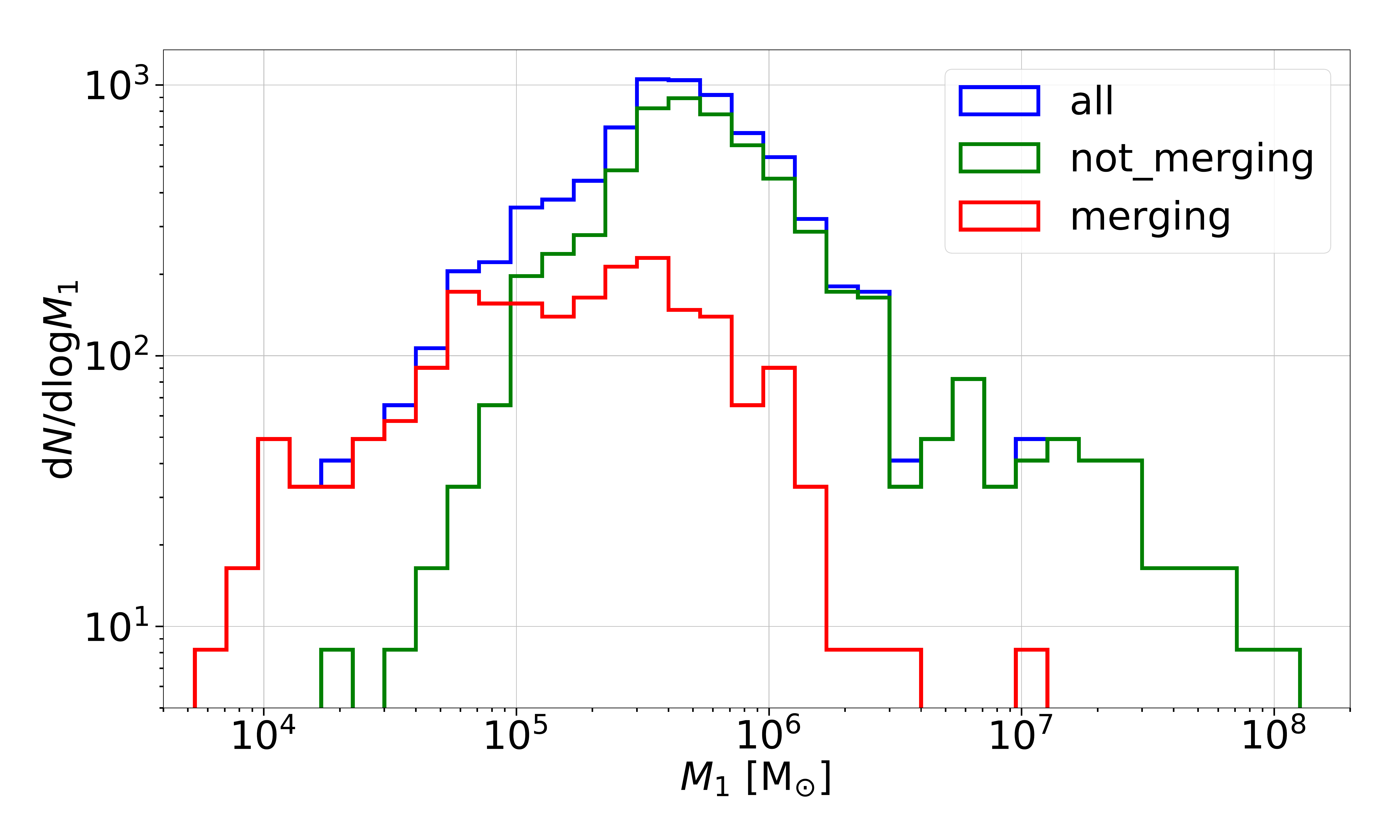}
 \end{minipage}
 \ \hspace{3mm} \
 \begin{minipage}[b]{0.45\textwidth}
  \centering
   \includegraphics[scale=0.12]{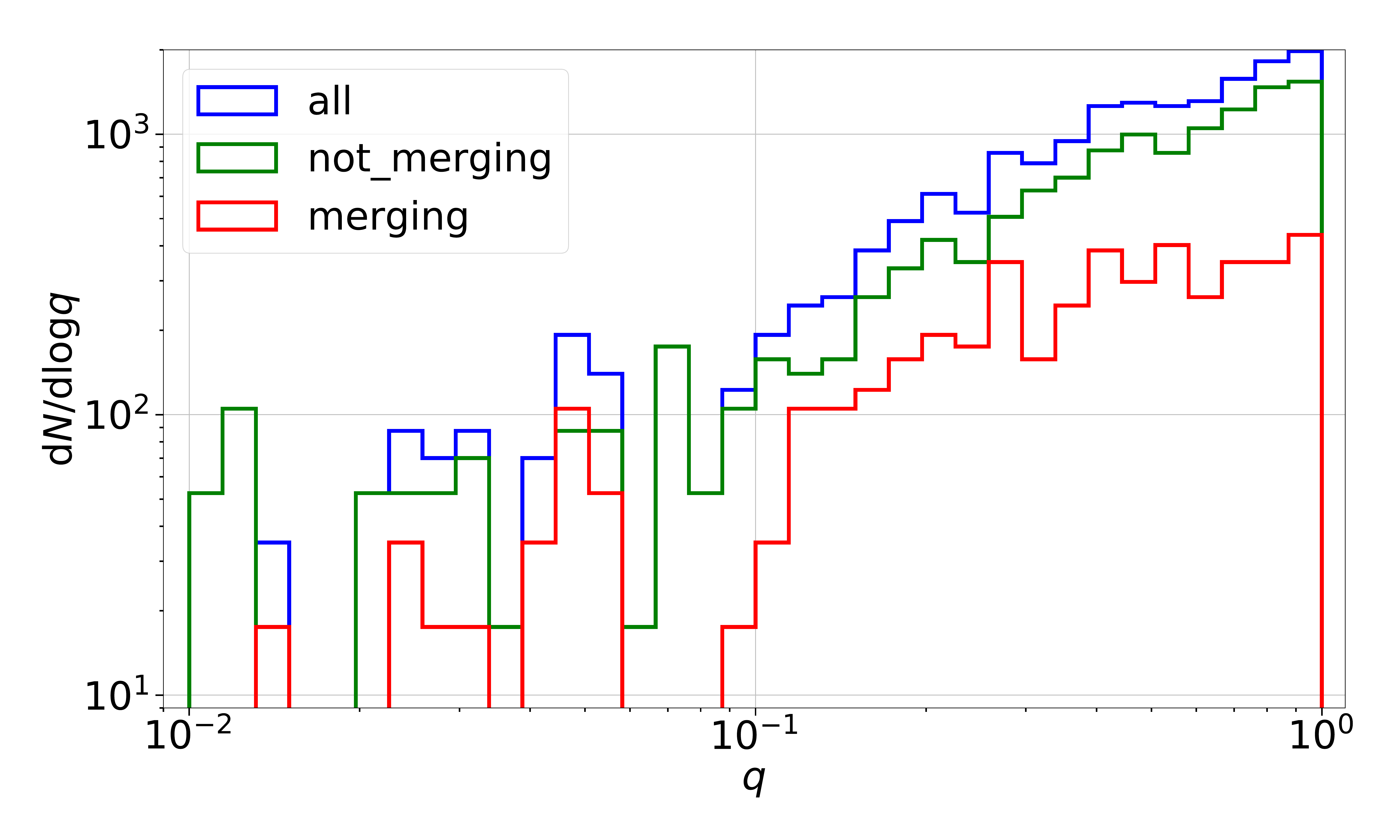}
 \end{minipage}
 
 \begin{minipage}[b]{0.45\textwidth}
   \centering
   \includegraphics[scale=0.12]{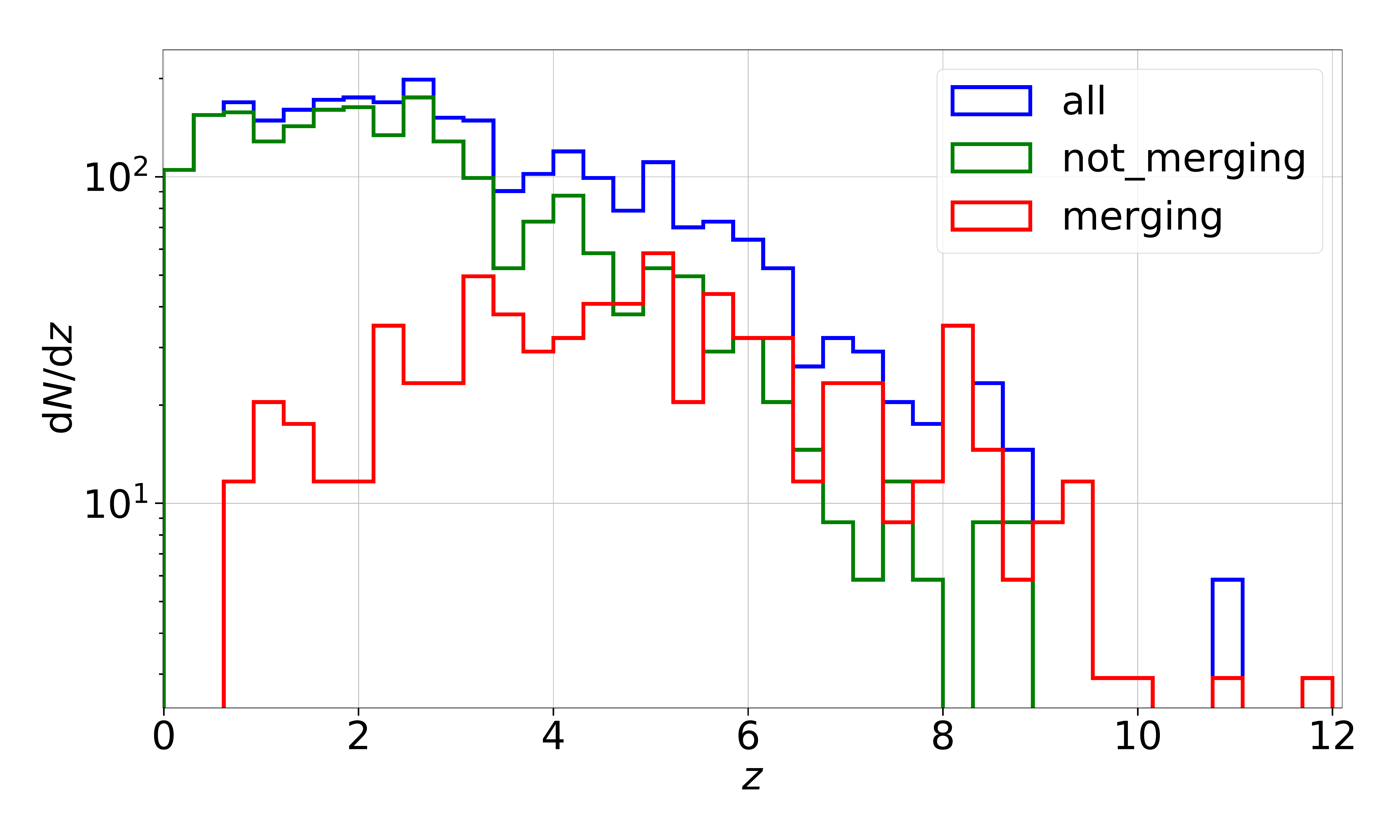}
 \end{minipage}
 \ \hspace{3mm} \
 \begin{minipage}[b]{0.45\textwidth}
  \centering
   \includegraphics[scale=0.12]{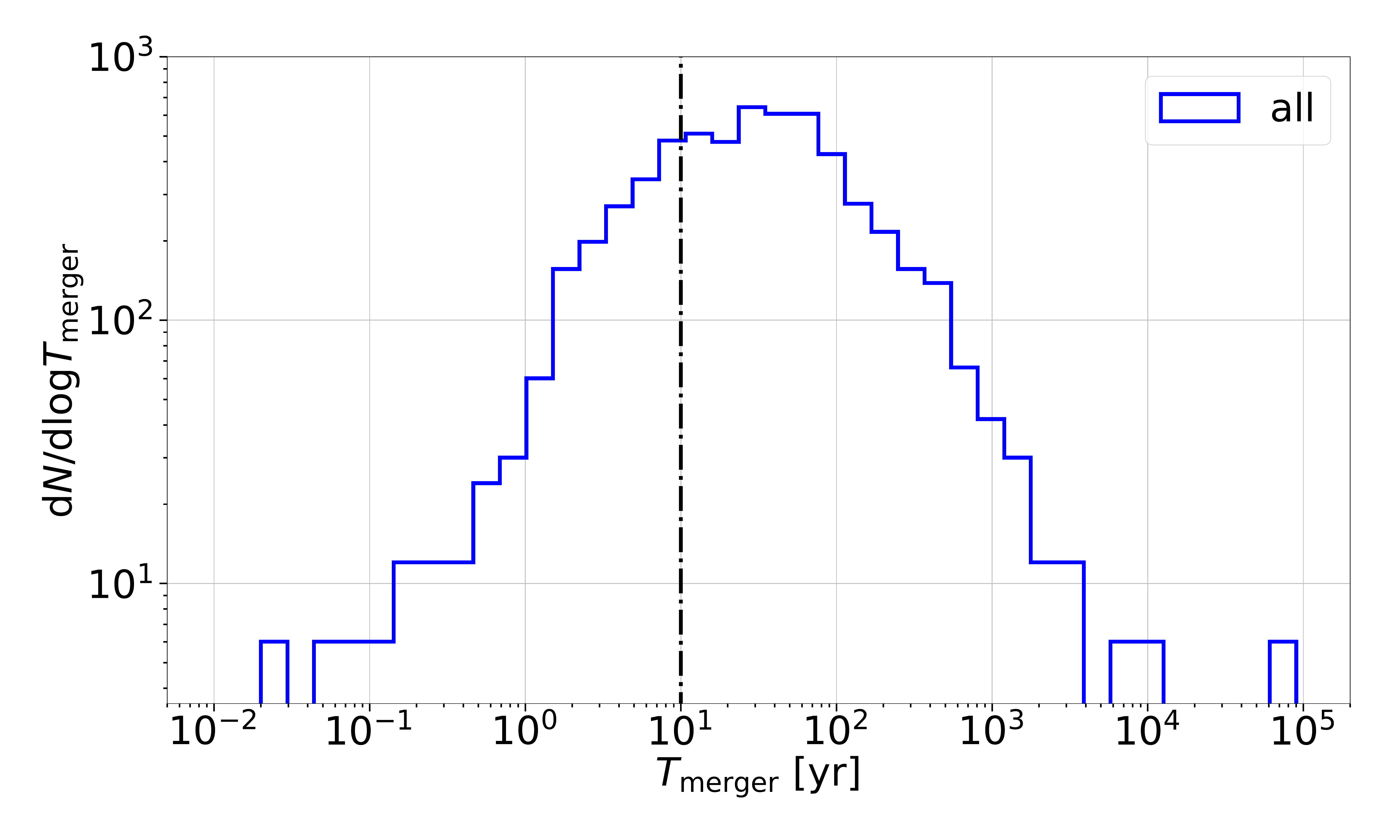}
 \end{minipage}
 \caption{\footnotesize{Example distributions of MBHBs detected with SNR $>8$ by $\upmu$Ares, assuming 10 years of observations. The sample comes from model {\it HS-delayed} of \cite{2019MNRAS.486.4044B}. A total of $\approx 10^3$ sources are detected (blue histograms); of those, $\approx 250$ merge in the detector band within the mission lifetime (red histograms) and $\approx 750$ are observed during the adiabatic inspiral over the whole mission duration (green histograms).}}
\label{fig:mbhbpop}
\end{figure*}

It is fundamental to extend the observing capabilities in GWs to cover the mass and redshift range between LISA and PTA: this range corresponds to the peak of the quasar epoch and probes the golden age of MBH growth. In fact, disentangling different growth mechanisms is a very difficult task, as their imprint is  blurred as time elapses.  By extending the observation window down to the $\upmu$-Hz range, {\it not only $\upmu$Ares will detect mergers of MBHBs at high SNR at $z \sim 7$--$15$, but it will also provide a large statistical sample of ${\cal O}(10^3)$ inspiralling MBHBs out to $z\approx 10$} (see Figure~\ref{fig:mbhbpop}).  This will allow us to infer the mass distribution of MBHs, as a function of redshift, when the signature of the different scenarios is dominant. For instance,  super-Eddington accretion onto light seeds, as opposed to Eddington-limited accretion on seeds that start 3--4 orders of magnitude larger in mass, could reveal itself through a different evolution of the high end of the mass distribution at such early epochs. Conversely, many of these scenarios, since they are designed to grow MBHs very quickly above $10^6$--$10^7$~M$_\odot$, would not be probed, and thus will not be disentangled, at the higher detection frequencies at which LISA will work. Hence, the power of a low-frequency GW detector is in its ability to probe the demographics of MBHs across masses and redshifts all the way up to the masses of high-redshift QSOs.

Host galaxy identification can provide even deeper insights into the emergence of QSOs and MBHs. In fact, the hosts of the $10^6$--$10^7$~M$_\odot$ MBHs are very dissimilar in different seeding models -- in particular, in ``stellar-seed'' models, the hosts at this BH mass stage are typically well-developed $10^{11}$--$10^{12}$~M$_\odot$ galaxies \cite{2017MNRAS.468.3935H}, whereas in the ``direct collapse'' models, the heavy seeds form in $10^7$--$10^8$~M$_\odot$ DM halos, which, after an order of magnitude growth in BH mass, still typically remain $<10^9$~M$_\odot$ sized dwarf galaxies \citep{2018ApJ...865L...9V}. Although LISA might provide enough information to identify the counterparts of relatively massive binaries at $z\lesssim 3$, {\it the superior low-frequency performance of $\upmu$Ares will greatly enhance EM-GW synergies}, as we further describe in Section~\ref{sec:multimulti}.



\subsubsection{Testing massive black hole formation scenarios}
\label{sec:seeds}

MBHs have grown from initial ``seeds'', formed at $z=10$--30, usually categorized as light (up to $10^2$~M$_\odot$ and formed as stellar remnants \cite{2001ApJ...551L..27M}) and heavy (up to $10^4$--$10^6$~M$_\odot$, formed via the collapse of a supermassive star or the runaway collapse of a dense star cluster -- see, e.g. \cite{2018arXiv181012310W} for a review) seeds.  Other, more exotic possibilities for the initial seeds range from primordial black holes (PBHs) (e.g. \cite{2002PhRvD..66f3505B}) to DM-powered massive stars \cite{2016RPPh...79f6902F}. The subsequent mass assembly is likely the result of a complex interplay between accretion and mergers, as mentioned above. Although LISA (in combination with 3G detectors) will be already well placed to disentangle several `standard' seeding models (e.g. the `Pop~III stars vs direct collapse' dichotomy, see \cite{2011PhRvD..83d4036S}), $\upmu$Ares will provide further insights onto alternative scenarios, involving particularly high-mass/high-redshift seeds that might be especially relevant to the formation of high-redshift QSOs.  



\paragraph{Bursts via direct collapse.}

One of the ways by which MBH seeds might form is via direct gas collapse. There are various pathways to BH formation in direct collapse \cite{2018arXiv181012310W}, the common aspect being that (1)  no conventional star (Pop~III) forms and (2) the resulting black hole seed is massive ($> 10^3$--$10^4$~$\msun$, with the upper limit highly dependent on the specific scenario). Some of the models involve the formation of  supermassive protostars/stars ($M > 10^5$~$\msun$) which later collapse into a BH once hydrostatic equilibrium cannot be maintained any longer or once the pulsational relativistic radial instability sets in \cite{2013ApJ...778..178H,2018MNRAS.474.2757H}. Early analytical work by \cite{1966ApJ...144..180F}, for example, showed that the progenitor mass must be $> 10^6$~$\msun$ in presence of rotation, which has been confirmed by further theoretical considerations \cite{2018arXiv181012310W}. In an even more `extreme' scenario, a $> 10^6$~$\msun$ MBH seed can directly form by relativistic collapse of a supermassive cloud formed following a gas-rich galaxy merger \cite{2015ApJ...810...51M}.

While for stellar progenitors up to $10^6$~$\msun$ the signal should be within the detection capabilities of LISA, for larger masses of the progenitors  the signal shifts to frequencies below $10^{-3}$~Hz, and the strain drops below $10^{-19}$ if the source is at $z > 6$ \cite{2009PhRvD..80f4001S}. Given that one needs to be able to detect such signals at very high redshift, when BH seeds should form, it follows that higher sensitivity at and below the lowest frequencies accessible to LISA will be needed.  In particular, the emergence of supermassive stars, which undergo direct collapse to form MBHs, requires special conditions of rapid mass inflow in pristine, very metal-poor gas.  This phase likely occurred very early on (with the typical such event at $z=15$--20) and will make a detection by LISA challenging.

An even richer sequence of signals might arise if, as suggested by some numerical relativity work \cite{2013PhRvL.111o1101R}, the progenitor undergoes a non-axisymmetric global instability (e.g. a bar instability) just before the radial instability becomes the dominant mode. In this case, like in standard protostellar collapse, fragmentation into two or more supermassive stars would arise, which would produce a collapse signal (burst + ringing) followed by a short duration inspiral + merger signal, as at least one binary might form at such small separation that merger will be prompt. Again, since this will be more likely for high progenitor masses, low-frequency sensitivity should be beneficial also for the detection of the inspiral + merger signal.

\paragraph{Build-up from  primordial black holes through EMRIs/IMRIs.}

The same inflationary mechanism that generates large-scale fluctuations in the CMB and the large-scale structure of the Universe (galaxies and clusters) also generates fluctuations on small scales. Some inflationary dynamics predict that thirty e-folds before the end of inflation large curvature fluctuations were generated, which then collapsed upon horizon re-entry to form black holes during the radiation era \cite{GarciaBellido:1996qt}. These PBHs are made out of photons which cannot escape the gravitational collapse of large-curvature perturbations generated from quantum fluctuations during inflation. They have masses ranging from values typical of planets ($10^{-6}$~$\msun$) to MBHs ($10^{6}$~$\msun$), and may have arisen due to sudden changes in the relativistic fluid's radiation pressure, as particles decouple or condense throughout the thermal history of the Universe \cite{2019arXiv190608217C}.


PBHs accreting from the intergalactic medium (IGM) would distort the temperature and polarization anisotropies of the CMB. This limits the total mass density in PBHs in the $10^5$--$10^6$~$\msun$ range to a small fraction of the DM density \cite{2017PhRvD..96h3524P,2017PhRvD..95d3534A}. However, depending on the details of the geometry of the accretion, the limit allows a PBH density up to that of present-day MBHs (e.g. \cite{2002MNRAS.335..965Y}).  Therefore, massive PBHs remain viable as seeds of rare early quasars. In the matter era, these MBHs act as seeds for structure formation \cite{Clesse:2015wea,2018MNRAS.478.3756C}. They could form disks very early and grow via gas accretion to gigantic sizes ($10^{9}$--$10^{11}$~$\msun$) during the age of the Universe. Their growth occurs mainly through gas accretion, but there is a possibility that they merge to form massive clusters of PBHs with a wide mass distribution, segregated in mass with the most massive at the centre due to dynamical friction, and the least massive orbiting or even evaporating from these clusters and thus constituting a diffuse uniform component of DM. In that case, one expects a large rate of hyperbolic encounters \cite{Garcia-Bellido:2017knh} as well as multiple EMRIs and IMRIs, at high redshifts, which are primary targets of the sub milli-Hz frequency band probed by $\upmu$Ares (see IMRI tracks in Figure~\ref{Fig1}). 


\paragraph{Population of MBHs in low-redshift dwarf galaxies.}

While MBHs are found to inhabit all massive galaxies, both at low and high redshift, their presence in low-mass systems has not been confirmed yet, with only some candidates detected to date as low-luminosity active galactic nuclei (AGN, e.g. \cite{2013ApJ...775..116R,2018MNRAS.478.2576M}). These objects represent the low-mass end of the MBH luminosity function, with masses of about $10^4$--$10^5$~$\msun$, and perfectly sample the halo mass regime where different seeding scenarios predict strong variations in the BH halo occupation fraction (e.g. \cite{2009MNRAS.400.1911V,2011ApJ...742...13B,2012NatCo...3.1304G}). 
\begin{wrapfigure}{R}{0.4\textwidth}
  \vspace{-0.5cm}
  \centering
    \includegraphics[width=0.4\textwidth]{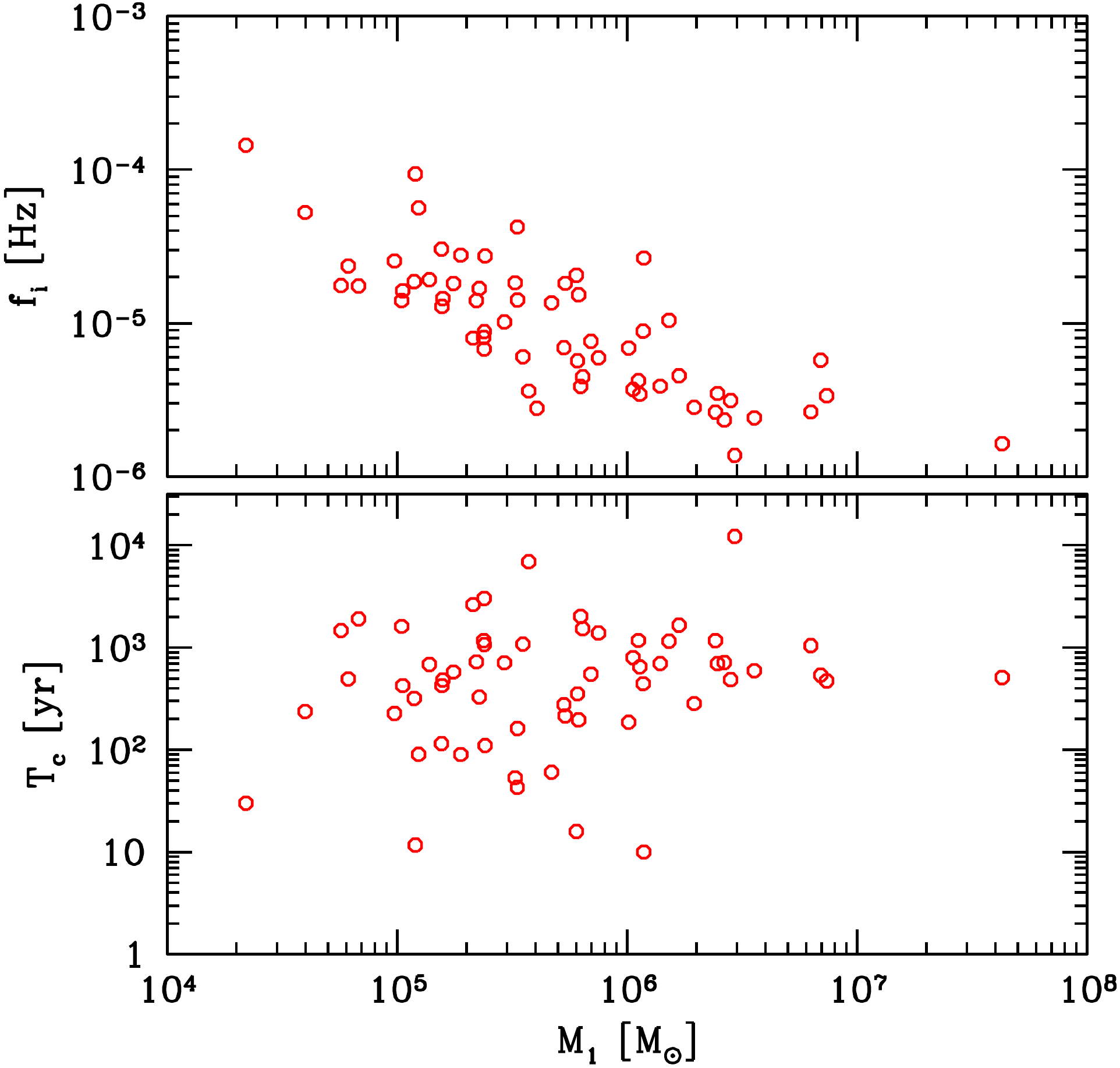}
  \vspace{-0.5cm}
  \caption{\footnotesize{Primary MBH mass vs observed frequency (top) and time to coalescence (bottom) for an illustrative local ($z<0.5$) inspiralling MBHBs population detected by $\upmu$Ares with SNR$>8$, from model {\it HS-delayed} of \cite{2019MNRAS.486.4044B}.}}
  \label{fig:localMBHBs}
\end{wrapfigure}

Because of the typically low gas densities and the low accretion rates onto the MBH in low-mass galaxies, those systems likely keep their memory of the initial seeding process, and their mass distribution nowadays would closely resemble the initial one. However, because of their mostly quiescent life, their detection via EM radiation would be very challenging. A unique opportunity to reveal their presence is represented by GW emission when they inspiral and coalesce with similar or more massive companions. Recently, \cite{2018ApJ...864L..19T} investigated the MBH pairing in dwarf galaxy mergers by means of extremely high-resolution numerical simulations, finding that the DM distribution plays a crucial role in the MBH evolution, with cored profiles resulting in the MBHs stalling at a few hundred pc separations. Nevertheless, we are still far away from a clear picture, especially because of the missing gas (and star formation) physics, that could strongly affect the inspiral (e.g. \cite{2013ApJ...777L..14F,2014MNRAS.439..474V,2015MNRAS.453.3437L,2017MNRAS.464.2952T,2017ApJ...838...13S}).

Given the relatively quiet evolution of dwarf galaxies, it is quite likely that the overall cosmic merger rate of the MBH they host is $\ll 1\,$yr$^{-1}$ at low redshifts, thus escaping detection at milli-Hz frequencies. Conversely, {\it $\upmu$Ares increased sensitivity at low frequency will allow detection of equal-mass inspiralling $10^5$~$(10^4)$~$\msun$ binaries at $z=0.1$ already $10^4$~$(10^3)$~years before the merger}, offering the unique opportunity of detecting several such systems in the local Universe even if they merge at a rate that is $\ll 1\,$yr$^{-1}$. A practical example is shown in Figure~\ref{fig:localMBHBs}. In this specific population model, $\approx50$~$(10)$ inspiralling MBHBs with primary mass $<10^5$~$(10^4)$~$\msun$ are detected at $z<0.5$. Note that the coalescence time is $>10$ years for all of them and none would be detected by LISA. The detection of such population will provide crucial information on the number of MBHBs in dwarf galaxies, putting constraints on the main seeding scenarios. Accurate measurements of the GW phase evolution will also allow to test the presence of DM spikes/cusps, which are predicted by cold DM models to gradually build-up undisturbed over an Hubble time in the quiet surroundings of MBHs hosted in dwarf galaxies \cite{1999PhRvL..83.1719G}.

\subsubsection{The physics of MBHB pairing}

The detailed physics of MBHB pairing is in itself very uncertain. It is generally accepted that following galaxy mergers, MBHs are delivered to the centre of the merger remnant by dynamical friction \cite{1943ApJ....97..255C}, eventually forming a Keplerian binary on $\sim$pc scales (\cite{1980Natur.287..307B}; for recent works, see, e.g. \cite{2009ApJ...696L..89C,2015MNRAS.447.2123C,2017MNRAS.471.3646P,2018ApJ...868...97K}). Further hardening of the binary, down to $\sim$millipc scales, is required to reach final coalescence via GW emission. The underlying physics driving this process is poorly understood (see \cite{2012AdAst2012E...3D}, for a review). In recent years, three-body scattering against the stellar background has been established as a viable route to coalescence (e.g. \cite{2011ApJ...732...89K,2011ApJ...732L..26P,2017ApJ...840...53R}). Similarly, in gas-rich environments, the formation of a massive circumbinary disk can extract energy and angular momentum from the binary, driving it to final coalescence (e.g. \cite{2009MNRAS.393.1423C}.) If the aforementioned mechanisms fail, subsequent galaxy mergers will eventually bring a third MBH into play, prompting the formation of MBH triplets, that have been shown to drive the pre-existing binary to the GW emission stage in a large ($\approx 30\%$) fraction of cases \cite{2010MNRAS.402.2308A,2018MNRAS.477.3910B,2018MNRAS.473.3410R}. MBHB pairing and coalescence can also be prompted by the inspiral of IMBHs hosted in star and globular clusters \cite{2006ApJ...641..319P}. The interplay and relative importance of the aforementioned mechanisms is hardly known, leaving a number of outstanding questions unanswered.

{\it How efficient are stellar and gas dynamics in driving the binary to coalescence?} Stellar hardening critically depends on the refilling rate of the binary loss cone and, especially in massive galaxies featuring low density cores, can proceed on characteristic time-scales up to Gyrs \cite{2015ApJ...810...49V,2015MNRAS.454L..66S,2018ApJ...864..113R}. On the other hand, MBHBs evolving in spirals and dwarf elliptical galaxies are expected to merge efficiently within few hundred Myrs due to rotation and higher central stellar densities of their hosts \cite{1803.11394}. When it comes to gas dynamics, even the understanding of the detailed physics establishing the sign of the gas torques exerted by the disk (i.e. whether the binary inspirals or outspirals), is incomplete. In fact, the net torque arises from a small asymmetry in the shape of the gas distribution near the edges of the minidisks ahead of and behind the MBHs \cite{2012A&A...545A.127R}, which are hard to resolve properly. 
According to recent studies, the binary inspiral time can be of a few million year for disk densities corresponding to an accretion rate normalized to 0.3 times the Eddington value \citep{2017MNRAS.469.4258T}.
However, given the dependence of this result on the small asymmetry in the gas distribution near the BHs, one has to keep in mind that 3D effects, radiation pressure, winds, etc. are likely to modify this time-scale significantly. Likewise, apart from additional physical effects, the subtlety of the torques also make them susceptible to numerical issues.  Several works recently found positive torques, resulting in widening, rather than shrinking, the MBHB orbit 
\citep{2017MNRAS.469.4258T,2019ApJ...871...84M,2019ApJ...875...66M}.

{\it How often do triple (and multiple) systems form and what is their role in the MBH build-up?} This depends critically on the ratio between the typical time elapsed in between galaxy mergers and the lifetime of MBHBs. If binary shrinking time-scales are long, formation of multiple MBH systems might be the norm, with consequences for galactic core scouring and MBH ejection in the galaxy outskirts and IGM \cite{2018MNRAS.473.3410R,2019MNRAS.486.4044B}. 

{\it Can IMBHs be effectively delivered to the centre of massive galaxies and what is their role in MBH build-up?} Galaxies are known to host hundreds --  Milky-Way (MW) sized -- to up to tens of thousands -- massive ellipticals -- star clusters. According to numerical simulations, up to 20\% of such clusters could harbour an IMBH \cite{2015MNRAS.454.3150G}. Clusters forming in the inner region of a galaxy, $\sim$1~kpc, can spiral toward the galactic centre via dynamical friction \cite{1975ApJ...196..407T}, releasing the central IMBH in the dense galactic nuclei. This mechanism has been advocated to explain MBH seeding and growth \cite{2001ApJ...562L..19E}. This scenario predicts the unavoidable formation of binaries, triplets, or even multiplets comprised of a MBH and several IMBHs (e.g. \cite{2014ApJ...796...40M,2018MNRAS.477.4423A,2019MNRAS.483..152A}). Clearly, the number of IMBHs that can get close to a MBH depends critically on several assumptions, like the clusters formation efficiency, the IMBH formation probability, and the galaxy structure. For a MW-sized galaxy, one can expect that, out of ${\cal O}(100)$ clusters, only $\sim$20--30 have the conditions to seed an IMBH \citep{2019arXiv190500902A}, and only 1--2 can reach the centre, undergoing efficient merger within $\sim$1~Gyr. In massive elliptical galaxies, instead, the number of IMBHs that can be packed in the centre of a galaxy can be as high as ${\cal O}(10$--$100)$. Massive galaxies are expected to host MBHs, with $M>10^7\,\msun$. Therefore, these systems might host the formation of MBH-IMBH pairs with a mass ratio $q<10^3$ . These EMRI/IMRI-like systems can form at relatively low redshift, and can constitute loud GW sources in the $10^{-5}$--$10^{-4}$~Hz frequency band.

{\it A $\upmu$-Hz GW detector like $\upmu$Ares is best placed to answer these questions, by catching MBHBs already during the early adiabatic inspiral (bottom right panel of figure~\ref{fig:mbhbpop}), where their coupling with the environment can still leave distinctive signatures}. First, each shrinking mechanism is expected to produce different eccentricity distributions \cite{2012JPhCS.363a2035R,2019MNRAS.486.4044B}, which can be easily measured during the inspiral with $10^{-4}$ precision \cite{2016PhRvD..94f4020N}. Second, in the case of gaseous drag, it might be possible to directly measure the effect of gas drag on the GW waveform. This should be the case especially for stellar-mass BHs or IMBHs merging with a MBH, for which the lighter companion results in a lower chirp mass and weaker GW torques, but stronger gas torques \cite{2019MNRAS.486.2754D}. At the high SNR measurements expected from $\upmu$Ares, the magnitude and frequency-dependence of the deviation from vacuum waveforms can be easily disentangled from uncertainties in the MBH binary parameters, allowing precise measurement of the effects of dynamical friction, disk migration and accretion \cite{2011PhRvD..84b4032K,2014PhRvD..89j4059B}.
Finally, MBH-IMBH of $10^8\,\msun+10^4\,\msun$ at $z<2$ can be in principle detected with high SNR thousands of years prior to coalescence (see tracks in figure~\ref{Fig1}). This leaves the intriguing possibility of observing directly in the waveform the complex dynamics generated by the inspiral of multiple IMBHs into a SMBH.


\vspace{0.5cm}
\fbox{\begin{minipage}{40em}
    {\bf Summary of MBH science goals:}
    \begin{itemize}[topsep=0pt,itemsep=-1pt,partopsep=0pt, parsep=0pt]
    \item probe the emergence of the high-redshift quasars;
    \item establish the relative importance of accretion vs mergers in growing MBHs;  
    \item disentangle seed formation models at the high-mass end of the seed spectrum;
    \item probe the population of inspiralling MBHs in low-redshift dwarf galaxies;
    \item pin down the physics of MBHB dynamics, including stellar hardening, gaseous drag, triplets, and multiplets MBH interactions;
    \item probe the formation and dynamics of IMBHs in galactic nuclei.
    \end{itemize}  
\end{minipage}}

\subsection{Multimessenger and multiband view}
\label{sec:multimulti}

One of the major strengths of a $\upmu$-Hz GW detector lies in its unparalleled potential for multimessenger and multiband science involving BHBs across the entire mass spectrum, as detailed below.

\subsubsection{Multimessenger observations of MBHBs}
\label{subsec:multimultiMBHB}
\begin{wrapfigure}{R}{0.5\textwidth}
  \vspace{-0.5cm}
\centering
 \includegraphics[width=0.5\textwidth]{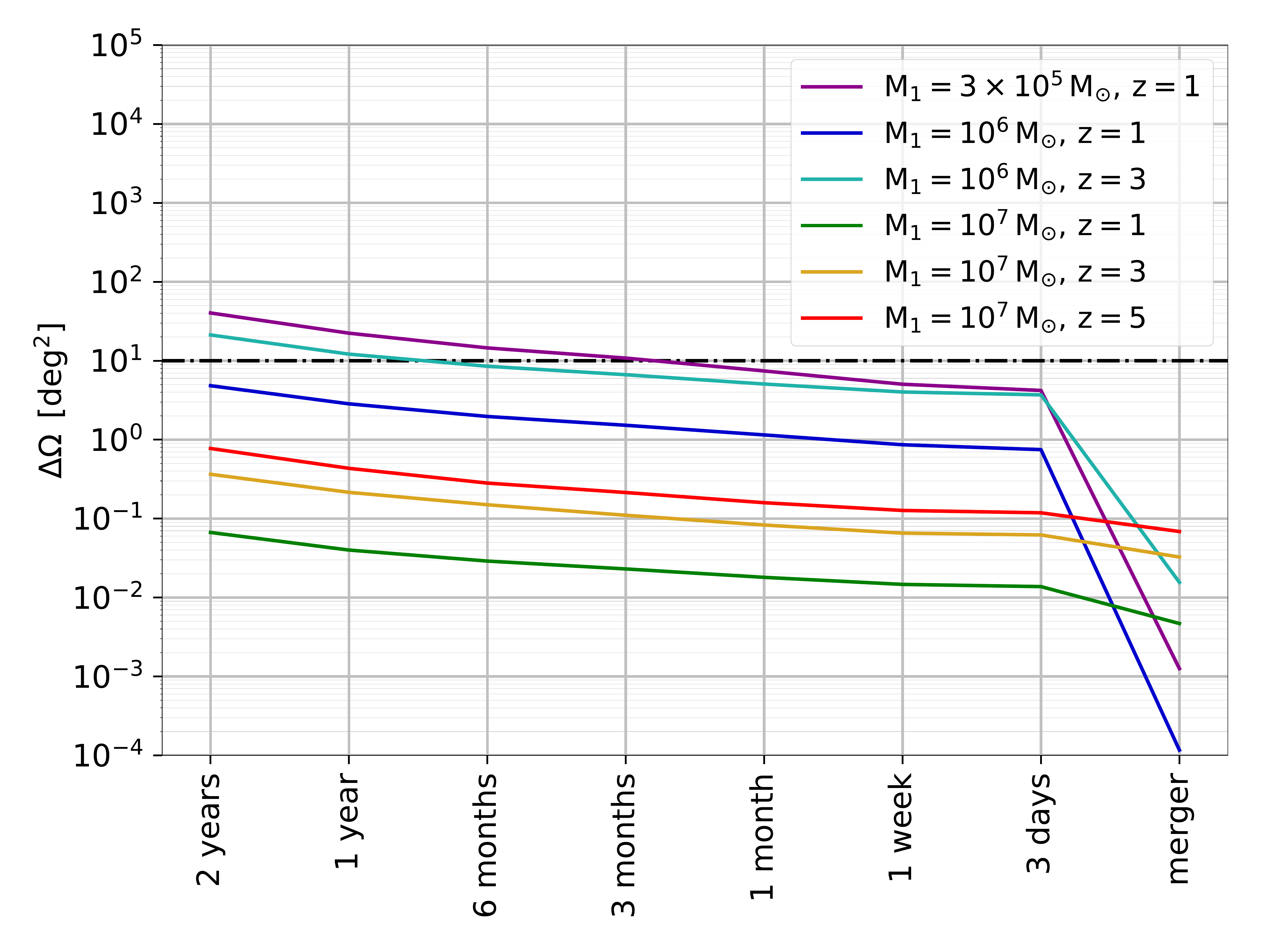} 
  \vspace{-0.5cm}
\caption{\footnotesize{Median sky location precision as a function of time to coalescence for selected systems, as labelled in the figure. Each curve is generated by simulating 1000 systems with random sky localization, inclination, polarization, and spin parameters. Results are obtained in the Fisher Matrix approximation, using the inspiral only waveform described in \cite{2014PhRvD..90l4029K}, which includes spin precession and higher harmonics. Sky localization at merger is rescaled with the gain in SNR squared computed by using IMR PhenomC waveforms \cite{2010PhRvD..82f4016S}. Caveat: these curves have to be considered as illustrative only. In fact, the Fisher Matrix calculations coupled the $\upmu$Ares sensitivity curve shown in Figure \ref{Fig1} with the LISA time dependent response function, as a consistent one for $\upmu$Ares was not yet available. We stress, however, that $\upmu$Ares two constellations at an angle with respect to each other allow to disentangle the two GW polarisations instantly regardless of the location of the source in the sky, and the Martian orbit offers a Doppler modulation comparable to that of LISA. We therefore expect the same level of information to be encoded in the $\upmu$Ares response function.}}
\label{fig:skyloc}
\end{wrapfigure}

Unlike stellar-origin BH mergers, mergers involving MBHBs are expected to occur in the gas-rich central regions of galaxy mergers, and to produce copious EM signatures.  Such EM signatures, combined with GWs, can deliver much more science than either one by itself.  Roughly, the gains can be divided for ``astrophysics" and for ``fundamental physics''. On the astrophysics side, tracing out the merger history of MBHs, together with their specific host galaxies, will shed light on the co-evolution of MBHs and galaxies -- a long-standing problem \cite{2013ARA&A..51..511K}. This requires unique identification of the host galaxy for each merger, for a large population of merger events, which is unlikely to be fully accomplished by LISA. Also on the astrophysics side, the GWs will yield precise system parameter estimates.  This will enable an unprecedented study of (binary) accretion onto MBHs whose masses, spins, and orbital parameters are known ab initio. On the fundamental physics side, a measurement of the redshift from the EM spectral observations will enable the cosmological use of chirping MBHBs as standard sirens, in a way that is highly synergistic with corresponding Type Ia supernova standard candles, see Section~\ref{sec:standardcandels}. A comparison between Hubble diagrams from standard sirens (with distances based on graviton measurements) and purely EM standard candles will serve to probe long-range deviations from standard GR (such as extra dimensions) since in models beyond GR, these Hubble diagrams can disagree (e.g.~\cite{2007ApJ...668L.143D,2019JCAP...07..024B}).  A determination of a time delay between the arrival of gravitons and photons will constrain massive gravity theories (see also Section~\ref{sec:gr}), since the graviton speed, and hence the time delay, depends on the graviton's rest mass. The frequency-dependence of this time delay would probe Lorentz invariance \cite{2008ApJ...684..870K}.

In principle, merging MBHs can have several distinct EM signatures, before, during, and after the merger. Prior to merger, a promising signal is a quasi-periodic EM ``chirp", tracking the phase of the GWs \cite{2017PhRvD..96b3004H,2018MNRAS.476.2249T}. Because of copious shock-heating, gas near the MBHs is expected to be unusually hot \cite{2010ApJ...715.1117B,2012ApJ...749..118S,2017ApJ...838...42B}. The corresponding UV/X-ray emission would have different (harder) spectra, with possible signatures of a disk cavity, and display periodicity on the orbital time-scale of days to minutes (years to days prior to merger). Doppler effects could increase variability over time, tracking the GW chirp \cite{2017PhRvD..96b3004H,2018ApJ...853..123S}.
At merger proper, the energetic burst of GWs may be accompanied by a luminous X-ray flare from the tidal ``squeezing" of gas (e.g. \cite{2002ApJ...567L...9A,2010MNRAS.407.2007C,2016MNRAS.457..939C}), EM signatures of the turn-on of a relativistic jet, or other flares from any direct coupling between the GWs and the surrounding plasma \cite{2017PhRvD..96l3003K}. Finally, in the coalescence aftermath, any circumbinary gas is expected to develop strong prompt shocks after the merger of the binary, due to the effectively instant mass-loss and centre-of-mass recoil of the remnant MBH. These should lead to a bright afterglow, whose nature depends on the amount and configuration of the nearby gas. In the case of a thin circumbinary disk, the afterglow should display a characteristic increase in both spectral hardness and overall brightness, on time-scales of weeks to years (e.g. \cite{2008ApJ...676L...5L,2008ApJ...684..835S,2010MNRAS.404..947C,2010MNRAS.401.2021R}).

In the absence of an ab-initio understanding of binary accretion and the corresponding spectral evolution properties of the above signatures, the most robust is likely to be the pre-merger periodic modulations, which are inevitably caused by the binary's orbital motion (the exception is if the binary is surrounded by optically thick gas, with the photosphere well outside the binary).  In order to allow a measurement of the coincident EM chirp, and phasing it relative to the GWs over several hundred cycles, the sources need to be localized on the sky weeks to months prior to merger (depending on the chirp mass). Since most merging binaries will also be at high redshift and faint, ideally the localization should be good enough to fit in the arcminute field of view of sensitive instruments (rather than the degree field of view of less sensitive survey telescopes).  This is beyond the capability of LISA as currently planned.

{\it By operating for a decade down to $\upmu$-Hz frequencies,  $\upmu$Ares can monitor up to thousands of inspiralling and hundreds of merging binaries with high SNR (see Figure~\ref{Fig1}).} Inspiralling systems can be observed slowly chirping for the whole mission duration, yielding precise information on $f$ and $\dot f$, besides measuring the binary masses and luminosity distance within $\approx 10\%$ and the location in the sky within $<$$100\,$deg$^2$. Conversely, merging systems with $M>10^5$~M$_{\odot}$ at $z<5$ can be localized within 1~deg$^2$ up to a year before final coalescence. Figure~\ref{fig:skyloc} shows the median sky localization accuracy that a $\upmu$Ares-class instrument can achieve on selected MBHBs as a function of time to coalescence. Most sources at $z<5$ will be localized at ${\cal O}(10\,$deg$^2)$ precision several months before mergers, allowing continuous monitoring with instruments such as the Large Synoptic Survey Telescope (LSST) or SKA. Typical sky location at merger are generally better than $100\,$arcmin$^2$, which will allow pointing with ultra-sensitive X-ray telescopes of the Athena and Lynx class. 

\subsubsection{Multiband at the two ends of the BH mass spectrum}

$\upmu$Ares will connect the milli-Hz window probed by the first generation of space-based detectors (i.e. LISA) with the nano-Hz regime probed by PTAs. By the late 2040s', it is conceivable that SKA will be operational at full sensitivity for about 20 years. SKA will pierce through the low frequency SGWB produced by MBHBs, constraining their merger rate and possibly their dynamical properties \cite{2017MNRAS.468..404C}. It will likely also allow the detection of several particularly massive and/or nearby systems, possibly allowing for a decent sky localization of ${\cal O}(100)\,$ deg$^2$ \cite{2010PhRvD..81j4008S}. A $\upmu$-Hz GW detector will complement these findings with a precise estimate of the SGWB at $3\times 10^{-7}<f<10^{-5}$~Hz, and by detecting several massive low-redshift systems down to $f<1$~$\mu$Hz (see Figure~\ref{Fig1}). {\it These complementary observations will provide a full characterization of the MBHB population and their environment including, in the case of counterpart detection,  how the host properties change across the mass spectrum and redshift.} There is even a small chance that a source observed by SKA will inspiral and merge into band within the detector lifetime \cite{2013ApJ...764..187S}.   

At the opposite end of the spectrum, a mission on a Mars orbit, will even allow a comparative better sensitivity at few milli-Hz compared to LISA (see Figure~\ref{Fig1}). This allows a 10-fold increase in the number of extragalactic stellar-origin BHBs detectable at those frequencies. A larger statistics of individually resolvable objects will facilitate the discrimination of different formation channel based on the eccentricity of the sources \cite{2016ApJ...830L..18B,2017MNRAS.465.4375N}, which can be combined with 3G findings to provide a complete view of the cosmic BHB population. We notice that few sources will also cross to the 3G band to become genuine multiband systems. The same holds true for any system involving IMBHs. This specific piece of multiband science, however, is better addressed with a deci-Hz detector, as exposed in a companion White Paper~\cite{wp3}, and is not further considered here.

\vspace{0.5cm}
\fbox{\begin{minipage}{40em}
    {\bf Summary of multimessenger and multiband science goals:}
    \begin{itemize}[topsep=0pt,itemsep=-1pt,partopsep=0pt, parsep=0pt]
    \item identify the characteristic signatures of inspiralling MBHBs, providing the key to search for them in millions of AGN spectra and lightcurves;
    \item investigate the interplay between gravity, matter, and light in the dynamical spacetime of inspiralling/merging MBHBs;  
    \item establish the connection between MBHBs and their hosts;
    \item provide a unique sample of standard sirens out to $z\gtrsim 5$;  
    \item fully characterize the MBHB population at low frequency, in connection with PTAs.
    \end{itemize}  
\end{minipage}}

\subsection{General relativity and beyond}
\label{sec:gr}

\paragraph{Strong lensing.}

\begin{wrapfigure}{R}{0.4\textwidth}
  \vspace{-0.5cm}
  \begin{center}
      \includegraphics[width=0.4\textwidth]{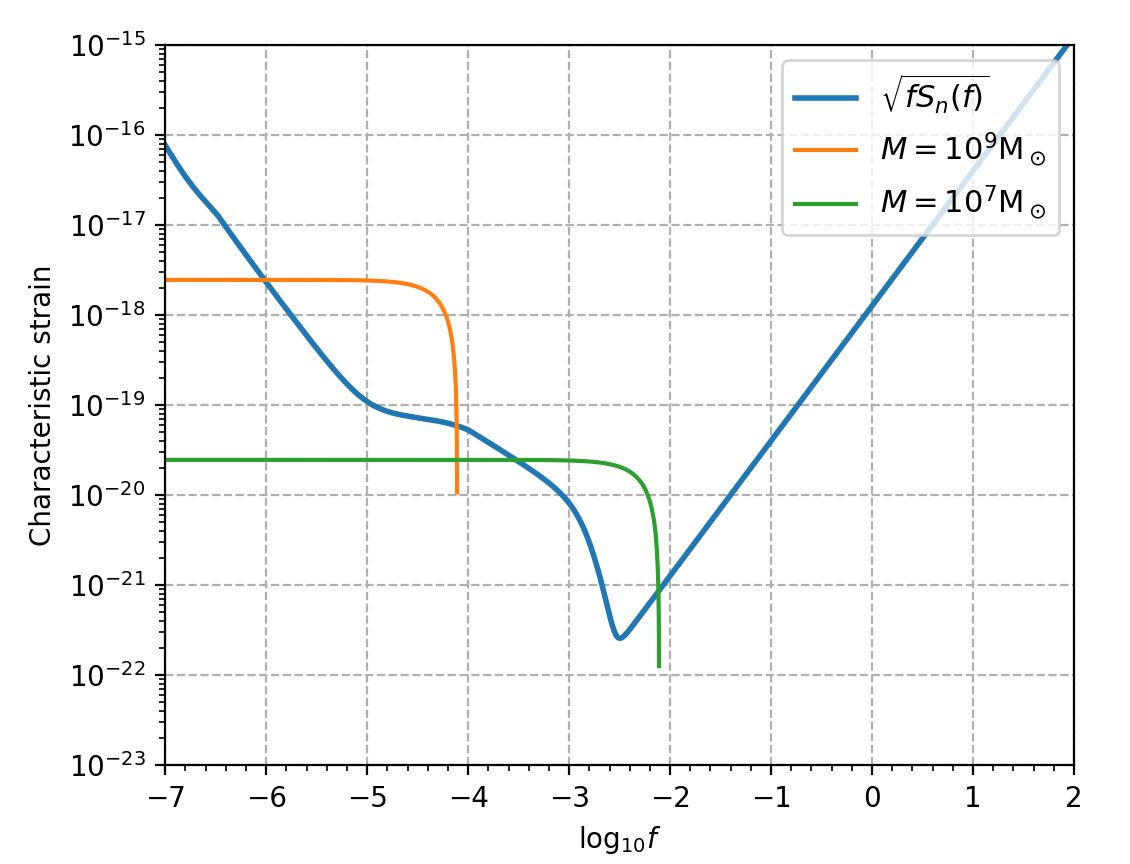} 
  \end{center}
  \vspace{-0.5cm}
  \caption{\footnotesize{Example signals from the memory effect compared to the $\upmu$Ares sensitivity curve. The spectrum of the signals are calculated based on the derivations done in~\cite{2019arXiv190611936I}}. The value of the redshift was chosen to be $z = 4$. Both MBHBs were chosen to have components of the equal mass. The green curve corresponds to the $M=10^7 \msun$ and the orange curve corresponds to $M=10^9 \msun$. The values for the characteristic timescale of the memory rise is inverse proportional to the frequency at coalescence and were chosen to be $\tau=100$ s and $\tau=10000$ s for green and orange curve respectively.}
  \label{fig:memory}
\end{wrapfigure}

Strongly lensed GWs may be detected via frequency-dependent effects, such as diffraction and interference ~\cite{1992grle.book.....S}. These wave effects can help identify the properties of the gravitational lens as the source's frequency increases toward coalescence~\cite{2003ApJ...595.1039T}. A $\upmu$-Hz mission like $\upmu$Ares will be able to detect wave effects for lenses with $10^{7} < M/M_{\odot} < 10^{10}$ (for $f \sim 1 \mu$Hz --10mHz), extending the range accessible to LISA towards higher masses. Lens characterization (redshifted mass and impact parameter) can be improved by at least a factor $\sim 2$ with respect to LISA based on the sensitivity curve. Based on the estimates for LISA, for which 1-few strongly lensed signals can be detected ~\cite{2010PhRvL.105y1101S}, {\it $\upmu$Ares could detect dozens or more strong lensed MBHBs including systems well before merger}, thanks also to the longer mission duration, and would have sensitivity to detect wave effects for more massive haloes due to the lower frequency range. Detecting strong-lenses has potential applications for probing the matter substructure and the expansion of the Universe  ~\cite{2011MNRAS.415.2773S}. 

Due to the motion of the detectors, multiple `images' from the same GW event arriving at different times will have different localisation patterns. Combining the information from different images can reduce the uncertainty in localisation significantly ~\cite{2004PhRvD..69b2002S}, potentially by three orders-of-magnitude. More stringent localisation constraints will enable better test of GR through polarization measurements~\cite{2018PhRvD..98b2008T}, and will help us to perform cosmography up to $z\approx 10$~\cite{2011MNRAS.415.2773S}.

\paragraph{Memory effect.}

Another interesting target for a low-frequency space-based detector is the so-called GW {\it memory}, i.e. the permanent relative displacement between two probe masses following the passage of GWs. There are two types of memory effects: linear  \cite{1974SvA....18...17Z,1985ZhETF..89..744B} and nonlinear  \cite{1991PhRvL..67.1486C,1992PhRvD..46.4304B}.  The nonlinear memory effect originates from the non-linearities in the Einstein field equations and therefore provides a direct test of GR in the strong regime.
The nonlinear memory effect can be physically interpreted as the signal generated by the gravitons radiated from the system  \cite{1992PhRvD..45..520T}.
It is possible to detect the build-up of the memory during the evolution of the MBHB  \cite{2009PhRvD..80b4002F}. A 4-year LISA mission may detect at most 1-10 memory events with SNR larger than 5 (see~\cite{2019arXiv190611936I}), but more numerous, higher-SNR events would be possible with a lower-frequency detector (see Figure~\ref{fig:memory}). A detection of the memory effect would have important implications for theoretical physics, because it has recently been shown that there are deep connections between gravitational memory, BMS supertranslations, Weinberg's formula for soft graviton production and the black hole information paradox (see e.g. \cite{2014arXiv1411.5745S,2016PhRvL.116w1301H,2019PhRvL.123b1101C}).

\paragraph{Dipole radiation/modification of GR.}

Low frequency detectors provide new opportunities for detecting deviations from GR in the propagation of GWs ~\cite{2018FrASS...5...44E}. $\upmu$Ares will be able to perform these tests beyond the capacity of LISA~\cite{2019JCAP...07..024B}. Graviton mass bounds from the modified dispersion relation would benefit both from the lower frequency and increased SNR \cite{2017RvMP...89b5004D}. A $\upmu$-Hz mission has the potential to constrain the graviton mass to a precision of $m_g\lesssim 10^{-29}$eV, two orders of magnitude better than LISA.  Many alternatives to GR predict GW oscillations, an effect similar to neutrino flavour oscillations \cite{2017PhRvL.119k1101M}. A lower frequency would improve tests of GW oscillations considerably, with a precision factor gain comparable to that of the graviton mass. There are other effects beyond GR that affect GW propagation in a frequency-independent manner (e.g. GW speed and damping). A $\upmu$-Hz detector will constrain those effects at a lower energy scale, well below the cutoff scale for dark energy effective theories \cite{2018PhRvL.121v1101D}.

In GR, quadrupole radiation is the leading-order GW effect for an inspiralling binary. In modified theories of gravity the strong equivalence principle is generally violated \cite{2014LRR....17....4W}, allowing for the existence of dipole radiation  \cite{2015CQGra..32x3001B,2016PhRvL.116x1104B}. For example, depending on the detailed field/curvature couplings in the scalar-tensor gravity, black holes  \cite{2015PhRvL.115u1105B,2016PhRvD..93b4010Y} and neutron stars  \cite{PhysRevD.87.081506,2014PhRvD..89d4024P,2018CQGra..35h5001W} can acquire scalar charges, even in a nonperturbative way  \cite{1993PhRvL..70.2220D,2017PhRvX...7d1025S,PhysRevLett.120.131103,Silva:2017uqg}. Because dipole radiation affects the waveform phase at a lower post-Newtonian order than the quadrupole radiation, very low-frequency observations with a space-based GW detector can either reveal it or yield very tight constraints. For a massive black hole binary, the improvement in the low-frequency band greatly improves parameter estimation, by providing much better sky localization, mass measurements, and so on  \cite{2005PhRvD..71h4025B}. This is important because strong-field effects (such as black hole scalarization, see e.g.  \cite{Silva:2017uqg}, depending on the theory parameters, in principle can happen in specific mass ranges, and dipolar radiation would mostly affect gravitational radiation at low frequencies. The prospects to obtain more accurate tests of modified gravity also apply to other types of gravitational-wave sources  \cite{2017PhRvD..96h4039C}, including DWDs \cite{Littenberg:2018xxx}.

\paragraph{Black-hole spectroscopy.}

GW emission during the post-merger ``ringdown" phase is a superposition of damped-oscillation modes called quasinormal modes (QNMs). The spectrum of these modes depends only on the geometry of the final BH, and therefore the QNM spectrum is a unique fingerprint of the remnant. While LISA can ``hear'' ringdown from $10^9 M_\odot$ BHs only for $z<1$  \cite{2019PhRvD..99b4005B}, $\upmu$Ares will see very large ringdown SNRs ($\sim 1000$) even at redshifts as large as $10$.  This will allow precise measurements of the remnant's spin and mass with fractional errors as low as $\mathcal{O}(10^{-3})$ at $z=10$. The quadrupole mode will still dominate the emission, but $\upmu$Ares will also see a large number of higher harmonics (up to $l=7$). These higher-order modes will significantly improve sky localization and distance estimation, and they will also provide incontrovertible evidence that these supermassive objects are indeed Kerr black holes: in GR the QNM frequencies depend only on the BH's mass and spin, and the consistency between different mode frequencies would break if the modes were to depend on any other parameters of the source.  

\vspace{0.5cm}
\fbox{\begin{minipage}{40em}
    {\bf Summary of General relativity and beyond science goals:}
    \begin{itemize}[topsep=0pt,itemsep=-1pt,partopsep=0pt, parsep=0pt]
    \item probe DM substructures and the Universe expansion via strong GW lensing;
    \item detect non-linear GW memory with high SNR from merging MBHBs;
    \item improve sensitivity to graviton mass and other deviations from GR by more than two orders of magnitude with respect to LISA.  
    \end{itemize}  
\end{minipage}}

\subsection{Cosmology and cosmography}

\subsubsection{Early-Universe cosmology}

GWs can carry unique information about the state of the Universe at energy scales far beyond the reach of EM cosmological observables. GW sources operating in the early Universe generate an SGWB. Its characteristic frequency today can be related to the Hubble factor at generation time $H_*$ \cite{Caprini:2018mtu}
\begin{equation}
\label{eq:fPT}
  	f=\frac{k}{2\pi}\frac{a_*}{a_0}=2.6 \times 10^{-8} \,{\rm Hz} \left(\frac{k}{H_*}\right)\left(\frac{g_*(T_*)}{100}\right)^{1/6}\frac{T_*}{\rm GeV}
\end{equation}
where $a_*$ and $a_0$ are the scale factors at generation and today, $T_*$ and $g_*(T_*)$ the Universe temperature and the number of relativistic degrees of freedom at generation (the second equality holds in the radiation dominated era). $k$ is the physical wave-number at the time of GW production, which for causality reasons must satisfy $k<H_*$. A part from the factor $k/H_*$ which, as we will see, depends on the details of the GW source, equation~\eqref{eq:fPT} relates the temperature of SGWB production epoch to the signal frequency, showing that GW observatories in different frequency ranges probe GW emission from different energy scales in the early Universe. There are mainly two classes of SGWB sources operating in the early Universe: those related to inflation and subsequent processes (such as reheating), and those related to primordial phase transitions.

\paragraph{Inflation.}
A SGWB is generically expected in the standard slow-roll inflationary scenario, extending in frequency with a slightly red-tilted spectrum from the horizon scale today to the one corresponding to the energy scale of inflation, i.e.  $10^{-19}<f<10^{11}$ Hz (assuming inflation is occurring at the highest energy scale allowed by CMB observations, i.e.~$10^{16}$ GeV -- for a review, see e.g.~\cite{Caprini:2018mtu}). Even though this signal interests the frequency detection range of all present and future GW detectors, measuring it is extremely challenging because of its low amplitude. At low frequencies $f<10^{-16}$ Hz, this SGWB is also the target of CMB experiments, through the measurement of the B-mode polarisation (for a complete treatment, see e.g.~\cite{Durrer:2008eom}). The present upper bound by the Planck satellite on the tensor to scalar ratio is $r < 0.07$ \cite{2018arXiv180706209P}, translating into $h^2 \Omega_{gw} < 3\cdot 10^{-16}$ (assuming no spectral tilt). This is expected to improve in the near future. The time-scale for CMB ground-based experiments is such that the Simons Array \cite{Ade:2018sbj} might bound $r < 2 \cdot10^{-3}$ by 2021 -- 2025 (the dates are start taking data -- results), and CMB stage IV \cite{2016arXiv161002743A} might reach $r < 10^{-3}$ by 2027 -- 2031. Concerning satellites, LiteBird \cite{Hazumi:2019lys} will reach  $r < 6 \cdot 10^{-4}$ by 2027 -- 2032, and proposed satellites such as PICO and COrE might reach $r < 10^{-4}$, which is the lowest bound CMB experiments can technically reach. In the case of no positive detection by CMB before 2050, future direct GW detection should therefore do better than $h^2 \Omega_{gw} \sim 2\cdot 10^{-19}$ (corresponding to $r = 10^{-4}$), which is far below the sensitivity of any GW mission under study. Note that there are scenarios, going beyond standard slow-roll inflation, in which the predicted SGWB spectral tilt becomes blue at high frequency, thereby opening up the possibility of a direct GW detection of the inflationary SGWB (e.g.~\cite{Cook:2011hg}; for a review, see \cite{Bartolo:2016ami}). This would constitute a major discovery, as it amounts to probing the inflationary potential near the end of inflation, which is observationally unconstrained. Consequently, it would bring extremely relevant information about inflation and the high energy physics model underlying it. On the other hand, among the proposed scenarios, none provides a signal specifically compelling  for $\upmu$Ares.

\begin{wrapfigure}{r}{0.5\textwidth}
  \vspace{-0.5cm}
  \begin{center}
    \includegraphics[width=0.5\textwidth]{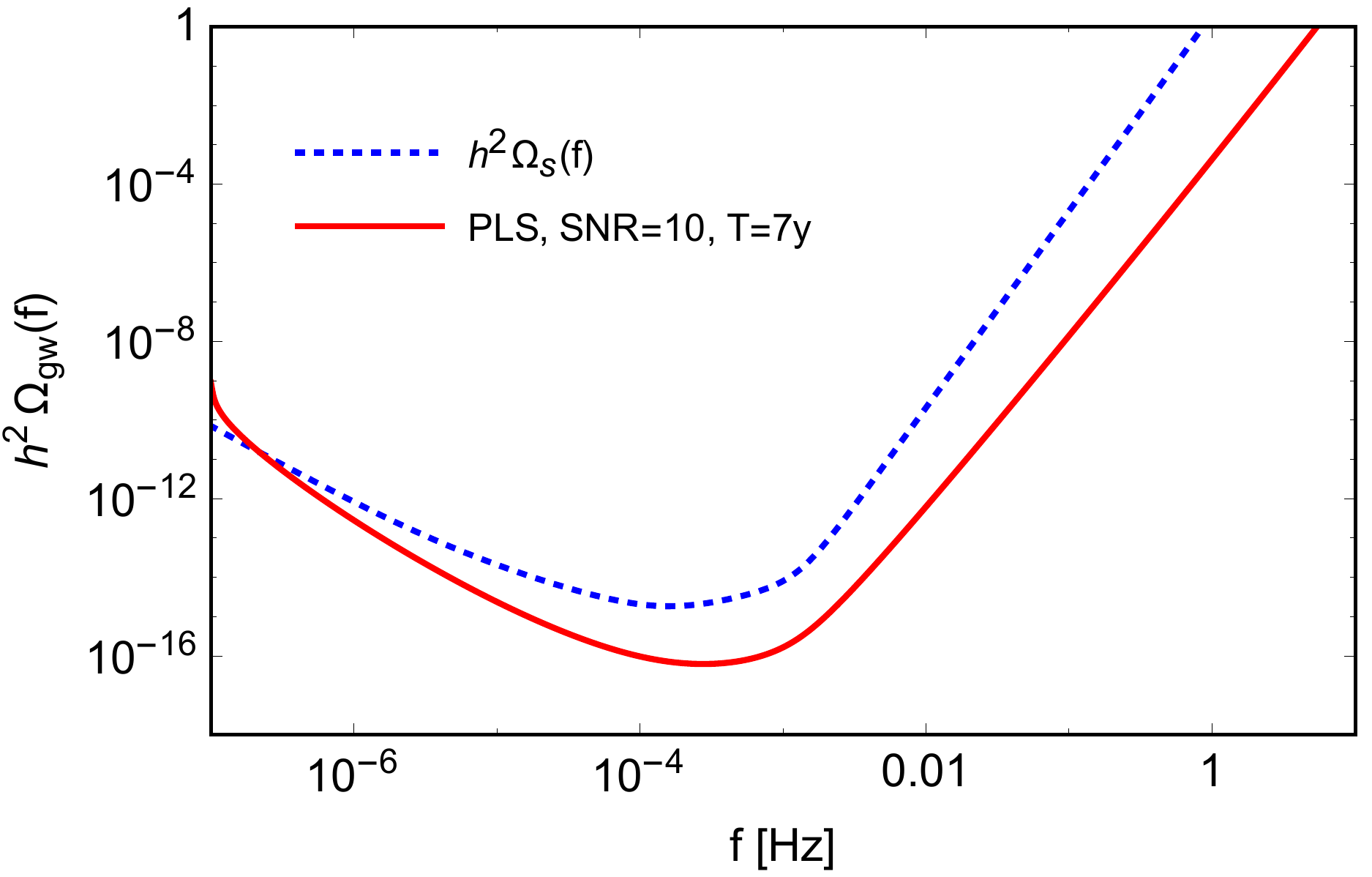}
  \end{center}
  \vspace{-0.5cm}
  \caption{\footnotesize{$\upmu$Ares power law sensitivity (PLS) curves to \mbox{SGWBs}, obtained following the procedure outlined in \cite{2019arXiv190609244C}, assuming a threshold SNR$=10$ for detection. $\upmu$Ares reach down to $h^2\Omega_{gw}\sim 6\cdot 10^{-17}$ at $2\cdot 10^{-4}\,$Hz, for comparison LISA gets to $h^2\Omega_{gw}\sim 2\cdot 10^{-13}$ at $3\cdot 10^{-3}\,$Hz.}}
  \label{fig:SGWBsens}
\end{wrapfigure}

\paragraph{Phase transitions.}
The situation is different for sources connected to primordial phase transitions. In particular, a first order phase transition in the early Universe can generate a SGWB via collisions of true-vacuum bubbles and the subsequent bulk motion in the Universe plasma (see e.g.~\cite{Kosowsky:1992vn,Kamionkowski:1993fg,Gogoberidze:2007an,Caprini:2009yp,Hindmarsh:2013xza,Hindmarsh:2015qta}). In this case, the SGWB is expected to peak at a frequency scale set by the size of the bubbles when they collide $R_*\sim v_w /\beta$ \cite{Caprini:2015zlo}, where $v_w$ is the bubble wall speed and $1/\beta$ denotes the duration of the phase transition (therefore ${k}/H_*\sim 2\pi \beta/(v_w\,H_*)$, c.f equation~\eqref{eq:fPT}).

Two phase transitions (PT) certainly occurred in the early Universe, the Quantum CromoDynamics (QCD) one at an energy scale of $T_*\sim 100$ MeV and the electroweak one at $T_*\sim 100$ GeV. In the context of the standard model of particle physics, they are both crossovers rather than first order~\cite{Kajantie:1995kf, Csikor:1998eu, Stephanov:2007fk}. However, the order of the QCD PT depends on the baryochemical potential, and there are hints that it might become first order if the lepton asymmetry is large in the early Universe, as expected when e.g.~a sterile neutrino is the DM~\cite{Boyarsky:2009ix, Schwarz:2009ii}. The SGWB from a first order QCD PT has been analysed in the context of PTA \cite{Caprini:2010xv}. Since the frequency range of PTA observatories corresponds to $k\sim H_*$ at the energy scale $T_*=100$ MeV (see equation ~\eqref{eq:fPT}), PTA can only probe very slow QCD PTs featuring  $\beta/H_*$ of $\mathcal{O}(1)$ or smaller. A weaker and briefer first order PT instead seems  more likely~\cite{Fang:2018axm, Capozziello:2018qjs, 2018arXiv181209676L, Bauswein:2019skm}.  A detector operating in the $\upmu$-Hz range would vastly improve our capability to investigate the cosmological QCD PT, as it would be sensitive to much wider regions of the PT parameter space, covering $10\lesssim \beta/H_* \lesssim 10^9$. With a power law sensitivity reaching down to $h^2\Omega_{gw}\sim 6\cdot 10^{-17}$ (see Figure \ref{fig:SGWBsens}), $\upmu$Ares has the potential to measure the SGWB from a weakly first order QCD PT with vacuum energy density of the order of 3\% of the radiation energy density in the Universe, happening e.g.~if $\beta/H_*=\mathcal{O}(100)$ and $v_w=\mathcal{O}(0.1)$.


Similarly, the EWPT is neither necessarily a crossover. Well motivated scenarios beyond the standard model (BSM) predict a first order EWPT, often together with baryogenesis processes~\cite{Caprini:2015zlo, Kozaczuk:2015owa, Kakizaki:2015wua, Chala:2016ykx, Dorsch:2016nrg, Bernon:2017jgv, Bruggisser:2018mrt, Bruggisser:2018mus, Megias:2018sxv, Chala:2018opy}. The detection of the EWPT SGWB would be an unquestionable proof that there exists BSM physics at the TeV scale. $\upmu$Ares may provide such a proof not necessarily after a BSM discovery at colliders. Both the Future Circular Collider and the International Linear Collider will be unable to fully test the parameter space leading to the first order EWPT~\cite{2017arXiv171007621F, Benedikt:2018csr}. So, independently of the collider findings, a strong EWPT will still be a compelling case in 2050.  

A first-order PT, not connected to the EWPT, can be also one of the few smoking guns of DM.
The ongoing DM searches are indeed cornering most of standard DM paradigms, motivating scenarios where the DM particle lies in a hidden sector. In this case, the interaction between DM and the SM particles is negligible, killing any hope to detect the DM particle via its non-gravitational effects. The detection of (or constraints on) a phase transition occurring in the hidden sectors would thus unveil some key features, e.g.~the mass scale of the DM sector. Interestingly, the first order phase transition occur both if the hidden sector has a QCD-like structure (with different parameters in the Columbian plot~\cite{Stephanov:2007fk,PhysRevLett.115.181101}), or a BSM Higgs-like potential (with the electroweak scale replaced by the hidden-sector scale)~\cite{Caprini:2015zlo}. $\upmu$Ares has the capabilities to probe hidden sectors when their energy scales are in the $10^{-2}$ -- $10^6$ GeV range. In view of the hidden PTs, $\upmu$Ares is a particularly valuable experiment succeeding SKA and LISA: if LISA observes a SGWB from PT, $\upmu$Ares reconstructs the signal in a broader frequency band and with higher accuracy (see e.g.~\cite{Figueroa:2018xtu, 2019arXiv190609244C}); if SKA detects the PT signal, $\upmu$Ares measures the high-frequency tail of the signal encoding the particle physics interactions in the plasma; and if neither LISA or SKA measure the PT signal, $\upmu$Ares has still the chance to detect it by accessing a wider parameter space (e.g.~around two orders of magnitude in $\beta/H$).  

A primordial phase transition can also lead to the formation of topological defects \cite{kibble_topology_1976}.
Among these, local cosmic strings are the most interesting in this context, as they can be powerful SGWB sources (see e.g.~\cite{vachaspati_gravitational_1985}). However, the SGWB spectrum peaks at a frequency scaling as $f_{\rm peak}\propto (G\mu)^{-1}$ ($\mu$ being the string linear energy density). From this scaling, it appears that low frequency detectors are less powerful in  constraining $G\mu$.

It is extremely important to remark, that above we have assumed the absence of a SGWB foreground of astrophysical origin, masking the cosmological signal. The detection of an astrophysical SGWB, for example from stellar-mass BBH and/or NSB, would also constitute a major discovery, but it is expected to occur on a shorter time-scale, by Earth-based interferometers and/or LISA. In the context of later detectors, it is important that the data processing provides an efficient ``cleaning" of astrophysical SGWB foregrounds. Otherwise, the above-mentioned science would not be lost, but the efficiency in constraining models and their parameter space would be correspondingly reduced.

\subsubsection{Late-Universe cosmology}
\label{sec:standardcandels}

The late time expansion of the Universe can be probed with standard sirens: compact binary coalescence for which an accurate measurement of both the luminosity distance (from GWs) and the associate redshift (from EM observations) can be obtained \cite{1986Natur.323..310S}. With these two quantities in hand, one can test the standard cosmological distance-redshift relation which in turn yields constraints on the cosmological parameters. Both stellar origin BHBs and BNSs are excellent examples of standard sirens. BNSs can be associated with an EM counterpart measurement, in which case a unique redshift can be associated to the GW event providing a single point in the distance-redshift diagram, as spectacularly demonstrated by GW170817 \cite{Abbott:2017xzu}. Stellar mass BHBs instead are believed to not produce EM counterparts, in which case redshift information can be recovered only by cross-correlating the localization volume of the GW event with galaxy catalogues. For these sources thus one can only construct posterior distribution in the distance-redshift diagram, describing the probability of the redshift associated with the GW event based on the distribution of galaxies within the GW localization volume, as first demonstrated by GW170817 \cite{Fishbach:2018gjp} and GW170814 \cite{Soares-Santos:2019irc}.

At sub milli-Hz frequencies instead, the most powerful standard sirens that one might hope to exploit are MBHB mergers. These are already the main cosmological sources for LISA, which will be able to detect up to few tens of them with EM counterpart, testing the expansion of the Universe at high redshift, up to $z\sim10$ \cite{2016JCAP...04..002T}. The results obtained with LISA shows that MBHBs will be ideal standard sirens to test deviations from $\Lambda$CDM at high redshift, specifically in the range $2<z<5$ \cite{2016JCAP...10..006C,2017JCAP...05..031C}. They will not, however, be detected in sufficient number to yield competitive constraints on the Hubble constant or the equation of state of dark energy, but will be extremely useful to test deviation in the cosmological propagation of GWs \cite{2019JCAP...07..024B}.

The low-frequency performance of $\upmu$Ares will improve the cosmological analysis performed with LISA mainly in two ways. On the one hand a large improvement with respect to LISA will come from the better measurements of the MBHBs parameters, due to the much longer observation of the inspiral phase (up to the mission duration of 10 years). Although the contribution of weak lensing will still dominate the distance error budget (3\,-\,5\% at $z=3$ \cite{2011ApJ...732...82P}), the determination of the sky position should be greatly enhanced, reducing the localization region of the GW event and thus increasing the probability of finding an EM counterpart (see Figure \ref{fig:skyloc}). In the optimistic case in which all merging MBHBs will have associated EM counterparts, $\upmu$Ares will increase the number of standard sirens accessible to LISA by a factor of ten, drastically improving the cosmological analyses. This will allow a measurement of the Hubble constant at the percent or sub-percent level and  the placement of tight constraints on deviations from $\Lambda$CDM and from GR at very high redshift. The improvement of the sensitivity at low frequency could moreover bring more detections at even higher redshift, specifically at $z>5$. This will help in expanding the cosmological investigations performed with LISA to larger distances, and thus in particular to test deviation from $\Lambda$CDM or from GR at even larger scales. On the other hand the number of non-merging MBHBs is expected to be much higher than the merging ones (see Figure \ref{fig:mbhbpop}). Although for these inspiralling MBHBs the sky localization should be of ${\mathcal O}$(10deg$^2$) only, with an associate luminosity distance relative error of $\approx 10\%$, continuous EM observations in the localization region could spot the hosting galaxy through some specific EM signatures, which might for example be constituted by some periodic variability or AGN transient activity, as envisaged in Section \ref{sec:multimulti}. Even though the true hosting galaxy is not identified, one might also be able to reduce the number of credible hosts within the localization region, possibly to few tens or less galaxies. This would allow us to use them as dark sirens and thus apply the statistical method to extract cosmological information by cross-correlating with galaxy catalogues. In any case these considerations lead us to the possibility to have an additional set of MBHB standard sirens at high redshift, possibly up to thousands. $\upmu$Ares would therefore greatly exceed the cosmological potential of LISA, leading to strong constraints on the standard cosmological model and accurate tests of GR at very high redshift.

Beside MBHBs, also stellar origin BHBs \cite{2018MNRAS.475.3485D} and EMRIs \cite{2008PhRvD..77d3512M} can be valuable standard sirens. Both types of sources are best observed at $f>10^{-3}$ Hz, where the $\upmu$Ares sensitivity can be $\approx3$ times better than LISA. It should be noted, however, that none of those sources is expected to generate a detectable EM counterpart and statistical methods to cross-correlate their localization volumes with galaxy catalogues in order to obtain redshift information, are needed \cite{2018MNRAS.475.3485D,Kyutoku:2016zxn}. Those methods are effective so long as localization volumes are significantly smaller than the typical scale over which the matter content of the Universe is approximately homogeneous, which applies only to relatively low $z$ sources detected at high SNRs. Therefore, although $\upmu$Ares can potentially do better than LISA also on this ground, detailed calculations are needed to quantify whether the improvement is significant.  
%

\vspace{0.5cm}
\fbox{\begin{minipage}{40em}
    {\bf Summary of cosmology and cosmography science goals:}
    \begin{itemize}[topsep=0pt,itemsep=-1pt,partopsep=0pt, parsep=0pt]
      
    \item investigate a vast region of the allowed parameter space of QCD phase transitions;      
    \item explore beyond standard physics by searching for an EW phase transition;
    \item explore first order phase transitions in the hidden sector;  
    \item probe the geometry of the Universe, constraining deviations from $\Lambda$CDM out to $z\approx 10$ by means of ${\cal O}(10^3)$ MBHB standard sirens.
    \end{itemize}  
\end{minipage}}


\subsection{The Milky Way}

\subsubsection{A panorama of stellar binaries}

The Milky Way is composed of roughly 100 billion stars, with a significant fraction close enough to a companion star to experience binary interactions throughout the evolution. Despite the ubiquity of binary systems, their signatures across the EM spectrum and recent detection of GW from mergers from compact binaries, major open questions remain unsolved (see \cite{2014LRR....17....3P} for a review). Current topics of research include the different mechanisms for mass transfer, its stability and outcomes, core-collapse supernova kicks, the nature of Type Ia supernova progenitors, mass and angular momentum loss and binary merger products \cite{wan12,iva13,lyn94}. The low frequency GW band is populated by a large variety of stellar-origin Galactic GW sources that will be crucial for understanding open issues in binary evolution. The $\upmu$-Hz GW band will open a new window to binaries composed of a CO with a non-degenerate (main sequence or brown dwarf) star, while also hosting binaries in which both stars are COs but remain undetectable in the LISA band (milli-Hz frequencies) due to longer orbital periods. Many of these are likely to have bright EM counterparts (those including NSs are even detectable in multiple EM wavebands) enabling new Galactic multimessenger studies. These studies will help to constrain binary evolution models and to pin down the formation of rare binary systems. Furthermore, this adds to the bigger open question of understanding the formation and evolution of COs and binaries in the global context of the evolution of galaxies, including the Milky Way. Contrary to EM surveys, the major advantage is that GWs are not hindered by dust extinction and crowding, making them ideal for driving Galactic studies of these sources as well as complementing EM studies at all latitudes \citep{kor19, lam19}. Moreover,  GWs from CO binaries can be observed throughout the entire Local Group facilitating multimessenger near-field cosmology \citep{kor18},
while circumbinary companions can be detected through GW radial velocity measurements \cite{Robson:2018svj,2019NatAs.tmp..381T,Randall:2018lnh,Wong:2019hsq,Tamanini:2019usx}.

Much like at the LISA milli-Hz band, the extreme number of sources detectable at $\upmu$-Hz will result in a strong Galactic foreground of unresolvable sources (see Figure \ref{Fig1}). This foreground can be useful as it contains an imprint of the overall properties and past history of the Galaxy.

\paragraph{Galactic double white dwarfs.}
The hundreds of millions of DWD that are currently present in the Milky Way emit across the entire GW spectrum. Population synthesis models predict DWD systems, that undergo binary interactions during their evolution, span a vast frequency range from $10^{-7}$ to $10^{-2}\,$Hz \citep{too12}. While optical observations typically cover $10^{-5} - 10^{-2}\,$Hz frequencies \cite[e.g.][]{bro16,mao17}. At frequencies below $10^{-5}\,$Hz (corresponding to binary orbital periods longer than a few days) the intrinsic faintness of WD stars makes DWDs hardly detectable even by the future generation large telescopes \citep{reb19}. Thus, low-frequency GW observations are crucial for constraining DWD formation models and relevant physical processes involved with mass transfer and the common envelope phases. These systems will also be one of the main contributors to the Galactic foreground in the $\upmu$-Hz band \citep{nel01a,rui10}. Although the LISA mission is expected to provide tens of thousands new DWD in the milli-Hz band \cite[e.g.][]{kor17,lam19}, $\upmu$Ares will greatly improve our understanding of these systems by leveraging on the individual identification of ${\cal O}(10^5)$ DWD and on the detailed characterization of the unresolved foreground down to $\lesssim 10^{-5}\,$Hz (see Figure~\ref{Fig1}).

\paragraph{Binary systems with WD companions.}
Among binary systems including WD companion, WD+MS are expected to be the most numerous. Many systems have been discovered thanks to the bright main sequence (MS) star \citep[e.g.][]{2011A&A...536A..43N}. However, optical telescopes are biased towards binaries with a low-mass WD, in which the MS star does not outshine the WD companion. Thus, being more sensitive towards massive WDs, low frequencies GW observations are the optimal method to study the full mass range of all systems, and will complement the optical surveys. The period and mass distributions of WD+MS binaries provides the most direct link to common-envelope evolution \citep[e.g.][]{too13}.
Cataclysmic variables (CVs) are semi-detached close binary systems in which a WD accretes material from a Roche-lobe-filling secondary, typically a MS star (but can also be a brown dwarf) \citep{war95}. These systems represent outstanding laboratories for studying accretion onto COs. Their accretion flows are unaffected by relativistic effects or ultra-strong magnetic fields, making them ideal test systems for our understanding of more complex/compact systems, such as accreting NSs and BHs \citep[e.g.][]{Knigge10}.
WD+Helium star systems are less abundant compared to WD + MS stars, nonetheless they can be detected in the optical band out to a few kpc \citep{2019A&A...621A..38G}. Recently, one of these systems, CD--30$^\circ$11223 \citep{Geier13} has been confirmed to be a detectable GW source \citep{kup18}.  WD + Helium star systems represent direct progenitors of DWDs and AM Cvn stars because the Helium stars is expected to become a WD without going through any more giant stages. Furthermore, these binaries also qualify as double degenerate supernova type Ia progenitors \citep[e.g.][]{Geier06}. Orbital periods of CD--30$^\circ$11223-like systems are typically longer than $40\,$min, thus to fully characterise this population an instrument covering lower GW frequencies is required.  In particular, because WD components are significantly less bright when compared to the Helium star companions, measuring the mass of the WDs is practically impossible with optical observations alone. The combination of EM observations to GWs in the $10^{-5}-10^{-4}\,$Hz enabled by $\upmu$Ares represents a unique opportunity for recovering the intrinsic parameters of these systems. 

\paragraph{Galactic double neutron stars (DNSs).}
Since the discovery of PSR J1913+16 \cite{hul75}, DNSs have been the prime GW sources \citep[e.g.][]{phi91}. The measurement of the change in orbital period of PSR J1913+16 due to GW radiation \citep{tay82} started a process that recently reached a spectacular highlight.  GW170817 was the first confirmed extragalactic DNS GW source discovered, which has been subsequently followed-up with observations in gamma-rays, at the optical/infrared, X-rays and at radio wavelengths \citep{abb17a,abb17b}. Although only a handful of Galactic DNS can be detected at milli-Hz frequencies, several 100s of thousands are expected to reside in the $\upmu$-Hz band \cite{nel01a,vig18}. The promise of $\upmu$-Hz GW observations for the study of NS binaries lies mainly in the determination of Galactic DNS merger rates and for establishing the properties of systems that can be related to previous binary evolution stages. However, if the EM signals of these systems are also detected, studies can be extended to investigate the details of the parent population and the ages of the mergers, probing the delay time between the formation of the double NS and the merger. Combined $\upmu$-Hz GW and EM observations will produce unprecedented large data sets and will enable increasingly accurate tests of the massive stellar binary evolution models.

\paragraph{Galactic black holes binaries (BHBs).} 
Current stellar population synthesis models predict that about a million BHBs are likely to be present in the MW \cite{nel01a,2018MNRAS.480.2704L,2018MNRAS.473.1186E}. The most massive systems, which stem from lowest metallicity progenitors, will be present in the galactic halo, as they were formed in satellite galaxies that were then accreted by the MW. Several hundreds of these systems are likely to have $\upmu$-Hz GW emission \cite{2018MNRAS.480.2704L}. However, given the probably lack of measurable $\dot{f}$ (unless eccentricity is present) properly measuring the chirp mass and identifying these objects may be challenging. These objects are unlikely to be confused with DWDs as those latter have lower SNR due to their low masses, thus occupying a different region of the $\upmu$Ares GW strain-frequency space (see Figure \ref{Fig1}). Confusion with stellar binaries is possible, but EM follow-up will likely be able to distinguish those sources.  If sky localisation is good enough, these could be prime candidates for EM follow-up (with facilities like SKA). This will give us an unique view on the formation  conditions of these systems (low-metallicity environment, stellar cluster, etc) and accretion onto isolated BHs, which is yet to be found.

\paragraph{NSs/BHs + stellar companion (possible X-ray binaries).}
Tight binaries with a stellar and CO will be observable, particularly for high mass systems, which are the direct progenitors of mergers observed by LIGO/Virgo. To achieve periods $<1\,$day, these systems must have undergone significant shrinkage of their initial orbit, likely through a common envelope evolution. Some of these systems will display EM emission, and multimessenger detections will allow to measure the distance component masses (e.g. with SKA). In comparison to EM observations, which often present biases, GW will produce a more complete view of these systems, allowing to constrain binary evolution processes based on a population.
NS accreting from a companion WD are also expected to be bright X-ray and loud GW  sources~\cite{2018PhRvL.121m1105T} emitting at sub-milli-Hz frequencies accessible to $\upmu$Ares.

\subsubsection{An unparalleled bird's eye view of SgrA$^*$.} 
\begin{figure}
\centering
 \includegraphics[scale=0.75]{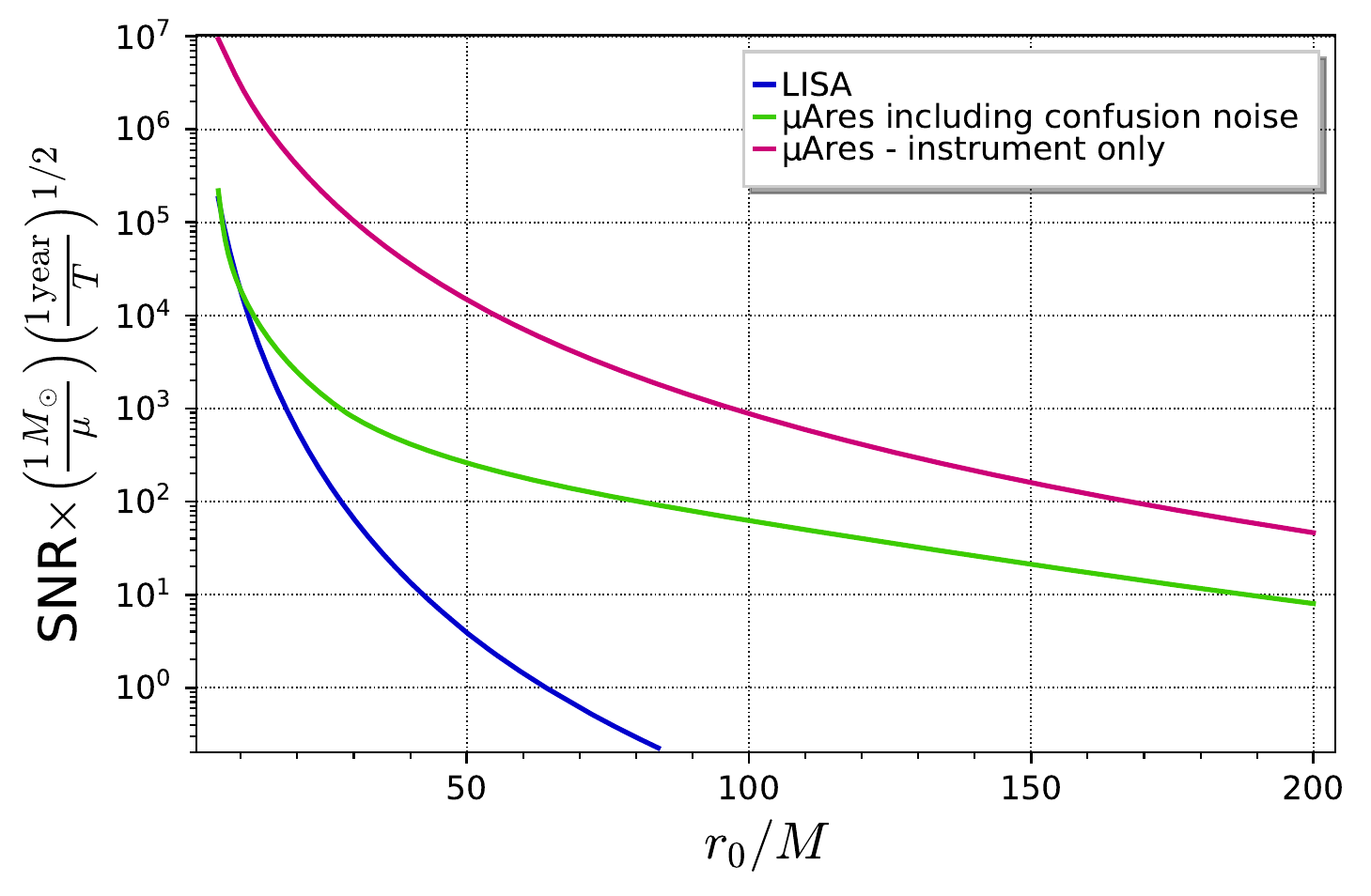} 
\caption{\footnotesize{SNR accumulated over 1 year of observation for a $1\msun$ object on a circular orbit near SgrA$^*$. For SgrA$^*$, $r_0 = 100M = 4.05AU$. This plot was made using the SageMath {\tt kerrgeodesic\_gw} package in the Black Hole Perturbation Toolkit \cite{BHPToolkit}.}}
\label{fig:MW_center}
\end{figure}

What is the nature of the environment surrounding MBHs? Our own galactic centre provides a unique insight into these question \cite{Genzel:2010} as it is close enough to allow for the direct observation of individual stars orbiting our MBH SgrA$^*$ (the celebrated S-stars \cite{Chu:2017gvt}) and even direct imaging of the BH itself \cite{Issaoun:2019afg}. This proximity also provides opportunities for GW observations. Theoretical models suggest that the inner 0.04 parsecs surrounding SgrA$^*$ should have a `dark cluster' of compact remnants and faint low-mass stars \cite{Alexander2017}. Recent work suggests that many of the latter, in the form of brown dwarfs, could be on highly relativistic orbits \cite{Amaro-Seoane:2019umn}. Fortuitously, the mass of SgrA$^*$ is just right so that if any of these bodies orbit sufficiently close to it they will emit GWs in the $\mu$-to-milli-Hz band \cite{Freitag:2002nm, Gourgoulhon:2019iyu}. This will allow us to measure the orbital properties of any faint or dark objects that orbit very close to the MBHs that are invisible or near invisible to other techniques. This will provide invaluable information about the dense stellar cluster near SgrA$^*$ (and, by extrapolation, other Milky-way like galaxies). These observations will also be complementary to efforts in the EM astronomy community which is also seeking to observe faint low-mass stars near SgrA$^*$ \cite{GC_Astro2020}. Searching for GWs from the galactic centre could also allow detection of more exotic objects, such as PBH \cite{Gourgoulhon:2019iyu}.

Whereas LISA will only be able to observe objects in the very strong-field of SgrA$^*$, the low frequency capabilities of the $\upmu$Ares mission greatly increases the ability to detect objects orbiting further out. From Figure~\ref{fig:MW_center} we see that a $1 M_\odot$ object could be detected with an SNR $\sim 10$ as far out as 8AU from the black hole with 1 year of observing time. A spectacular example of $\upmu$Ares capabilities, as shown in Figure \ref{Fig1}, is given by the possibility of detecting an inspiralling $10\msun$ BH onto SgrA$^*$ more than 100 Myr before the final plunge. Using a rough EMRI estimate of about $300$ Gyr$^{-1}$ for a MW-like nucleus (see \cite{2017PhRvD..95j3012B}, and references therein), this means that $\upmu$Ares might detect ${\cal O}(100)$ EMRI progenitors in the Galactic centre, revolutionizing our understanding of relativistic dynamics in galactic nuclei and dense stellar systems. Perhaps even more intriguing would be the detection of a pulsar orbiting SgrA$^*$. An inspiralling NS can also be seen around SgrA$^*$ out to several AUs, many Myr before final plunge. If this happens to be an active pulsar detectable with SKA, precise timing will yield exquisite tests of gravity and of the nature of SgrA$^*$ \cite{2015aska.confE..42S}. 

\vspace{0.5cm}
\fbox{\begin{minipage}{40em}
    {\bf Summary of Milky Way science goals:}
    \begin{itemize}[topsep=0pt,itemsep=-1pt,partopsep=0pt, parsep=0pt]
    \item understand common envelope physics via detection of mixed (CO + MS star) binaries and the distribution of DWD at $f<10^{-4}\,$Hz;
    \item physics of contact and over-contact binaries via joint GW + EM detection;  
    \item characterization of BHB, NSB and BH-NS population in the Galaxy, synergies with PTAs and SKA;  
    \item unveil the dynamics of stars and COs around SgrA$^*$.
    \end{itemize}  
\end{minipage}}

\section{Strawman mission concept}
\label{sec:strawman}

Surprisingly, a GW observatory with the desired low-frequency sensitivity as outlined in Section~\ref{sec:sources} is feasible with technology realistically available within the next decades. However, such an observatory requires an arm length on the order of 100 million kilometers. The three LISA spacecrafts will trail the Earth in an equilateral triangular formation separated by 2.5 million kilometers. The sheer size of the $\upmu$Ares constellation makes similar orbits impossible. The only viable alternative is a constellation with the Sun in its center as outlined in \cite{1907.11305}. Such a formation requires at least three spacecraft (triangular formation) although a higher number is easy to implement. Four spacecraft in a quadrilateral formation would not only increase redundancy but also improve sensitivity by means of orthogonal arms.

\paragraph{Orbits.}
For the purpose of this paper, we limit ourselves to a traditional triangular formation. As mentioned in Section~\ref{sec:sources}, two non-coplanar constellations are necessary for source localization. Table~\ref{tab:orbits} shows arm length variations and line-of-sight velocities for three exemplary missions with orbit radii of 0.7 AU (Venus), 1.0 AU (Earth), and 1.5 AU (Mars) obtained from \cite{pc_oliver2019}. We chose a 10 year mission lifetime. The first constellation is within the ecliptic plane while the second one is tilted by 90 degrees for optimal source localization.\\

\begin{table}[hbtp]
\centering
\settowidth\tymin{\textbf{Activities}}
\setlength\extrarowheight{2pt}
\begin{tabulary}{1.0\textwidth}{|LRRR|}
\hline
\textbf{Average orbit radius}                                                                      & \textbf{0.7\,AU} & \textbf{1.0\,AU} & \textbf{1.5\,AU} \\ \hline
Average arm length                                                                                 & 187 million\,km  & 259 million\,km  & 395 million\,km  \\ \hline
Total arm length variations     & \multirow{2}{*}{185,000\,km}      & \multirow{2}{*}{105,000\,km}      & \multirow{2}{*}{492,000\,km}      \\
(in plane constellation)     &       &       &       \\ \hline
Total arm length variations  & \multirow{2}{*}{176,000\,km}    & \multirow{2}{*}{217,000\,km}      & \multirow{2}{*}{65,700\,km}      \\
(out of plane constellation) &     &       &       \\ \hline
Maximum line-of-sight velocities                                                                     & 6\,m/s         & 4\,m/s          & 12\,m/s          \\ \hline
Doppler frequencies at 1064\,nm                                                       & 5.5\,MHz         & 4.0\,MHz          & 11.0\,MHz         \\ \hline
\end{tabulary}
\caption{\footnotesize{Arm length variations, line-of-sight velocities and resultant Doppler frequencies for three exemplary orbit radii over a 10 year mission lifetime.}}
\label{tab:orbits}
\end{table}

\begin{figure}[hbtp]
\centering
\includegraphics[scale=0.6]{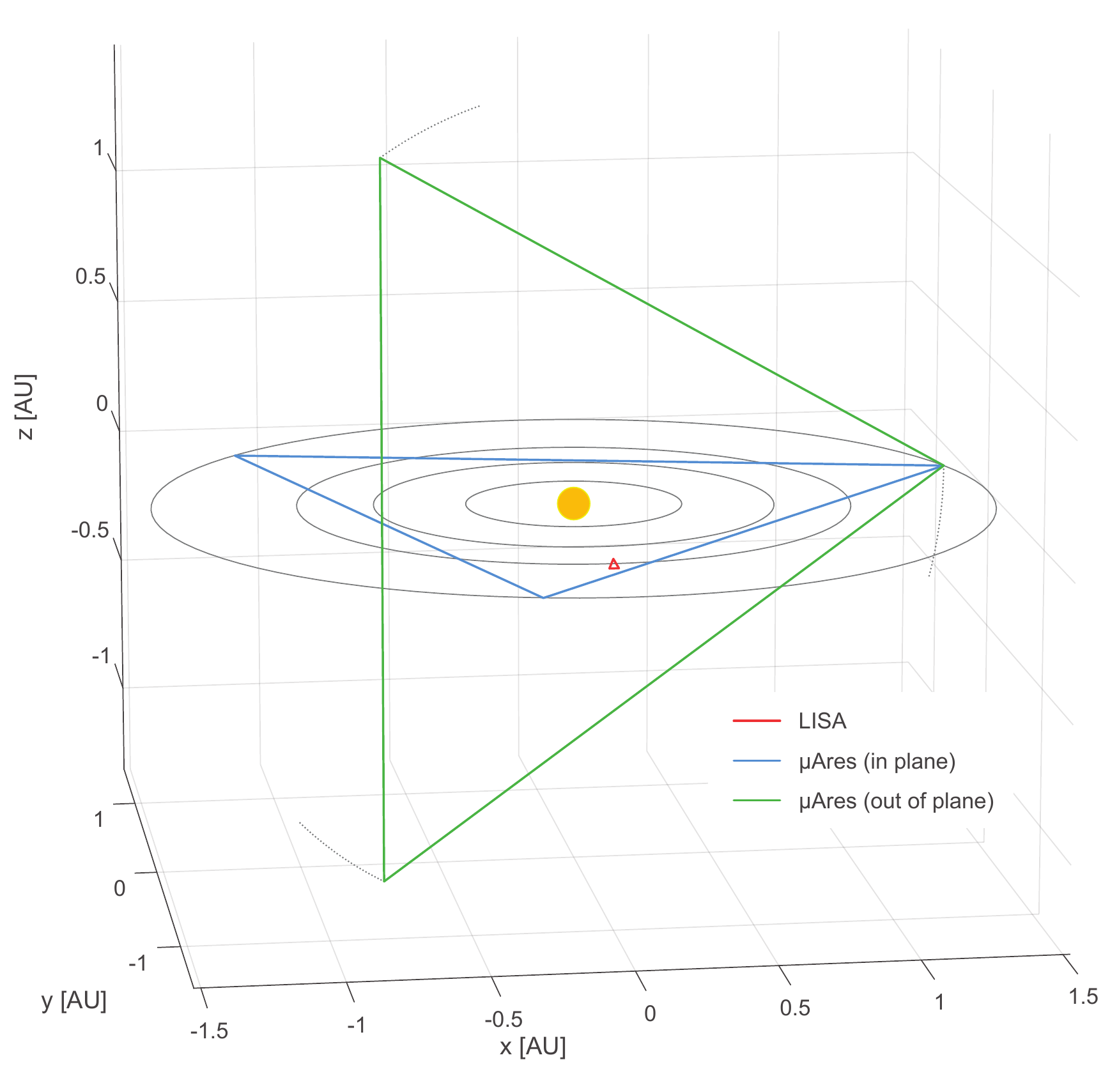} 
\caption{\footnotesize{The two $\upmu$Ares constellations (blue and green, 395 million kilometer arm length) alongside a to-scale depiction of LISA (red, 2.5 million kilometer arm length). One $\upmu$Ares constellation is trailing Mars within the ecliptic plane while the other is in an equal orbit 90 degrees tilted with respect to the ecliptic.}}
\label{fig:orbits}
\end{figure}

It is imperative to keep the line-of-sight velocity within reasonable limits as it directly impacts the heterodyne frequency of the interferometric readout. However, even for constellations as large as 395 million kilometers, a detailed analysis of the Doppler frequencies using tools described in \citep[p.~76]{barke2015inter} results in a very reasonable heterodyne frequency range of $2...12$\,MHz for all six spacecraft. 
Since longer arms increase the low-frequency sensitivity of the observatory, we decided to analyze the overall feasibility of these Mars-trailing orbits. These two $\upmu$Ares constellations are shown in Figure~\ref{fig:orbits} alongside a to-scale depiction of LISA (2.5 million kilometer arms). Within 10 years the spacecraft will complete 5.3 revolutions around the Sun. Despite the long arms the point-ahead angle stability of $\pm$\,$0.3$\,$\upmu$rad is even superior to LISA.

\paragraph{Mission parameters.}
To match the sensitivity outlined in Figure~\ref{Fig1} the total read-out noise in the interferometric signal must not exceed 50\,pm/$\sqrt{\text{Hz}}$ with an acceleration noise of $1 \times 10^{-15}$\,$\text{ms}^{-1}/\sqrt{\text{Hz}}$. Such values are perfectly realistic:
\begin{itemize}[topsep=4pt,itemsep=1pt,partopsep=0pt, parsep=0pt]
    \item The required acceleration noise is only 2 times lower than what has already been shown during the 2016 LISA Pathfinder mission  \cite{PhysRevLett.120.061101}.
    \item The read-out noise level can be achieved with a received laser power on the order of 15\,pW which results in a shotnoise limit of 35\,pm. This leaves room for spurious influences from laser relative intensity noise (estimated at $5 \times 10^{-8}$ at 2\,MHz) and electronic noise of the photodetector (estimated at 1\,pA/$\sqrt{\text{Hz}}$ and 1\,pV/$\sqrt{\text{Hz}}$) given a local oscillator laser power of $\approx 0.2$\,mW. \cite{Barke_2015}
\end{itemize}
All these parameters are very similar to the LISA requirements. However, in order to receive 15\,pW of laser power at the far spacecraft over a distance of 430 million kilometers, a significant increase in laser power and telescope diameter is necessary. One possible combination would be a 1 meter telescope diameter paired with 10 watts of laser output power. This could be made possible by 
\begin{itemize}[topsep=4pt,itemsep=1pt,partopsep=0pt, parsep=0pt]
    \item a relaxed dimensional stability requirement of the opto-mechanical architecture (from 1 (LISA) to 10\,pm/$\sqrt{\text{Hz}}$) due to the higher read-out noise limit compared to LISA, and
    \item a relaxed phase fidelity requirement for the laser amplifier \cite{Trobs:10} (from 0.6 (LISA) to 6\,$\upmu$rad/$\sqrt{\text{Hz}}$) due to the lower maximum heterodyne frequency which results in an increase in tolerable timing jitter. This requirement is given with respect to 2.4\,GHz clock-tone sidebands for the inter-spacecraft frequency distribution as described in \cite{Heinzel_2011}.
\end{itemize}
Additionally, while an inter-spacecraft frequency distribution might still be necessary, the relaxed timing noise requirements may lead to the omission of a pilot-tone correction scheme for the analog to digital converters of the phasemeter. Power requirements for the increase laser power could be satisfied by either installing larger solar panels which in contrast to LISA can face the Sun directly, or alternatively an radioisotope power systems.

\paragraph{Feasibility.} 
The outlined $\upmu$Ares mission concept would be less complex in many ways due to more stable orbits that could make some of the required subsystems in LISA obsolete (e.g. point-ahead angle mechanism, pilot tone correction). Furthermore, all dimensional and timing stability requirements are relaxed due to the lower maximum heterodyne frequency and a higher read-out noise level. This means that most other hardware developed for LISA will meet the $\upmu$Ares requirements with a margin. However, we identified a number of challenges that would require further research:
\begin{itemize}[topsep=4pt,itemsep=1pt,partopsep=0pt, parsep=0pt]
    \item Acceleration noise: LISA Pathfinder did measurements down to Fourier frequencies of $2 \times 10^{-5}$. It is not yet obvious that a slightly improved acceleration noise compared to LISA Pathfinder can be achieved over the entire $\upmu$-Hz range.
    \item Laser power: 10 watt laser fiber amplifiers are commercially available. It needs to be demonstrated that such amplifiers can meet a the relaxed phase fidelity requirement at full output power.
    \item Laser power stability: LISA has a relative stability requirement of $1 \times 10^{-8}$ at 5\,MHz. This noise increases towards lower frequencies and it needs to be shown that $5 \times 10^{-8}$ can be achieved at 2\,MHz.
    \item Orbit insertion: Upcoming launch vehicles like ``New Glenn'' by \emph{Blue Origin} (7 meter fairing) or ``Space Launch System'' by \emph{NASA} (8.4 meter fairing) are certainly large and powerful enough for the outlined $\upmu$Ares spacecraft. The planned ``Starship'' by \emph{SpaceX} (9 meter fairing) should even be capable to transport all spacecraft directly to a heliocentric Mars orbit. However, the delta-v required for 90 degree out of ecliptic plane orbit insertion could be extremely challenging.
    \item Ranging: The $\upmu$Ares constellation requires knowledge about the absolute spacecraft distance of $< 10$ meters. Depending on the Deep Space Network positioning accuracy, the on-board ranging algorithm might need to be improved.
\end{itemize}
    The overall concept benefits greatly from LISA technology heritage and upcoming cheap and powerful heavy-lift orbital launch vehicles. The compelling scientific benefits of the $\upmu$Ares mission concept demand a more detailed feasibility study which should include a more detailed look at the benefits and drawbacks of different orbits. Even only a single $\upmu$Ares-like constellation within the ecliptic plane will allow us to observe the very rich landscape of $\upmu$-Hz gravitational wave sources in great detail.

\vspace{0.5cm}
\fbox{\begin{minipage}{40em}
    {\bf Summary of strawman mission concept:}
    \begin{itemize}[topsep=0pt,itemsep=-1pt,partopsep=0pt, parsep=0pt]
    \item two equilateral triangle constellations with an arm length of 430 million km;
    \item heterodyne frequency range: $2...10$\,MHz (line-of-sight velocity $< 10$\,m/s);
    \item 1 meter telescope diameter, 10 watts laser output power: 35\,pm read-out noise limit;
    \item acceleration noise: $1 \times 10^{-15}$\,$\text{ms}^{-1}/\sqrt{\text{Hz}}$ down to $1\times10^{-7}$\,Hz.
    \end{itemize}  
\end{minipage}}

\pagebreak

\footnotesize{
\bibliographystyle{mnras}
\bibliography{biblio}

\begin{thebibliography}{}
\makeatletter
\relax
\def\mn@urlcharsother{\let\do\@makeother \do\$\do\&\do\#\do\^\do\_\do\%\do\~}
\def\mn@doi{\begingroup\mn@urlcharsother \@ifnextchar [ {\mn@doi@}
  {\mn@doi@[]}}
\def\mn@doi@[#1]#2{\def\@tempa{#1}\ifx\@tempa\@empty \href
  {http://dx.doi.org/#2} {doi:#2}\else \href {http://dx.doi.org/#2} {#1}\fi
  \endgroup}
\def\mn@eprint#1#2{\mn@eprint@#1:#2::\@nil}
\def\mn@eprint@arXiv#1{\href {http://arxiv.org/abs/#1} {{\tt arXiv:#1}}}
\def\mn@eprint@dblp#1{\href {http://dblp.uni-trier.de/rec/bibtex/#1.xml}
  {dblp:#1}}
\def\mn@eprint@#1:#2:#3:#4\@nil{\def\@tempa {#1}\def\@tempb {#2}\def\@tempc
  {#3}\ifx \@tempc \@empty \let \@tempc \@tempb \let \@tempb \@tempa \fi \ifx
  \@tempb \@empty \def\@tempb {arXiv}\fi \@ifundefined
  {mn@eprint@\@tempb}{\@tempb:\@tempc}{\expandafter \expandafter \csname
  mn@eprint@\@tempb\endcsname \expandafter{\@tempc}}}

\bibitem[\protect\citeauthoryear{Abada et~al.}{Abada
  et~al.}{2019}]{Benedikt:2018csr}
Abada A.,  et~al., 2019, \mn@doi [Eur. Phys. J. ST]
  {10.1140/epjst/e2019-900087-0}, 228, 755

\bibitem[\protect\citeauthoryear{{Abazajian} et~al.,}{{Abazajian}
  et~al.}{2016}]{2016arXiv161002743A}
{Abazajian} K.~N.,  et~al., 2016, arXiv e-prints, \href
  {https://ui.adsabs.harvard.edu/abs/2016arXiv161002743A} {p. arXiv:1610.02743}

\bibitem[\protect\citeauthoryear{Abbott et~al.}{Abbott
  et~al.}{2016a}]{TheLIGOScientific:2016pea}
Abbott B.~P.,  et~al., 2016a, \mn@doi [Phys. Rev.] {10.1103/PhysRevX.6.041015,
  10.1103/PhysRevX.8.039903}, X6, 041015

\bibitem[\protect\citeauthoryear{Abbott et~al.}{Abbott
  et~al.}{2016b}]{TheLIGOScientific:2016src}
Abbott B.~P.,  et~al., 2016b, \mn@doi [Phys. Rev. Lett.]
  {10.1103/PhysRevLett.116.221101, 10.1103/PhysRevLett.121.129902}, 116, 221101

\bibitem[\protect\citeauthoryear{{Abbott} et~al.,}{{Abbott}
  et~al.}{2017a}]{abb17a}
{Abbott} B.~P.,  et~al., 2017a, \mn@doi [\prl]
  {10.1103/PhysRevLett.119.161101}, \href
  {https://ui.adsabs.harvard.edu/abs/2017PhRvL.119p1101A} {119, 161101}

\bibitem[\protect\citeauthoryear{Abbott et~al.}{Abbott
  et~al.}{2017b}]{Abbott:2017xzu}
Abbott B.~P.,  et~al., 2017b, \mn@doi [Nature] {10.1038/nature24471}, 551, 85

\bibitem[\protect\citeauthoryear{{Abbott} et~al.,}{{Abbott}
  et~al.}{2017c}]{abb17b}
{Abbott} B.~P.,  et~al., 2017c, \mn@doi [\apjl] {10.3847/2041-8213/aa91c9},
  \href {https://ui.adsabs.harvard.edu/abs/2017ApJ...848L..12A} {848, L12}

\bibitem[\protect\citeauthoryear{Abbott et~al.}{Abbott
  et~al.}{2018}]{LIGOScientific:2018mvr}
Abbott B.~P.,  et~al., 2018

\bibitem[\protect\citeauthoryear{Aguirre et~al.}{Aguirre
  et~al.}{2019}]{Ade:2018sbj}
Aguirre J.,  et~al., 2019, \mn@doi [JCAP] {10.1088/1475-7516/2019/02/056},
  1902, 056

\bibitem[\protect\citeauthoryear{{Alexander}}{{Alexander}}{2017}]{Alexander2017}
{Alexander} T.,  2017, \mn@doi [\araa] {10.1146/annurev-astro-091916-055306},
  \href {https://ui.adsabs.harvard.edu/abs/2017ARA&A..55...17A} {55, 17}

\bibitem[\protect\citeauthoryear{{Ali-Ha{\"\i}moud} \&
  {Kamionkowski}}{{Ali-Ha{\"\i}moud} \&
  {Kamionkowski}}{2017}]{2017PhRvD..95d3534A}
{Ali-Ha{\"\i}moud} Y.,  {Kamionkowski} M.,  2017, \mn@doi [\prd]
  {10.1103/PhysRevD.95.043534}, \href
  {https://ui.adsabs.harvard.edu/abs/2017PhRvD..95d3534A} {95, 043534}

\bibitem[\protect\citeauthoryear{Amaro-Seoane}{Amaro-Seoane}{2019}]{Amaro-Seoane:2019umn}
Amaro-Seoane P.,  2019, \mn@doi [Phys. Rev.] {10.1103/PhysRevD.99.123025}, D99,
  123025

\bibitem[\protect\citeauthoryear{{Amaro-Seoane}, {Sesana}, {Hoffman},
  {Benacquista}, {Eichhorn}, {Makino}  \& {Spurzem}}{{Amaro-Seoane}
  et~al.}{2010}]{2010MNRAS.402.2308A}
{Amaro-Seoane} P.,  {Sesana} A.,  {Hoffman} L.,  {Benacquista} M.,  {Eichhorn}
  C.,  {Makino} J.,   {Spurzem} R.,  2010, \mn@doi [\mnras]
  {10.1111/j.1365-2966.2009.16104.x}, \href
  {https://ui.adsabs.harvard.edu/abs/2010MNRAS.402.2308A} {402, 2308}

\bibitem[\protect\citeauthoryear{{Amaro-Seoane} et~al.,}{{Amaro-Seoane}
  et~al.}{2012}]{2012CQGra..29l4016A}
{Amaro-Seoane} P.,  et~al., 2012, \mn@doi [Classical and Quantum Gravity]
  {10.1088/0264-9381/29/12/124016}, \href
  {https://ui.adsabs.harvard.edu/abs/2012CQGra..29l4016A} {29, 124016}

\bibitem[\protect\citeauthoryear{{Amaro-Seoane} et~al.,}{{Amaro-Seoane}
  et~al.}{2017}]{2017arXiv170200786A}
{Amaro-Seoane} P.,  et~al., 2017, arXiv e-prints, \href
  {https://ui.adsabs.harvard.edu/abs/2017arXiv170200786A} {p. arXiv:1702.00786}

\bibitem[\protect\citeauthoryear{{Arca-Sedda} \&
  {Capuzzo-Dolcetta}}{{Arca-Sedda} \&
  {Capuzzo-Dolcetta}}{2019}]{2019MNRAS.483..152A}
{Arca-Sedda} M.,  {Capuzzo-Dolcetta} R.,  2019, \mn@doi [\mnras]
  {10.1093/mnras/sty3096}, \href
  {https://ui.adsabs.harvard.edu/abs/2019MNRAS.483..152A} {483, 152}

\bibitem[\protect\citeauthoryear{{Arca-Sedda} \& {Gualandris}}{{Arca-Sedda} \&
  {Gualandris}}{2018}]{2018MNRAS.477.4423A}
{Arca-Sedda} M.,  {Gualandris} A.,  2018, \mn@doi [\mnras]
  {10.1093/mnras/sty922}, \href
  {https://ui.adsabs.harvard.edu/abs/2018MNRAS.477.4423A} {477, 4423}

\bibitem[\protect\citeauthoryear{{Arca Sedda}, {Askar}  \& {Giersz}}{{Arca
  Sedda} et~al.}{2019}]{2019arXiv190500902A}
{Arca Sedda} M.,  {Askar} A.,   {Giersz} M.,  2019, arXiv e-prints, \href
  {https://ui.adsabs.harvard.edu/abs/2019arXiv190500902A} {}

\bibitem[\protect\citeauthoryear{Armano et~al.,}{Armano
  et~al.}{2018}]{PhysRevLett.120.061101}
Armano M.,  et~al., 2018, \mn@doi [Phys. Rev. Lett.]
  {10.1103/PhysRevLett.120.061101}, 120, 061101

\bibitem[\protect\citeauthoryear{{Armitage} \& {Natarajan}}{{Armitage} \&
  {Natarajan}}{2002}]{2002ApJ...567L...9A}
{Armitage} P.~J.,  {Natarajan} P.,  2002, \mn@doi [\apjl] {10.1086/339770},
  \href {https://ui.adsabs.harvard.edu/abs/2002ApJ...567L...9A} {567, L9}

\bibitem[\protect\citeauthoryear{{Tauris}}{BHP}{}]{BHPToolkit}
{Black Hole Perturbation Toolkit},
  (\href{http://bhptoolkit.org/}{bhptoolkit.org})

\bibitem[\protect\citeauthoryear{{Ba{\~n}ados} et~al.,}{{Ba{\~n}ados}
  et~al.}{2018}]{2018Natur.553..473B}
{Ba{\~n}ados} E.,  et~al., 2018, \mn@doi [\nat] {10.1038/nature25180}, \href
  {https://ui.adsabs.harvard.edu/abs/2018Natur.553..473B} {553, 473}

\bibitem[\protect\citeauthoryear{{Babak} et~al.,}{{Babak}
  et~al.}{2017}]{2017PhRvD..95j3012B}
{Babak} S.,  et~al., 2017, \mn@doi [\prd] {10.1103/PhysRevD.95.103012}, \href
  {https://ui.adsabs.harvard.edu/abs/2017PhRvD..95j3012B} {95, 103012}

\bibitem[\protect\citeauthoryear{{Baibhav} \& {Berti}}{{Baibhav} \&
  {Berti}}{2019}]{2019PhRvD..99b4005B}
{Baibhav} V.,  {Berti} E.,  2019, \mn@doi [\prd] {10.1103/PhysRevD.99.024005},
  \href {https://ui.adsabs.harvard.edu/abs/2019PhRvD..99b4005B} {99, 024005}

\bibitem[\protect\citeauthoryear{Baker et~al.,}{Baker
  et~al.}{2019}]{1907.11305}
Baker J.,  et~al., 2019, Space Based Gravitational Wave Astronomy Beyond LISA
  (\mn@eprint {} {arXiv:1907.11305})

\bibitem[\protect\citeauthoryear{{Barausse} \& {Yagi}}{{Barausse} \&
  {Yagi}}{2015}]{2015PhRvL.115u1105B}
{Barausse} E.,  {Yagi} K.,  2015, \mn@doi [\prl]
  {10.1103/PhysRevLett.115.211105}, \href
  {https://ui.adsabs.harvard.edu/abs/2015PhRvL.115u1105B} {115, 211105}

\bibitem[\protect\citeauthoryear{Barausse, Palenzuela, Ponce  \&
  Lehner}{Barausse et~al.}{2013}]{PhysRevD.87.081506}
Barausse E.,  Palenzuela C.,  Ponce M.,   Lehner L.,  2013, \mn@doi [Phys. Rev.
  D] {10.1103/PhysRevD.87.081506}, 87, 081506

\bibitem[\protect\citeauthoryear{{Barausse}, {Cardoso}  \& {Pani}}{{Barausse}
  et~al.}{2014}]{2014PhRvD..89j4059B}
{Barausse} E.,  {Cardoso} V.,   {Pani} P.,  2014, \mn@doi [\prd]
  {10.1103/PhysRevD.89.104059}, \href
  {https://ui.adsabs.harvard.edu/abs/2014PhRvD..89j4059B} {89, 104059}

\bibitem[\protect\citeauthoryear{{Barausse}, {Yunes}  \&
  {Chamberlain}}{{Barausse} et~al.}{2016}]{2016PhRvL.116x1104B}
{Barausse} E.,  {Yunes} N.,   {Chamberlain} K.,  2016, \mn@doi [\prl]
  {10.1103/PhysRevLett.116.241104}, \href
  {https://ui.adsabs.harvard.edu/abs/2016PhRvL.116x1104B} {116, 241104}

\bibitem[\protect\citeauthoryear{{Barkana} \& {Loeb}}{{Barkana} \&
  {Loeb}}{2001}]{2001PhR...349..125B}
{Barkana} R.,  {Loeb} A.,  2001, \mn@doi [\physrep]
  {10.1016/S0370-1573(01)00019-9}, \href
  {https://ui.adsabs.harvard.edu/abs/2001PhR...349..125B} {349, 125}

\bibitem[\protect\citeauthoryear{Barke}{Barke}{2015}]{barke2015inter}
Barke S.,  2015, {Inter-spacecraft frequency distribution for future
  gravitational wave observatories}.
\url {http://simonbarke.com/download/Simon_Barke-PhD_Thesis_2015-vorgelegt.pdf}

\bibitem[\protect\citeauthoryear{Barke, Wang, Delgado, Tröbs, Heinzel  \&
  Danzmann}{Barke et~al.}{2015}]{Barke_2015}
Barke S.,  Wang Y.,  Delgado J. J.~E.,  Tröbs M.,  Heinzel G.,   Danzmann K.,
  2015, \mn@doi [Classical and Quantum Gravity]
  {10.1088/0264-9381/32/9/095004}, 32, 095004

\bibitem[\protect\citeauthoryear{Bartolo et~al.}{Bartolo
  et~al.}{2016a}]{Bartolo:2016ami}
Bartolo N.,  et~al., 2016a, \mn@doi [JCAP] {10.1088/1475-7516/2016/12/026},
  1612, 026

\bibitem[\protect\citeauthoryear{{Bartolo} et~al.,}{{Bartolo}
  et~al.}{2016b}]{2016JCAP...12..026B}
{Bartolo} N.,  et~al., 2016b, \mn@doi [\jcap] {10.1088/1475-7516/2016/12/026},
  \href {https://ui.adsabs.harvard.edu/abs/2016JCAP...12..026B} {2016, 026}

\bibitem[\protect\citeauthoryear{Bauswein et~al.,}{Bauswein
  et~al.}{2019}]{Bauswein:2019skm}
Bauswein A.,  et~al., 2019, \mn@doi [AIP Conf. Proc.] {10.1063/1.5117803},
  2127, 020013

\bibitem[\protect\citeauthoryear{{Bean} \& {Magueijo}}{{Bean} \&
  {Magueijo}}{2002}]{2002PhRvD..66f3505B}
{Bean} R.,  {Magueijo} J.,  2002, \mn@doi [\prd] {10.1103/PhysRevD.66.063505},
  \href {https://ui.adsabs.harvard.edu/abs/2002PhRvD..66f3505B} {66, 063505}

\bibitem[\protect\citeauthoryear{{Begelman}}{{Begelman}}{1979}]{1979MNRAS.187..237B}
{Begelman} M.~C.,  1979, \mn@doi [\mnras] {10.1093/mnras/187.2.237}, \href
  {https://ui.adsabs.harvard.edu/abs/1979MNRAS.187..237B} {187, 237}

\bibitem[\protect\citeauthoryear{{Begelman}, {Blandford}  \& {Rees}}{{Begelman}
  et~al.}{1980}]{1980Natur.287..307B}
{Begelman} M.~C.,  {Blandford} R.~D.,   {Rees} M.~J.,  1980, \mn@doi [\nat]
  {10.1038/287307a0}, \href
  {https://ui.adsabs.harvard.edu/abs/1980Natur.287..307B} {287, 307}

\bibitem[\protect\citeauthoryear{{Belgacem} et~al.,}{{Belgacem}
  et~al.}{2019}]{2019JCAP...07..024B}
{Belgacem} E.,  et~al., 2019, \mn@doi [\jcap] {10.1088/1475-7516/2019/07/024},
  \href {https://ui.adsabs.harvard.edu/abs/2019JCAP...07..024B} {2019, 024}

\bibitem[\protect\citeauthoryear{{Bellovary}, {Volonteri}, {Governato}, {Shen},
  {Quinn}  \& {Wadsley}}{{Bellovary} et~al.}{2011}]{2011ApJ...742...13B}
{Bellovary} J.,  {Volonteri} M.,  {Governato} F.,  {Shen} S.,  {Quinn} T.,
  {Wadsley} J.,  2011, \mn@doi [\apj] {10.1088/0004-637X/742/1/13}, \href
  {https://ui.adsabs.harvard.edu/abs/2011ApJ...742...13B} {742, 13}

\bibitem[\protect\citeauthoryear{Bernon, Bian  \& Jiang}{Bernon
  et~al.}{2018}]{Bernon:2017jgv}
Bernon J.,  Bian L.,   Jiang Y.,  2018, \mn@doi [JHEP]
  {10.1007/JHEP05(2018)151}, 05, 151

\bibitem[\protect\citeauthoryear{{Berti}, {Buonanno}  \& {Will}}{{Berti}
  et~al.}{2005}]{2005PhRvD..71h4025B}
{Berti} E.,  {Buonanno} A.,   {Will} C.~M.,  2005, \mn@doi [\prd]
  {10.1103/PhysRevD.71.084025}, \href
  {https://ui.adsabs.harvard.edu/abs/2005PhRvD..71h4025B} {71, 084025}

\bibitem[\protect\citeauthoryear{{Berti} et~al.,}{{Berti}
  et~al.}{2015}]{2015CQGra..32x3001B}
{Berti} E.,  et~al., 2015, \mn@doi [Classical and Quantum Gravity]
  {10.1088/0264-9381/32/24/243001}, \href
  {https://ui.adsabs.harvard.edu/abs/2015CQGra..32x3001B} {32, 243001}

\bibitem[\protect\citeauthoryear{{Blanchet} \& {Damour}}{{Blanchet} \&
  {Damour}}{1992}]{1992PhRvD..46.4304B}
{Blanchet} L.,  {Damour} T.,  1992, \mn@doi [\prd] {10.1103/PhysRevD.46.4304},
  \href {https://ui.adsabs.harvard.edu/abs/1992PhRvD..46.4304B} {46, 4304}

\bibitem[\protect\citeauthoryear{Blanchet, Spallicci  \& Whiting}{Blanchet
  et~al.}{2011}]{blspwh11}
Blanchet L.,  Spallicci A.,   Whiting B.,  eds, 2011, Mass and motion in
  general relativity.
 Fundamental Theories of Physics Vol. 162, Springer, Berlin

\bibitem[\protect\citeauthoryear{{Bode}, {Haas}, {Bogdanovi{\'c}}, {Laguna}  \&
  {Shoemaker}}{{Bode} et~al.}{2010}]{2010ApJ...715.1117B}
{Bode} T.,  {Haas} R.,  {Bogdanovi{\'c}} T.,  {Laguna} P.,   {Shoemaker} D.,
  2010, \mn@doi [\apj] {10.1088/0004-637X/715/2/1117}, \href
  {https://ui.adsabs.harvard.edu/abs/2010ApJ...715.1117B} {715, 1117}

\bibitem[\protect\citeauthoryear{{Bonetti}, {Haardt}, {Sesana}  \&
  {Barausse}}{{Bonetti} et~al.}{2018}]{2018MNRAS.477.3910B}
{Bonetti} M.,  {Haardt} F.,  {Sesana} A.,   {Barausse} E.,  2018, \mn@doi
  [\mnras] {10.1093/mnras/sty896}, \href
  {https://ui.adsabs.harvard.edu/abs/2018MNRAS.477.3910B} {477, 3910}

\bibitem[\protect\citeauthoryear{{Bonetti}, {Sesana}, {Haardt}, {Barausse}  \&
  {Colpi}}{{Bonetti} et~al.}{2019}]{2019MNRAS.486.4044B}
{Bonetti} M.,  {Sesana} A.,  {Haardt} F.,  {Barausse} E.,   {Colpi} M.,  2019,
  \mn@doi [\mnras] {10.1093/mnras/stz903}, \href
  {https://ui.adsabs.harvard.edu/abs/2019MNRAS.486.4044B} {486, 4044}

\bibitem[\protect\citeauthoryear{{Bowen}, {Campanelli}, {Krolik}, {Mewes}  \&
  {Noble}}{{Bowen} et~al.}{2017}]{2017ApJ...838...42B}
{Bowen} D.~B.,  {Campanelli} M.,  {Krolik} J.~H.,  {Mewes} V.,   {Noble} S.~C.,
   2017, \mn@doi [\apj] {10.3847/1538-4357/aa63f3}, \href
  {https://ui.adsabs.harvard.edu/abs/2017ApJ...838...42B} {838, 42}

\bibitem[\protect\citeauthoryear{{Bowman}, {Rogers}, {Monsalve}, {Mozdzen}  \&
  {Mahesh}}{{Bowman} et~al.}{2018}]{2018Natur.555...67B}
{Bowman} J.~D.,  {Rogers} A.~E.~E.,  {Monsalve} R.~A.,  {Mozdzen} T.~J.,
  {Mahesh} N.,  2018, \mn@doi [\nat] {10.1038/nature25792}, \href
  {https://ui.adsabs.harvard.edu/abs/2018Natur.555...67B} {555, 67}

\bibitem[\protect\citeauthoryear{Boyarsky, Ruchayskiy  \&
  Shaposhnikov}{Boyarsky et~al.}{2009}]{Boyarsky:2009ix}
Boyarsky A.,  Ruchayskiy O.,   Shaposhnikov M.,  2009, \mn@doi [Ann. Rev. Nucl.
  Part. Sci.] {10.1146/annurev.nucl.010909.083654}, 59, 191

\bibitem[\protect\citeauthoryear{{Braginskii} \& {Grishchuk}}{{Braginskii} \&
  {Grishchuk}}{1985}]{1985ZhETF..89..744B}
{Braginskii} V.~B.,  {Grishchuk} L.~P.,  1985, Zhurnal Eksperimentalnoi i
  Teoreticheskoi Fiziki, \href
  {https://ui.adsabs.harvard.edu/abs/1985ZhETF..89..744B} {89, 744}

\bibitem[\protect\citeauthoryear{{Bramberger}, {Brandenberger}, {Jreidini}  \&
  {Quintin}}{{Bramberger} et~al.}{2015}]{2015JCAP...06..007B}
{Bramberger} S.~F.,  {Brandenberger} R.~H.,  {Jreidini} P.,   {Quintin} J.,
  2015, \mn@doi [\jcap] {10.1088/1475-7516/2015/06/007}, \href
  {https://ui.adsabs.harvard.edu/abs/2015JCAP...06..007B} {2015, 007}

\bibitem[\protect\citeauthoryear{{Breivik}, {Rodriguez}, {Larson}, {Kalogera}
  \& {Rasio}}{{Breivik} et~al.}{2016}]{2016ApJ...830L..18B}
{Breivik} K.,  {Rodriguez} C.~L.,  {Larson} S.~L.,  {Kalogera} V.,   {Rasio}
  F.~A.,  2016, \mn@doi [\apjl] {10.3847/2041-8205/830/1/L18}, \href
  {https://ui.adsabs.harvard.edu/abs/2016ApJ...830L..18B} {830, L18}

\bibitem[\protect\citeauthoryear{{Brown}, {Gianninas}, {Kilic}, {Kenyon}  \&
  {Allende Prieto}}{{Brown} et~al.}{2016}]{bro16}
{Brown} W.~R.,  {Gianninas} A.,  {Kilic} M.,  {Kenyon} S.~J.,   {Allende
  Prieto} C.,  2016, \mn@doi [\apj] {10.3847/0004-637X/818/2/155}, \href
  {https://ui.adsabs.harvard.edu/abs/2016ApJ...818..155B} {818, 155}

\bibitem[\protect\citeauthoryear{Bruggisser, Von~Harling, Matsedonskyi  \&
  Servant}{Bruggisser et~al.}{2018a}]{Bruggisser:2018mrt}
Bruggisser S.,  Von~Harling B.,  Matsedonskyi O.,   Servant G.,  2018a, \mn@doi
  [JHEP] {10.1007/JHEP12(2018)099}, 12, 099

\bibitem[\protect\citeauthoryear{Bruggisser, Von~Harling, Matsedonskyi  \&
  Servant}{Bruggisser et~al.}{2018b}]{Bruggisser:2018mus}
Bruggisser S.,  Von~Harling B.,  Matsedonskyi O.,   Servant G.,  2018b, \mn@doi
  [Phys. Rev. Lett.] {10.1103/PhysRevLett.121.131801}, 121, 131801

\bibitem[\protect\citeauthoryear{{Cai}, {Tamanini}  \& {Yang}}{{Cai}
  et~al.}{2017}]{2017JCAP...05..031C}
{Cai} R.-G.,  {Tamanini} N.,   {Yang} T.,  2017, \mn@doi [\jcap]
  {10.1088/1475-7516/2017/05/031}, \href
  {https://ui.adsabs.harvard.edu/abs/2017JCAP...05..031C} {2017, 031}

\bibitem[\protect\citeauthoryear{{Callegari}, {Mayer}, {Kazantzidis}, {Colpi},
  {Governato}, {Quinn}  \& {Wadsley}}{{Callegari}
  et~al.}{2009}]{2009ApJ...696L..89C}
{Callegari} S.,  {Mayer} L.,  {Kazantzidis} S.,  {Colpi} M.,  {Governato} F.,
  {Quinn} T.,   {Wadsley} J.,  2009, \mn@doi [\apjl]
  {10.1088/0004-637X/696/1/L89}, \href
  {https://ui.adsabs.harvard.edu/abs/2009ApJ...696L..89C} {696, L89}

\bibitem[\protect\citeauthoryear{{Capelo}, {Volonteri}, {Dotti}, {Bellovary},
  {Mayer}  \& {Governato}}{{Capelo} et~al.}{2015}]{2015MNRAS.447.2123C}
{Capelo} P.~R.,  {Volonteri} M.,  {Dotti} M.,  {Bellovary} J.~M.,  {Mayer} L.,
   {Governato} F.,  2015, \mn@doi [\mnras] {10.1093/mnras/stu2500}, \href
  {https://ui.adsabs.harvard.edu/abs/2015MNRAS.447.2123C} {447, 2123}

\bibitem[\protect\citeauthoryear{Capozziello, Khodadi  \& Lambiase}{Capozziello
  et~al.}{2019}]{Capozziello:2018qjs}
Capozziello S.,  Khodadi M.,   Lambiase G.,  2019, \mn@doi [Phys. Lett.]
  {10.1016/j.physletb.2019.01.004}, B789, 626

\bibitem[\protect\citeauthoryear{Caprini \& Figueroa}{Caprini \&
  Figueroa}{2018}]{Caprini:2018mtu}
Caprini C.,  Figueroa D.~G.,  2018, \mn@doi [Class. Quant. Grav.]
  {10.1088/1361-6382/aac608}, 35, 163001

\bibitem[\protect\citeauthoryear{{Caprini} \& {Tamanini}}{{Caprini} \&
  {Tamanini}}{2016}]{2016JCAP...10..006C}
{Caprini} C.,  {Tamanini} N.,  2016, \mn@doi [\jcap]
  {10.1088/1475-7516/2016/10/006}, \href
  {https://ui.adsabs.harvard.edu/abs/2016JCAP...10..006C} {2016, 006}

\bibitem[\protect\citeauthoryear{Caprini, Durrer  \& Servant}{Caprini
  et~al.}{2009}]{Caprini:2009yp}
Caprini C.,  Durrer R.,   Servant G.,  2009, \mn@doi [JCAP]
  {10.1088/1475-7516/2009/12/024}, 0912, 024

\bibitem[\protect\citeauthoryear{Caprini, Durrer  \& Siemens}{Caprini
  et~al.}{2010}]{Caprini:2010xv}
Caprini C.,  Durrer R.,   Siemens X.,  2010, \mn@doi [Phys. Rev.]
  {10.1103/PhysRevD.82.063511}, D82, 063511

\bibitem[\protect\citeauthoryear{Caprini et~al.}{Caprini
  et~al.}{2016}]{Caprini:2015zlo}
Caprini C.,  et~al., 2016, \mn@doi [JCAP] {10.1088/1475-7516/2016/04/001},
  1604, 001

\bibitem[\protect\citeauthoryear{{Caprini}, {Figueroa}, {Flauger}, {Nardini},
  {Peloso}, {Pieroni}, {Ricciardone}  \& {Tasinato}}{{Caprini}
  et~al.}{2019}]{2019arXiv190609244C}
{Caprini} C.,  {Figueroa} D.~G.,  {Flauger} R.,  {Nardini} G.,  {Peloso} M.,
  {Pieroni} M.,  {Ricciardone} A.,   {Tasinato} G.,  2019, arXiv e-prints,
  \href {https://ui.adsabs.harvard.edu/abs/2019arXiv190609244C} {p.
  arXiv:1906.09244}

\bibitem[\protect\citeauthoryear{{Carr} \& {Silk}}{{Carr} \&
  {Silk}}{2018}]{2018MNRAS.478.3756C}
{Carr} B.,  {Silk} J.,  2018, \mn@doi [\mnras] {10.1093/mnras/sty1204}, \href
  {https://ui.adsabs.harvard.edu/abs/2018MNRAS.478.3756C} {478, 3756}

\bibitem[\protect\citeauthoryear{{Carr}, {Clesse}, {Garcia-Bellido}  \&
  {Kuhnel}}{{Carr} et~al.}{2019}]{2019arXiv190608217C}
{Carr} B.,  {Clesse} S.,  {Garcia-Bellido} J.,   {Kuhnel} F.,  2019, arXiv
  e-prints, \href {https://ui.adsabs.harvard.edu/abs/2019arXiv190608217C} {p.
  arXiv:1906.08217}

\bibitem[\protect\citeauthoryear{{Cerioli}, {Lodato}  \& {Price}}{{Cerioli}
  et~al.}{2016}]{2016MNRAS.457..939C}
{Cerioli} A.,  {Lodato} G.,   {Price} D.~J.,  2016, \mn@doi [\mnras]
  {10.1093/mnras/stw034}, \href
  {https://ui.adsabs.harvard.edu/abs/2016MNRAS.457..939C} {457, 939}

\bibitem[\protect\citeauthoryear{Chala, Nardini  \& Sobolev}{Chala
  et~al.}{2016}]{Chala:2016ykx}
Chala M.,  Nardini G.,   Sobolev I.,  2016, \mn@doi [Phys. Rev.]
  {10.1103/PhysRevD.94.055006}, D94, 055006

\bibitem[\protect\citeauthoryear{Chala, Ramos  \& Spannowsky}{Chala
  et~al.}{2019}]{Chala:2018opy}
Chala M.,  Ramos M.,   Spannowsky M.,  2019, \mn@doi [Eur. Phys. J.]
  {10.1140/epjc/s10052-019-6655-1}, C79, 156

\bibitem[\protect\citeauthoryear{{Chamberlain} \& {Yunes}}{{Chamberlain} \&
  {Yunes}}{2017}]{2017PhRvD..96h4039C}
{Chamberlain} K.,  {Yunes} N.,  2017, \mn@doi [\prd]
  {10.1103/PhysRevD.96.084039}, \href
  {https://ui.adsabs.harvard.edu/abs/2017PhRvD..96h4039C} {96, 084039}

\bibitem[\protect\citeauthoryear{{Chandrasekhar}}{{Chandrasekhar}}{1943}]{1943ApJ....97..255C}
{Chandrasekhar} S.,  1943, \mn@doi [\apj] {10.1086/144517}, \href
  {https://ui.adsabs.harvard.edu/abs/1943ApJ....97..255C} {97, 255}

\bibitem[\protect\citeauthoryear{{Chang}, {Strubbe}, {Menou}  \&
  {Quataert}}{{Chang} et~al.}{2010}]{2010MNRAS.407.2007C}
{Chang} P.,  {Strubbe} L.~E.,  {Menou} K.,   {Quataert} E.,  2010, \mn@doi
  [\mnras] {10.1111/j.1365-2966.2010.17056.x}, \href
  {https://ui.adsabs.harvard.edu/abs/2010MNRAS.407.2007C} {407, 2007}

\bibitem[\protect\citeauthoryear{{Chen}, {Middleton}, {Sesana}, {Del Pozzo}  \&
  {Vecchio}}{{Chen} et~al.}{2017}]{2017MNRAS.468..404C}
{Chen} S.,  {Middleton} H.,  {Sesana} A.,  {Del Pozzo} W.,   {Vecchio} A.,
  2017, \mn@doi [\mnras] {10.1093/mnras/stx475}, \href
  {https://ui.adsabs.harvard.edu/abs/2017MNRAS.468..404C} {468, 404}

\bibitem[\protect\citeauthoryear{{Christodoulou}}{{Christodoulou}}{1991}]{1991PhRvL..67.1486C}
{Christodoulou} D.,  1991, \mn@doi [\prl] {10.1103/PhysRevLett.67.1486}, \href
  {https://ui.adsabs.harvard.edu/abs/1991PhRvL..67.1486C} {67, 1486}

\bibitem[\protect\citeauthoryear{Chu et~al.}{Chu et~al.}{2018}]{Chu:2017gvt}
Chu D.~S.,  et~al., 2018, \mn@doi [Astrophys. J.] {10.3847/1538-4357/,
  10.3847/1538-4357/aaa3eb, 10.3847/1538-4357/aad734}, 854, 12

\bibitem[\protect\citeauthoryear{Clesse \& Garcia-Bellido}{Clesse \&
  Garcia-Bellido}{2015}]{Clesse:2015wea}
Clesse S.,  Garcia-Bellido J.,  2015, \mn@doi [Phys. Rev.]
  {10.1103/PhysRevD.92.023524}, D92, 023524

\bibitem[\protect\citeauthoryear{{Comp{\`e}re}}{{Comp{\`e}re}}{2019}]{2019PhRvL.123b1101C}
{Comp{\`e}re} G.,  2019, \mn@doi [\prl] {10.1103/PhysRevLett.123.021101}, \href
  {https://ui.adsabs.harvard.edu/abs/2019PhRvL.123b1101C} {123, 021101}

\bibitem[\protect\citeauthoryear{Cook \& Sorbo}{Cook \&
  Sorbo}{2012}]{Cook:2011hg}
Cook J.~L.,  Sorbo L.,  2012, \mn@doi [Phys. Rev.] {10.1103/PhysRevD.86.069901,
  10.1103/PhysRevD.85.023534}, D85, 023534

\bibitem[\protect\citeauthoryear{{Corrales}, {Haiman}  \&
  {MacFadyen}}{{Corrales} et~al.}{2010}]{2010MNRAS.404..947C}
{Corrales} L.~R.,  {Haiman} Z.,   {MacFadyen} A.,  2010, \mn@doi [\mnras]
  {10.1111/j.1365-2966.2010.16324.x}, \href
  {https://ui.adsabs.harvard.edu/abs/2010MNRAS.404..947C} {404, 947}

\bibitem[\protect\citeauthoryear{Csikor, Fodor  \& Heitger}{Csikor
  et~al.}{1999}]{Csikor:1998eu}
Csikor F.,  Fodor Z.,   Heitger J.,  1999, \mn@doi [Phys. Rev. Lett.]
  {10.1103/PhysRevLett.82.21}, 82, 21

\bibitem[\protect\citeauthoryear{{Cuadra}, {Armitage}, {Alexander}  \&
  {Begelman}}{{Cuadra} et~al.}{2009}]{2009MNRAS.393.1423C}
{Cuadra} J.,  {Armitage} P.~J.,  {Alexander} R.~D.,   {Begelman} M.~C.,  2009,
  \mn@doi [\mnras] {10.1111/j.1365-2966.2008.14147.x}, \href
  {https://ui.adsabs.harvard.edu/abs/2009MNRAS.393.1423C} {393, 1423}

\bibitem[\protect\citeauthoryear{{Damour} \& {Esposito-Farese}}{{Damour} \&
  {Esposito-Farese}}{1993}]{1993PhRvL..70.2220D}
{Damour} T.,  {Esposito-Farese} G.,  1993, \mn@doi [\prl]
  {10.1103/PhysRevLett.70.2220}, \href
  {https://ui.adsabs.harvard.edu/abs/1993PhRvL..70.2220D} {70, 2220}

\bibitem[\protect\citeauthoryear{{Deffayet} \& {Menou}}{{Deffayet} \&
  {Menou}}{2007}]{2007ApJ...668L.143D}
{Deffayet} C.,  {Menou} K.,  2007, \mn@doi [\apjl] {10.1086/522931}, \href
  {https://ui.adsabs.harvard.edu/abs/2007ApJ...668L.143D} {668, L143}

\bibitem[\protect\citeauthoryear{{Del Pozzo}, {Sesana}  \& {Klein}}{{Del Pozzo}
  et~al.}{2018}]{2018MNRAS.475.3485D}
{Del Pozzo} W.,  {Sesana} A.,   {Klein} A.,  2018, \mn@doi [\mnras]
  {10.1093/mnras/sty057}, \href
  {https://ui.adsabs.harvard.edu/abs/2018MNRAS.475.3485D} {475, 3485}

\bibitem[\protect\citeauthoryear{{Derdzinski}, {D'Orazio}, {Duffell}, {Haiman}
  \& {MacFadyen}}{{Derdzinski} et~al.}{2019}]{2019MNRAS.486.2754D}
{Derdzinski} A.~M.,  {D'Orazio} D.,  {Duffell} P.,  {Haiman} Z.,   {MacFadyen}
  A.,  2019, \mn@doi [\mnras] {10.1093/mnras/stz1026}, \href
  {https://ui.adsabs.harvard.edu/abs/2019MNRAS.486.2754D} {486, 2754}

\bibitem[\protect\citeauthoryear{{Dewdney}, {Hall}, {Schilizzi}  \&
  {Lazio}}{{Dewdney} et~al.}{2009}]{2009IEEEP..97.1482D}
{Dewdney} P.~E.,  {Hall} P.~J.,  {Schilizzi} R.~T.,   {Lazio} T.~J.~L.~W.,
  2009, \mn@doi [IEEE Proceedings] {10.1109/JPROC.2009.2021005}, \href
  {https://ui.adsabs.harvard.edu/abs/2009IEEEP..97.1482D} {97, 1482}

\bibitem[\protect\citeauthoryear{{Do} et~al.,}{{Do}
  et~al.}{2019}]{GC_Astro2020}
{Do} T.,  et~al., 2019, in \baas. p.~530 (\mn@eprint {arXiv} {1903.05293})

\bibitem[\protect\citeauthoryear{Doneva \& Yazadjiev}{Doneva \&
  Yazadjiev}{2018}]{PhysRevLett.120.131103}
Doneva D.~D.,  Yazadjiev S.~S.,  2018, \mn@doi [Phys. Rev. Lett.]
  {10.1103/PhysRevLett.120.131103}, 120, 131103

\bibitem[\protect\citeauthoryear{Dorsch, Huber, Konstandin  \& No}{Dorsch
  et~al.}{2017}]{Dorsch:2016nrg}
Dorsch G.~C.,  Huber S.~J.,  Konstandin T.,   No J.~M.,  2017, \mn@doi [JCAP]
  {10.1088/1475-7516/2017/05/052}, 1705, 052

\bibitem[\protect\citeauthoryear{{Dotti}, {Sesana}  \& {Decarli}}{{Dotti}
  et~al.}{2012}]{2012AdAst2012E...3D}
{Dotti} M.,  {Sesana} A.,   {Decarli} R.,  2012, \mn@doi [Advances in
  Astronomy] {10.1155/2012/940568}, \href
  {https://ui.adsabs.harvard.edu/abs/2012AdAst2012E...3D} {2012, 940568}

\bibitem[\protect\citeauthoryear{{Dubois}, {Volonteri}  \& {Silk}}{{Dubois}
  et~al.}{2014}]{2014MNRAS.440.1590D}
{Dubois} Y.,  {Volonteri} M.,   {Silk} J.,  2014, \mn@doi [\mnras]
  {10.1093/mnras/stu373}, \href
  {https://ui.adsabs.harvard.edu/abs/2014MNRAS.440.1590D} {440, 1590}

\bibitem[\protect\citeauthoryear{Durrer}{Durrer}{2008}]{Durrer:2008eom}
Durrer R.,  2008, {The Cosmic Microwave Background}.
Cambridge University Press, Cambridge, \mn@doi{10.1017/CBO9780511817205}

\bibitem[\protect\citeauthoryear{{Ebisuzaki} et~al.,}{{Ebisuzaki}
  et~al.}{2001}]{2001ApJ...562L..19E}
{Ebisuzaki} T.,  et~al., 2001, \mn@doi [\apjl] {10.1086/338118}, \href
  {https://ui.adsabs.harvard.edu/abs/2001ApJ...562L..19E} {562, L19}

\bibitem[\protect\citeauthoryear{{Elbert}, {Bullock}  \& {Kaplinghat}}{{Elbert}
  et~al.}{2018}]{2018MNRAS.473.1186E}
{Elbert} O.~D.,  {Bullock} J.~S.,   {Kaplinghat} M.,  2018, \mn@doi [\mnras]
  {10.1093/mnras/stx1959}, \href
  {https://ui.adsabs.harvard.edu/abs/2018MNRAS.473.1186E} {473, 1186}

\bibitem[\protect\citeauthoryear{{Ezquiaga} \& {Zumalac{\'a}rregui}}{{Ezquiaga}
  \& {Zumalac{\'a}rregui}}{2018}]{2018FrASS...5...44E}
{Ezquiaga} J.~M.,  {Zumalac{\'a}rregui} M.,  2018, \mn@doi [Frontiers in
  Astronomy and Space Sciences] {10.3389/fspas.2018.00044}, \href
  {https://ui.adsabs.harvard.edu/abs/2018FrASS...5...44E} {5, 44}

\bibitem[\protect\citeauthoryear{{Fan} et~al.,}{{Fan}
  et~al.}{2006}]{2006AJ....132..117F}
{Fan} X.,  et~al., 2006, \mn@doi [\aj] {10.1086/504836}, \href
  {https://ui.adsabs.harvard.edu/abs/2006AJ....132..117F} {132, 117}

\bibitem[\protect\citeauthoryear{Fang, Wu  \& Zhang}{Fang
  et~al.}{2019}]{Fang:2018axm}
Fang Z.,  Wu Y.-L.,   Zhang L.,  2019, \mn@doi [Phys. Rev.]
  {10.1103/PhysRevD.99.034028}, D99, 034028

\bibitem[\protect\citeauthoryear{{Favata}}{{Favata}}{2009}]{2009PhRvD..80b4002F}
{Favata} M.,  2009, \mn@doi [\prd] {10.1103/PhysRevD.80.024002}, \href
  {https://ui.adsabs.harvard.edu/abs/2009PhRvD..80b4002F} {80, 024002}

\bibitem[\protect\citeauthoryear{{Fiacconi}, {Mayer}, {Ro{\v{s}}kar}  \&
  {Colpi}}{{Fiacconi} et~al.}{2013}]{2013ApJ...777L..14F}
{Fiacconi} D.,  {Mayer} L.,  {Ro{\v{s}}kar} R.,   {Colpi} M.,  2013, \mn@doi
  [\apjl] {10.1088/2041-8205/777/1/L14}, \href
  {https://ui.adsabs.harvard.edu/abs/2013ApJ...777L..14F} {777, L14}

\bibitem[\protect\citeauthoryear{Figueroa, Megias, Nardini, Pieroni, Quiros,
  Ricciardone  \& Tasinato}{Figueroa et~al.}{2018}]{Figueroa:2018xtu}
Figueroa D.~G.,  Megias E.,  Nardini G.,  Pieroni M.,  Quiros M.,  Ricciardone
  A.,   Tasinato G.,  2018, \mn@doi [PoS] {10.22323/1.325.0036}, GRASS2018, 036

\bibitem[\protect\citeauthoryear{Fishbach et~al.}{Fishbach
  et~al.}{2019}]{Fishbach:2018gjp}
Fishbach M.,  et~al., 2019, \mn@doi [Astrophys. J.] {10.3847/2041-8213/aaf96e},
  871, L13

\bibitem[\protect\citeauthoryear{{Fowler}}{{Fowler}}{1966}]{1966ApJ...144..180F}
{Fowler} W.~A.,  1966, \mn@doi [\apj] {10.1086/148594}, \href
  {https://ui.adsabs.harvard.edu/abs/1966ApJ...144..180F} {144, 180}

\bibitem[\protect\citeauthoryear{{Freese}, {Rindler-Daller}, {Spolyar}  \&
  {Valluri}}{{Freese} et~al.}{2016}]{2016RPPh...79f6902F}
{Freese} K.,  {Rindler-Daller} T.,  {Spolyar} D.,   {Valluri} M.,  2016,
  \mn@doi [Reports on Progress in Physics] {10.1088/0034-4885/79/6/066902},
  \href {https://ui.adsabs.harvard.edu/abs/2016RPPh...79f6902F} {79, 066902}

\bibitem[\protect\citeauthoryear{Freitag}{Freitag}{2003}]{Freitag:2002nm}
Freitag M.,  2003, \mn@doi [Astrophys. J.] {10.1086/367813}, 583, L21

\bibitem[\protect\citeauthoryear{{Fujii} et~al.,}{{Fujii}
  et~al.}{2017}]{2017arXiv171007621F}
{Fujii} K.,  et~al., 2017, arXiv e-prints, \href
  {https://ui.adsabs.harvard.edu/abs/2017arXiv171007621F} {p. arXiv:1710.07621}

\bibitem[\protect\citeauthoryear{Garcia-Bellido \& Nesseris}{Garcia-Bellido \&
  Nesseris}{2018}]{Garcia-Bellido:2017knh}
Garcia-Bellido J.,  Nesseris S.,  2018, \mn@doi [Phys. Dark Univ.]
  {10.1016/j.dark.2018.06.001}, 21, 61

\bibitem[\protect\citeauthoryear{Garcia-Bellido, Linde  \&
  Wands}{Garcia-Bellido et~al.}{1996}]{GarciaBellido:1996qt}
Garcia-Bellido J.,  Linde A.~D.,   Wands D.,  1996, \mn@doi [Phys. Rev.]
  {10.1103/PhysRevD.54.6040}, D54, 6040

\bibitem[\protect\citeauthoryear{{Gardner} et~al.,}{{Gardner}
  et~al.}{2006}]{2006SSRv..123..485G}
{Gardner} J.~P.,  et~al., 2006, \mn@doi [\ssr] {10.1007/s11214-006-8315-7},
  \href {https://ui.adsabs.harvard.edu/abs/2006SSRv..123..485G} {123, 485}

\bibitem[\protect\citeauthoryear{{Gaskin} et~al.,}{{Gaskin}
  et~al.}{2017}]{2017SPIE10397E..0SG}
{Gaskin} J.~A.,  et~al., 2017, in \procspie. p. 103970S,
  \mn@doi{10.1117/12.2273911}

\bibitem[\protect\citeauthoryear{{Geier}, {Nesslinger}, {Heber}, {Przybilla},
  {Napiwotzki}  \& {Kudritzki}}{{Geier} et~al.}{2007}]{Geier06}
{Geier} S.,  {Nesslinger} S.,  {Heber} U.,  {Przybilla} N.,  {Napiwotzki} R.,
  {Kudritzki} R.~P.,  2007, in {Napiwotzki} R.,  {Burleigh} M.~R.,  eds,
  Astronomical Society of the Pacific Conference Series Vol. 372, 15th European
  Workshop on White Dwarfs. p.~393 (\mn@eprint {arXiv} {astro-ph/0612532})

\bibitem[\protect\citeauthoryear{{Geier} et~al.,}{{Geier}
  et~al.}{2013}]{Geier13}
{Geier} S.,  et~al., 2013, \mn@doi [\aap] {10.1051/0004-6361/201321395}, \href
  {http://adsabs.harvard.edu/abs/2013A%26A...554A..54G} {554, A54}

\bibitem[\protect\citeauthoryear{{Geier}, {Raddi}, {Gentile Fusillo}  \&
  {Marsh}}{{Geier} et~al.}{2019}]{2019A&A...621A..38G}
{Geier} S.,  {Raddi} R.,  {Gentile Fusillo} N.~P.,   {Marsh} T.~R.,  2019,
  \mn@doi [\aap] {10.1051/0004-6361/201834236}, \href
  {https://ui.adsabs.harvard.edu/abs/2019A&A...621A..38G} {621, A38}

\bibitem[\protect\citeauthoryear{{Genzel}, {Eisenhauer}  \&
  {Gillessen}}{{Genzel} et~al.}{2010}]{Genzel:2010}
{Genzel} R.,  {Eisenhauer} F.,   {Gillessen} S.,  2010, \mn@doi [Reviews of
  Modern Physics] {10.1103/RevModPhys.82.3121}, \href
  {https://ui.adsabs.harvard.edu/abs/2010RvMP...82.3121G} {82, 3121}

\bibitem[\protect\citeauthoryear{{Giersz}, {Leigh}, {Hypki}, {L{\"u}tzgendorf}
  \& {Askar}}{{Giersz} et~al.}{2015}]{2015MNRAS.454.3150G}
{Giersz} M.,  {Leigh} N.,  {Hypki} A.,  {L{\"u}tzgendorf} N.,   {Askar} A.,
  2015, \mn@doi [\mnras] {10.1093/mnras/stv2162}, \href
  {https://ui.adsabs.harvard.edu/abs/2015MNRAS.454.3150G} {454, 3150}

\bibitem[\protect\citeauthoryear{{Gilmozzi} \& {Spyromilio}}{{Gilmozzi} \&
  {Spyromilio}}{2007}]{2007Msngr.127...11G}
{Gilmozzi} R.,  {Spyromilio} J.,  2007, The Messenger, \href
  {https://ui.adsabs.harvard.edu/abs/2007Msngr.127...11G} {127, 11}

\bibitem[\protect\citeauthoryear{Gogoberidze, Kahniashvili  \&
  Kosowsky}{Gogoberidze et~al.}{2007}]{Gogoberidze:2007an}
Gogoberidze G.,  Kahniashvili T.,   Kosowsky A.,  2007, \mn@doi [Phys. Rev.]
  {10.1103/PhysRevD.76.083002}, D76, 083002

\bibitem[\protect\citeauthoryear{{Gondolo} \& {Silk}}{{Gondolo} \&
  {Silk}}{1999}]{1999PhRvL..83.1719G}
{Gondolo} P.,  {Silk} J.,  1999, \mn@doi [Physical Review Letters]
  {10.1103/PhysRevLett.83.1719}, \href
  {https://ui.adsabs.harvard.edu/abs/1999PhRvL..83.1719G} {83, 1719}

\bibitem[\protect\citeauthoryear{Gourgoulhon, Le~Tiec, Vincent  \&
  Warburton}{Gourgoulhon et~al.}{2019}]{Gourgoulhon:2019iyu}
Gourgoulhon E.,  Le~Tiec A.,  Vincent F.~H.,   Warburton N.,  2019, \mn@doi
  [Astron. Astrophys.] {10.1051/0004-6361/201935406}, 627, A92

\bibitem[\protect\citeauthoryear{{Greene}}{{Greene}}{2012}]{2012NatCo...3.1304G}
{Greene} J.~E.,  2012, \mn@doi [Nature Communications] {10.1038/ncomms2314},
  \href {https://ui.adsabs.harvard.edu/abs/2012NatCo...3.1304G} {3, 1304}

\bibitem[\protect\citeauthoryear{{Habouzit}, {Volonteri}  \&
  {Dubois}}{{Habouzit} et~al.}{2017}]{2017MNRAS.468.3935H}
{Habouzit} M.,  {Volonteri} M.,   {Dubois} Y.,  2017, \mn@doi [\mnras]
  {10.1093/mnras/stx666}, \href
  {https://ui.adsabs.harvard.edu/abs/2017MNRAS.468.3935H} {468, 3935}

\bibitem[\protect\citeauthoryear{{Haemmerl{\'e}}, {Woods}, {Klessen}, {Heger}
  \& {Whalen}}{{Haemmerl{\'e}} et~al.}{2018}]{2018MNRAS.474.2757H}
{Haemmerl{\'e}} L.,  {Woods} T.~E.,  {Klessen} R.~S.,  {Heger} A.,   {Whalen}
  D.~J.,  2018, \mn@doi [\mnras] {10.1093/mnras/stx2919}, \href
  {https://ui.adsabs.harvard.edu/abs/2018MNRAS.474.2757H} {474, 2757}

\bibitem[\protect\citeauthoryear{{Haiman}}{{Haiman}}{2004}]{2004ApJ...613...36H}
{Haiman} Z.,  2004, \mn@doi [\apj] {10.1086/422910}, \href
  {https://ui.adsabs.harvard.edu/abs/2004ApJ...613...36H} {613, 36}

\bibitem[\protect\citeauthoryear{{Haiman}}{{Haiman}}{2017}]{2017PhRvD..96b3004H}
{Haiman} Z.,  2017, \mn@doi [\prd] {10.1103/PhysRevD.96.023004}, \href
  {https://ui.adsabs.harvard.edu/abs/2017PhRvD..96b3004H} {96, 023004}

\bibitem[\protect\citeauthoryear{{Haiman} et~al.,}{{Haiman}
  et~al.}{2019}]{2019BAAS...51c.557H}
{Haiman} Z.,  et~al., 2019, in Bulletin of the American Astronomical Society.
  p.~557 (\mn@eprint {arXiv} {1903.08579})

\bibitem[\protect\citeauthoryear{{Hanany} et~al.,}{{Hanany}
  et~al.}{2019}]{2019arXiv190210541H}
{Hanany} S.,  et~al., 2019, arXiv e-prints, \href
  {https://ui.adsabs.harvard.edu/abs/2019arXiv190210541H} {p. arXiv:1902.10541}

\bibitem[\protect\citeauthoryear{{Hawking}, {Perry}  \& {Strominger}}{{Hawking}
  et~al.}{2016}]{2016PhRvL.116w1301H}
{Hawking} S.~W.,  {Perry} M.~J.,   {Strominger} A.,  2016, \mn@doi [\prl]
  {10.1103/PhysRevLett.116.231301}, \href
  {https://ui.adsabs.harvard.edu/abs/2016PhRvL.116w1301H} {116, 231301}

\bibitem[\protect\citeauthoryear{Hazumi et~al.}{Hazumi
  et~al.}{2019}]{Hazumi:2019lys}
Hazumi M.,  et~al., 2019, \mn@doi [J. Low. Temp. Phys.]
  {10.1007/s10909-019-02150-5}, 194, 443

\bibitem[\protect\citeauthoryear{Heinzel, Esteban, Barke, Otto, Wang, Garcia
  \& Danzmann}{Heinzel et~al.}{2011}]{Heinzel_2011}
Heinzel G.,  Esteban J.~J.,  Barke S.,  Otto M.,  Wang Y.,  Garcia A.~F.,
  Danzmann K.,  2011, \mn@doi [Classical and Quantum Gravity]
  {10.1088/0264-9381/28/9/094008}, 28, 094008

\bibitem[\protect\citeauthoryear{Hindmarsh, Huber, Rummukainen  \&
  Weir}{Hindmarsh et~al.}{2014}]{Hindmarsh:2013xza}
Hindmarsh M.,  Huber S.~J.,  Rummukainen K.,   Weir D.~J.,  2014, \mn@doi
  [Phys. Rev. Lett.] {10.1103/PhysRevLett.112.041301}, 112, 041301

\bibitem[\protect\citeauthoryear{Hindmarsh, Huber, Rummukainen  \&
  Weir}{Hindmarsh et~al.}{2015}]{Hindmarsh:2015qta}
Hindmarsh M.,  Huber S.~J.,  Rummukainen K.,   Weir D.~J.,  2015, \mn@doi
  [Phys. Rev.] {10.1103/PhysRevD.92.123009}, D92, 123009

\bibitem[\protect\citeauthoryear{{Hosokawa}, {Omukai}  \& {Yorke}}{{Hosokawa}
  et~al.}{2012}]{2012ApJ...756...93H}
{Hosokawa} T.,  {Omukai} K.,   {Yorke} H.~W.,  2012, \mn@doi [\apj]
  {10.1088/0004-637X/756/1/93}, \href
  {https://ui.adsabs.harvard.edu/abs/2012ApJ...756...93H} {756, 93}

\bibitem[\protect\citeauthoryear{{Hosokawa}, {Yorke}, {Inayoshi}, {Omukai}  \&
  {Yoshida}}{{Hosokawa} et~al.}{2013}]{2013ApJ...778..178H}
{Hosokawa} T.,  {Yorke} H.~W.,  {Inayoshi} K.,  {Omukai} K.,   {Yoshida} N.,
  2013, \mn@doi [\apj] {10.1088/0004-637X/778/2/178}, \href
  {https://ui.adsabs.harvard.edu/abs/2013ApJ...778..178H} {778, 178}

\bibitem[\protect\citeauthoryear{{Hulse} \& {Taylor}}{{Hulse} \&
  {Taylor}}{1975}]{hul75}
{Hulse} R.~A.,  {Taylor} J.~H.,  1975, \mn@doi [\apjl] {10.1086/181708}, \href
  {https://ui.adsabs.harvard.edu/abs/1975ApJ...195L..51H} {195, L51}

\bibitem[\protect\citeauthoryear{{Inayoshi}, {Haiman}  \&
  {Ostriker}}{{Inayoshi} et~al.}{2016}]{2016MNRAS.459.3738I}
{Inayoshi} K.,  {Haiman} Z.,   {Ostriker} J.~P.,  2016, \mn@doi [\mnras]
  {10.1093/mnras/stw836}, \href
  {https://ui.adsabs.harvard.edu/abs/2016MNRAS.459.3738I} {459, 3738}

\bibitem[\protect\citeauthoryear{{Islo}, {Simon}, {Burke-Spolaor}  \&
  {Siemens}}{{Islo} et~al.}{2019}]{2019arXiv190611936I}
{Islo} K.,  {Simon} J.,  {Burke-Spolaor} S.,   {Siemens} X.,  2019, arXiv
  e-prints, \href {https://ui.adsabs.harvard.edu/abs/2019arXiv190611936I} {p.
  arXiv:1906.11936}

\bibitem[\protect\citeauthoryear{Issaoun et~al.}{Issaoun
  et~al.}{2019}]{Issaoun:2019afg}
Issaoun S.,  et~al., 2019, \mn@doi [Astrophys. J.] {10.3847/1538-4357/aaf732},
  871, 30

\bibitem[\protect\citeauthoryear{{Ivanova} et~al.,}{{Ivanova}
  et~al.}{2013}]{iva13}
{Ivanova} N.,  et~al., 2013, \mn@doi [\aapr] {10.1007/s00159-013-0059-2}, \href
  {http://adsabs.harvard.edu/abs/2013A%26ARv..21...59I} {21, 59}

\bibitem[\protect\citeauthoryear{{Janssen} et~al.,}{{Janssen}
  et~al.}{2015}]{2015aska.confE..37J}
{Janssen} G.,  et~al., 2015, in Advancing Astrophysics with the Square
  Kilometre Array (AASKA14). p.~37 (\mn@eprint {arXiv} {1501.00127})

\bibitem[\protect\citeauthoryear{Jennrich}{Jennrich}{2019}]{pc_oliver2019}
Jennrich O.,  2019, {Private Communication}

\bibitem[\protect\citeauthoryear{Kajantie, Laine, Rummukainen  \&
  Shaposhnikov}{Kajantie et~al.}{1996}]{Kajantie:1995kf}
Kajantie K.,  Laine M.,  Rummukainen K.,   Shaposhnikov M.~E.,  1996, \mn@doi
  [Nucl. Phys.] {10.1016/0550-3213(96)00052-1}, B466, 189

\bibitem[\protect\citeauthoryear{Kakizaki, Kanemura  \& Matsui}{Kakizaki
  et~al.}{2015}]{Kakizaki:2015wua}
Kakizaki M.,  Kanemura S.,   Matsui T.,  2015, \mn@doi [Phys. Rev.]
  {10.1103/PhysRevD.92.115007}, D92, 115007

\bibitem[\protect\citeauthoryear{Kamionkowski, Kosowsky  \&
  Turner}{Kamionkowski et~al.}{1994}]{Kamionkowski:1993fg}
Kamionkowski M.,  Kosowsky A.,   Turner M.~S.,  1994, \mn@doi [Phys. Rev.]
  {10.1103/PhysRevD.49.2837}, D49, 2837

\bibitem[\protect\citeauthoryear{{Kelly}, {Baker}, {Etienne}, {Giacomazzo}  \&
  {Schnittman}}{{Kelly} et~al.}{2017}]{2017PhRvD..96l3003K}
{Kelly} B.~J.,  {Baker} J.~G.,  {Etienne} Z.~B.,  {Giacomazzo} B.,
  {Schnittman} J.,  2017, \mn@doi [\prd] {10.1103/PhysRevD.96.123003}, \href
  {https://ui.adsabs.harvard.edu/abs/2017PhRvD..96l3003K} {96, 123003}

\bibitem[\protect\citeauthoryear{{Khan}, {Just}  \& {Merritt}}{{Khan}
  et~al.}{2011}]{2011ApJ...732...89K}
{Khan} F.~M.,  {Just} A.,   {Merritt} D.,  2011, \mn@doi [\apj]
  {10.1088/0004-637X/732/2/89}, \href
  {https://ui.adsabs.harvard.edu/abs/2011ApJ...732...89K} {732, 89}

\bibitem[\protect\citeauthoryear{{Khan}, {Berczik}  \& {Just}}{{Khan}
  et~al.}{2018a}]{1803.11394}
{Khan} F.~M.,  {Berczik} P.,   {Just} A.,  2018a, \mn@doi [\aap]
  {10.1051/0004-6361/201730489}, \href
  {http://adsabs.harvard.edu/abs/2018A%26A...615A..71K} {615, A71}

\bibitem[\protect\citeauthoryear{{Khan}, {Capelo}, {Mayer}  \&
  {Berczik}}{{Khan} et~al.}{2018b}]{2018ApJ...868...97K}
{Khan} F.~M.,  {Capelo} P.~R.,  {Mayer} L.,   {Berczik} P.,  2018b, \mn@doi
  [\apj] {10.3847/1538-4357/aae77b}, \href
  {https://ui.adsabs.harvard.edu/abs/2018ApJ...868...97K} {868, 97}

\bibitem[\protect\citeauthoryear{Kibble}{Kibble}{1976}]{kibble_topology_1976}
Kibble T. W.~B.,  1976, \mn@doi [Journal of Physics A: Mathematical and
  General] {10.1088/0305-4470/9/8/029}, 9, 1387

\bibitem[\protect\citeauthoryear{{Klein}, {Cornish}  \& {Yunes}}{{Klein}
  et~al.}{2014}]{2014PhRvD..90l4029K}
{Klein} A.,  {Cornish} N.,   {Yunes} N.,  2014, \mn@doi [\prd]
  {10.1103/PhysRevD.90.124029}, \href
  {https://ui.adsabs.harvard.edu/abs/2014PhRvD..90l4029K} {90, 124029}

\bibitem[\protect\citeauthoryear{{Klein} et~al.,}{{Klein}
  et~al.}{2016}]{2016PhRvD..93b4003K}
{Klein} A.,  et~al., 2016, \mn@doi [\prd] {10.1103/PhysRevD.93.024003}, \href
  {https://ui.adsabs.harvard.edu/abs/2016PhRvD..93b4003K} {93, 024003}

\bibitem[\protect\citeauthoryear{{Knigge}}{{Knigge}}{2010}]{Knigge10}
{Knigge} C.,  2010, in {Kalogera} V.,  {van der Sluys} M.,  eds,  American
  Institute of Physics Conference Series Vol. 1314, American Institute of
  Physics Conference Series. pp 171--180, \mn@doi{10.1063/1.3536361}

\bibitem[\protect\citeauthoryear{{Kocsis}, {Haiman}  \& {Menou}}{{Kocsis}
  et~al.}{2008}]{2008ApJ...684..870K}
{Kocsis} B.,  {Haiman} Z.,   {Menou} K.,  2008, \mn@doi [\apj]
  {10.1086/590230}, \href
  {https://ui.adsabs.harvard.edu/abs/2008ApJ...684..870K} {684, 870}

\bibitem[\protect\citeauthoryear{{Kocsis}, {Yunes}  \& {Loeb}}{{Kocsis}
  et~al.}{2011}]{2011PhRvD..84b4032K}
{Kocsis} B.,  {Yunes} N.,   {Loeb} A.,  2011, \mn@doi [\prd]
  {10.1103/PhysRevD.84.024032}, \href
  {https://ui.adsabs.harvard.edu/abs/2011PhRvD..84b4032K} {84, 024032}

\bibitem[\protect\citeauthoryear{{Kormendy} \& {Ho}}{{Kormendy} \&
  {Ho}}{2013}]{2013ARA&A..51..511K}
{Kormendy} J.,  {Ho} L.~C.,  2013, \mn@doi [\araa]
  {10.1146/annurev-astro-082708-101811}, \href
  {https://ui.adsabs.harvard.edu/abs/2013ARA&A..51..511K} {51, 511}

\bibitem[\protect\citeauthoryear{{Korol}, {Rossi}, {Groot}, {Nelemans},
  {Toonen}  \& {Brown}}{{Korol} et~al.}{2017}]{kor17}
{Korol} V.,  {Rossi} E.~M.,  {Groot} P.~J.,  {Nelemans} G.,  {Toonen} S.,
  {Brown} A. G.~A.,  2017, \mn@doi [\mnras] {10.1093/mnras/stx1285}, \href
  {https://ui.adsabs.harvard.edu/abs/2017MNRAS.470.1894K} {470, 1894}

\bibitem[\protect\citeauthoryear{{Korol}, {Koop}  \& {Rossi}}{{Korol}
  et~al.}{2018}]{kor18}
{Korol} V.,  {Koop} O.,   {Rossi} E.~M.,  2018, \mn@doi [\apjl]
  {10.3847/2041-8213/aae587}, \href
  {https://ui.adsabs.harvard.edu/abs/2018ApJ...866L..20K} {866, L20}

\bibitem[\protect\citeauthoryear{{Korol}, {Rossi}  \& {Barausse}}{{Korol}
  et~al.}{2019}]{kor19}
{Korol} V.,  {Rossi} E.~M.,   {Barausse} E.,  2019, \mn@doi [\mnras]
  {10.1093/mnras/sty3440}, \href
  {https://ui.adsabs.harvard.edu/abs/2019MNRAS.483.5518K} {483, 5518}

\bibitem[\protect\citeauthoryear{Kosowsky \& Turner}{Kosowsky \&
  Turner}{1993}]{Kosowsky:1992vn}
Kosowsky A.,  Turner M.~S.,  1993, \mn@doi [Phys. Rev.]
  {10.1103/PhysRevD.47.4372}, D47, 4372

\bibitem[\protect\citeauthoryear{Kozaczuk}{Kozaczuk}{2015}]{Kozaczuk:2015owa}
Kozaczuk J.,  2015, \mn@doi [JHEP] {10.1007/JHEP10(2015)135}, 10, 135

\bibitem[\protect\citeauthoryear{{Kulier}, {Ostriker}, {Natarajan}, {Lackner}
  \& {Cen}}{{Kulier} et~al.}{2015}]{2015ApJ...799..178K}
{Kulier} A.,  {Ostriker} J.~P.,  {Natarajan} P.,  {Lackner} C.~N.,   {Cen} R.,
  2015, \mn@doi [\apj] {10.1088/0004-637X/799/2/178}, \href
  {https://ui.adsabs.harvard.edu/abs/2015ApJ...799..178K} {799, 178}

\bibitem[\protect\citeauthoryear{{Kupfer} et~al.,}{{Kupfer}
  et~al.}{2018}]{kup18}
{Kupfer} T.,  et~al., 2018, preprint, \href
  {http://adsabs.harvard.edu/abs/2018arXiv180500482K} {} (\mn@eprint {arXiv}
  {1805.00482})

\bibitem[\protect\citeauthoryear{Kyutoku \& Seto}{Kyutoku \&
  Seto}{2017}]{Kyutoku:2016zxn}
Kyutoku K.,  Seto N.,  2017, \mn@doi [Phys. Rev.] {10.1103/PhysRevD.95.083525},
  D95, 083525

\bibitem[\protect\citeauthoryear{{Lamberts} et~al.,}{{Lamberts}
  et~al.}{2018}]{2018MNRAS.480.2704L}
{Lamberts} A.,  et~al., 2018, \mn@doi [\mnras] {10.1093/mnras/sty2035}, \href
  {https://ui.adsabs.harvard.edu/abs/2018MNRAS.480.2704L} {480, 2704}

\bibitem[\protect\citeauthoryear{{Lamberts}, {Blunt}, {Littenberg},
  {Garrison-Kimmel}, {Kupfer}  \& {Sanderson}}{{Lamberts} et~al.}{2019}]{lam19}
{Lamberts} A.,  {Blunt} S.,  {Littenberg} T.,  {Garrison-Kimmel} S.,  {Kupfer}
  T.,   {Sanderson} R.,  2019, arXiv e-prints, \href
  {https://ui.adsabs.harvard.edu/abs/2019arXiv190700014L} {p. arXiv:1907.00014}

\bibitem[\protect\citeauthoryear{{Li}, {Yang}  \& {Yuan}}{{Li}
  et~al.}{2018}]{2018arXiv181209676L}
{Li} M.-W.,  {Yang} Y.,   {Yuan} P.-H.,  2018, arXiv e-prints, \href
  {https://ui.adsabs.harvard.edu/abs/2018arXiv181209676L} {p. arXiv:1812.09676}

\bibitem[\protect\citeauthoryear{{Lippai}, {Frei}  \& {Haiman}}{{Lippai}
  et~al.}{2008}]{2008ApJ...676L...5L}
{Lippai} Z.,  {Frei} Z.,   {Haiman} Z.,  2008, \mn@doi [\apjl]
  {10.1086/587034}, \href
  {https://ui.adsabs.harvard.edu/abs/2008ApJ...676L...5L} {676, L5}

\bibitem[\protect\citeauthoryear{Littenberg \& Yunes}{Littenberg \&
  Yunes}{2019}]{Littenberg:2018xxx}
Littenberg T.~B.,  Yunes N.,  2019, \mn@doi [Class. Quant. Grav.]
  {10.1088/1361-6382/ab0a3d}, 36, 095017

\bibitem[\protect\citeauthoryear{{Lupi}, {Haardt}, {Dotti}  \& {Colpi}}{{Lupi}
  et~al.}{2015}]{2015MNRAS.453.3437L}
{Lupi} A.,  {Haardt} F.,  {Dotti} M.,   {Colpi} M.,  2015, \mn@doi [\mnras]
  {10.1093/mnras/stv1920}, \href
  {https://ui.adsabs.harvard.edu/abs/2015MNRAS.453.3437L} {453, 3437}

\bibitem[\protect\citeauthoryear{{Lupi}, {Haardt}, {Dotti}, {Fiacconi}, {Mayer}
   \& {Madau}}{{Lupi} et~al.}{2016}]{2016MNRAS.456.2993L}
{Lupi} A.,  {Haardt} F.,  {Dotti} M.,  {Fiacconi} D.,  {Mayer} L.,   {Madau}
  P.,  2016, \mn@doi [\mnras] {10.1093/mnras/stv2877}, \href
  {https://ui.adsabs.harvard.edu/abs/2016MNRAS.456.2993L} {456, 2993}

\bibitem[\protect\citeauthoryear{{Lyne} \& {Lorimer}}{{Lyne} \&
  {Lorimer}}{1994}]{lyn94}
{Lyne} A.~G.,  {Lorimer} D.~R.,  1994, \mn@doi [\nat] {10.1038/369127a0}, \href
  {https://ui.adsabs.harvard.edu/abs/1994Natur.369..127L} {369, 127}

\bibitem[\protect\citeauthoryear{{MacLeod} \& {Hogan}}{{MacLeod} \&
  {Hogan}}{2008}]{2008PhRvD..77d3512M}
{MacLeod} C.~L.,  {Hogan} C.~J.,  2008, \mn@doi [\prd]
  {10.1103/PhysRevD.77.043512}, \href
  {https://ui.adsabs.harvard.edu/abs/2008PhRvD..77d3512M} {77, 043512}

\bibitem[\protect\citeauthoryear{{Madau} \& {Rees}}{{Madau} \&
  {Rees}}{2001}]{2001ApJ...551L..27M}
{Madau} P.,  {Rees} M.~J.,  2001, \mn@doi [\apjl] {10.1086/319848}, \href
  {https://ui.adsabs.harvard.edu/abs/2001ApJ...551L..27M} {551, L27}

\bibitem[\protect\citeauthoryear{{Malbon}, {Baugh}, {Frenk}  \&
  {Lacey}}{{Malbon} et~al.}{2007}]{2007MNRAS.382.1394M}
{Malbon} R.~K.,  {Baugh} C.~M.,  {Frenk} C.~S.,   {Lacey} C.~G.,  2007, \mn@doi
  [\mnras] {10.1111/j.1365-2966.2007.12317.x}, \href
  {https://ui.adsabs.harvard.edu/abs/2007MNRAS.382.1394M} {382, 1394}

\bibitem[\protect\citeauthoryear{{Maoz} \& {Hallakoun}}{{Maoz} \&
  {Hallakoun}}{2017}]{mao17}
{Maoz} D.,  {Hallakoun} N.,  2017, \mn@doi [\mnras] {10.1093/mnras/stx102},
  \href {https://ui.adsabs.harvard.edu/abs/2017MNRAS.467.1414M} {467, 1414}

\bibitem[\protect\citeauthoryear{{Mastrobuono-Battisti}, {Perets}  \&
  {Loeb}}{{Mastrobuono-Battisti} et~al.}{2014}]{2014ApJ...796...40M}
{Mastrobuono-Battisti} A.,  {Perets} H.~B.,   {Loeb} A.,  2014, \mn@doi [\apj]
  {10.1088/0004-637X/796/1/40}, \href
  {https://ui.adsabs.harvard.edu/abs/2014ApJ...796...40M} {796, 40}

\bibitem[\protect\citeauthoryear{{Max}, {Platscher}  \& {Smirnov}}{{Max}
  et~al.}{2017}]{2017PhRvL.119k1101M}
{Max} K.,  {Platscher} M.,   {Smirnov} J.,  2017, \mn@doi [\prl]
  {10.1103/PhysRevLett.119.111101}, \href
  {https://ui.adsabs.harvard.edu/abs/2017PhRvL.119k1101M} {119, 111101}

\bibitem[\protect\citeauthoryear{{Mayer}, {Fiacconi}, {Bonoli}, {Quinn},
  {Ro{\v{s}}kar}, {Shen}  \& {Wadsley}}{{Mayer}
  et~al.}{2015}]{2015ApJ...810...51M}
{Mayer} L.,  {Fiacconi} D.,  {Bonoli} S.,  {Quinn} T.,  {Ro{\v{s}}kar} R.,
  {Shen} S.,   {Wadsley} J.,  2015, \mn@doi [\apj]
  {10.1088/0004-637X/810/1/51}, \href
  {https://ui.adsabs.harvard.edu/abs/2015ApJ...810...51M} {810, 51}

\bibitem[\protect\citeauthoryear{Megías, Nardini  \& Quirós}{Megías
  et~al.}{2018}]{Megias:2018sxv}
Megías E.,  Nardini G.,   Quirós M.,  2018, \mn@doi [JHEP]
  {10.1007/JHEP09(2018)095}, 09, 095

\bibitem[\protect\citeauthoryear{{Mezcua}, {Civano}, {Marchesi}, {Suh},
  {Fabbiano}  \& {Volonteri}}{{Mezcua} et~al.}{2018}]{2018MNRAS.478.2576M}
{Mezcua} M.,  {Civano} F.,  {Marchesi} S.,  {Suh} H.,  {Fabbiano} G.,
  {Volonteri} M.,  2018, \mn@doi [\mnras] {10.1093/mnras/sty1163}, \href
  {https://ui.adsabs.harvard.edu/abs/2018MNRAS.478.2576M} {478, 2576}

\bibitem[\protect\citeauthoryear{{Moody}, {Shi}  \& {Stone}}{{Moody}
  et~al.}{2019}]{2019ApJ...875...66M}
{Moody} M. S.~L.,  {Shi} J.-M.,   {Stone} J.~M.,  2019, \mn@doi [\apj]
  {10.3847/1538-4357/ab09ee}, \href
  {https://ui.adsabs.harvard.edu/abs/2019ApJ...875...66M} {875, 66}

\bibitem[\protect\citeauthoryear{{Mortlock} et~al.,}{{Mortlock}
  et~al.}{2011}]{2011Natur.474..616M}
{Mortlock} D.~J.,  et~al., 2011, \mn@doi [\nat] {10.1038/nature10159}, \href
  {https://ui.adsabs.harvard.edu/abs/2011Natur.474..616M} {474, 616}

\bibitem[\protect\citeauthoryear{{Mu{\~n}oz}, {Miranda}  \& {Lai}}{{Mu{\~n}oz}
  et~al.}{2019}]{2019ApJ...871...84M}
{Mu{\~n}oz} D.~J.,  {Miranda} R.,   {Lai} D.,  2019, \mn@doi [\apj]
  {10.3847/1538-4357/aaf867}, \href
  {https://ui.adsabs.harvard.edu/abs/2019ApJ...871...84M} {871, 84}

\bibitem[\protect\citeauthoryear{{Nandra} et~al.,}{{Nandra}
  et~al.}{2013}]{2013arXiv1306.2307N}
{Nandra} K.,  et~al., 2013, arXiv e-prints, \href
  {https://ui.adsabs.harvard.edu/abs/2013arXiv1306.2307N} {p. arXiv:1306.2307}

\bibitem[\protect\citeauthoryear{{Nebot G{\'o}mez-Mor{\'a}n} et~al.,}{{Nebot
  G{\'o}mez-Mor{\'a}n} et~al.}{2011}]{2011A&A...536A..43N}
{Nebot G{\'o}mez-Mor{\'a}n} A.,  et~al., 2011, \mn@doi [\aap]
  {10.1051/0004-6361/201117514}, \href
  {https://ui.adsabs.harvard.edu/abs/2011A&A...536A..43N} {536, A43}

\bibitem[\protect\citeauthoryear{{Nelemans}, {Yungelson}  \& {Portegies
  Zwart}}{{Nelemans} et~al.}{2001}]{nel01a}
{Nelemans} G.,  {Yungelson} L.~R.,   {Portegies Zwart} S.~F.,  2001, \mn@doi
  [\aap] {10.1051/0004-6361:20010683}, \href
  {https://ui.adsabs.harvard.edu/abs/2001A&A...375..890N} {375, 890}

\bibitem[\protect\citeauthoryear{{Nishizawa}, {Berti}, {Klein}  \&
  {Sesana}}{{Nishizawa} et~al.}{2016}]{2016PhRvD..94f4020N}
{Nishizawa} A.,  {Berti} E.,  {Klein} A.,   {Sesana} A.,  2016, \mn@doi [\prd]
  {10.1103/PhysRevD.94.064020}, \href
  {https://ui.adsabs.harvard.edu/abs/2016PhRvD..94f4020N} {94, 064020}

\bibitem[\protect\citeauthoryear{{Nishizawa}, {Sesana}, {Berti}  \&
  {Klein}}{{Nishizawa} et~al.}{2017}]{2017MNRAS.465.4375N}
{Nishizawa} A.,  {Sesana} A.,  {Berti} E.,   {Klein} A.,  2017, \mn@doi
  [\mnras] {10.1093/mnras/stw2993}, \href
  {https://ui.adsabs.harvard.edu/abs/2017MNRAS.465.4375N} {465, 4375}

\bibitem[\protect\citeauthoryear{{Pacucci} \& {Ferrara}}{{Pacucci} \&
  {Ferrara}}{2015}]{2015MNRAS.448..104P}
{Pacucci} F.,  {Ferrara} A.,  2015, \mn@doi [\mnras] {10.1093/mnras/stv018},
  \href {https://ui.adsabs.harvard.edu/abs/2015MNRAS.448..104P} {448, 104}

\bibitem[\protect\citeauthoryear{{Pacucci} et~al.,}{{Pacucci}
  et~al.}{2019}]{2019BAAS...51c.117P}
{Pacucci} F.,  et~al., 2019, in Bulletin of the American Astronomical Society.
  p.~117 (\mn@eprint {arXiv} {1903.07623})

\bibitem[\protect\citeauthoryear{{Palenzuela}, {Barausse}, {Ponce}  \&
  {Lehner}}{{Palenzuela} et~al.}{2014}]{2014PhRvD..89d4024P}
{Palenzuela} C.,  {Barausse} E.,  {Ponce} M.,   {Lehner} L.,  2014, \mn@doi
  [Physical Review D] {10.1103/PhysRevD.89.044024}, \href
  {https://ui.adsabs.harvard.edu/abs/2014PhRvD..89d4024P} {89, 044024}

\bibitem[\protect\citeauthoryear{{Petiteau}, {Babak}  \& {Sesana}}{{Petiteau}
  et~al.}{2011}]{2011ApJ...732...82P}
{Petiteau} A.,  {Babak} S.,   {Sesana} A.,  2011, \mn@doi [The Astrophysical
  Journal] {10.1088/0004-637X/732/2/82}, \href
  {https://ui.adsabs.harvard.edu/abs/2011ApJ...732...82P} {732, 82}

\bibitem[\protect\citeauthoryear{{Pfister}, {Lupi}, {Capelo}, {Volonteri},
  {Bellovary}  \& {Dotti}}{{Pfister} et~al.}{2017}]{2017MNRAS.471.3646P}
{Pfister} H.,  {Lupi} A.,  {Capelo} P.~R.,  {Volonteri} M.,  {Bellovary} J.~M.,
    {Dotti} M.,  2017, \mn@doi [\mnras] {10.1093/mnras/stx1853}, \href
  {https://ui.adsabs.harvard.edu/abs/2017MNRAS.471.3646P} {471, 3646}

\bibitem[\protect\citeauthoryear{{Phinney}}{{Phinney}}{1991}]{phi91}
{Phinney} E.~S.,  1991, \mn@doi [\apjl] {10.1086/186163}, \href
  {https://ui.adsabs.harvard.edu/abs/1991ApJ...380L..17P} {380, L17}

\bibitem[\protect\citeauthoryear{{Planck Collaboration} et~al.,}{{Planck
  Collaboration} et~al.}{2018}]{2018arXiv180706209P}
{Planck Collaboration} et~al., 2018, arXiv e-prints, \href
  {https://ui.adsabs.harvard.edu/abs/2018arXiv180706209P} {p. arXiv:1807.06209}

\bibitem[\protect\citeauthoryear{{Portegies Zwart}, {Baumgardt}, {McMillan},
  {Makino}, {Hut}  \& {Ebisuzaki}}{{Portegies Zwart}
  et~al.}{2006}]{2006ApJ...641..319P}
{Portegies Zwart} S.~F.,  {Baumgardt} H.,  {McMillan} S.~L.~W.,  {Makino} J.,
  {Hut} P.,   {Ebisuzaki} T.,  2006, \mn@doi [\apj] {10.1086/500361}, \href
  {https://ui.adsabs.harvard.edu/abs/2006ApJ...641..319P} {641, 319}

\bibitem[\protect\citeauthoryear{{Postnov} \& {Yungelson}}{{Postnov} \&
  {Yungelson}}{2014}]{2014LRR....17....3P}
{Postnov} K.~A.,  {Yungelson} L.~R.,  2014, \mn@doi [Living Reviews in
  Relativity] {10.12942/lrr-2014-3}, \href
  {https://ui.adsabs.harvard.edu/abs/2014LRR....17....3P} {17, 3}

\bibitem[\protect\citeauthoryear{{Poulin}, {Serpico}, {Calore}, {Clesse}  \&
  {Kohri}}{{Poulin} et~al.}{2017}]{2017PhRvD..96h3524P}
{Poulin} V.,  {Serpico} P.~D.,  {Calore} F.,  {Clesse} S.,   {Kohri} K.,  2017,
  \mn@doi [\prd] {10.1103/PhysRevD.96.083524}, \href
  {https://ui.adsabs.harvard.edu/abs/2017PhRvD..96h3524P} {96, 083524}

\bibitem[\protect\citeauthoryear{{Preto}, {Berentzen}, {Berczik}  \&
  {Spurzem}}{{Preto} et~al.}{2011}]{2011ApJ...732L..26P}
{Preto} M.,  {Berentzen} I.,  {Berczik} P.,   {Spurzem} R.,  2011, \mn@doi
  [\apjl] {10.1088/2041-8205/732/2/L26}, \href
  {https://ui.adsabs.harvard.edu/abs/2011ApJ...732L..26P} {732, L26}

\bibitem[\protect\citeauthoryear{{Punturo} et~al.,}{{Punturo}
  et~al.}{2010}]{2010CQGra..27s4002P}
{Punturo} M.,  et~al., 2010, \mn@doi [Classical and Quantum Gravity]
  {10.1088/0264-9381/27/19/194002}, \href
  {https://ui.adsabs.harvard.edu/abs/2010CQGra..27s4002P} {27, 194002}

\bibitem[\protect\citeauthoryear{Randall \& Xianyu}{Randall \&
  Xianyu}{2019}]{Randall:2018lnh}
Randall L.,  Xianyu Z.-Z.,  2019, \mn@doi [Astrophys. J.]
  {10.3847/1538-4357/ab20c6}, 878, 75

\bibitem[\protect\citeauthoryear{{Rantala}, {Pihajoki}, {Johansson}, {Naab},
  {Lah{\'e}n}  \& {Sawala}}{{Rantala} et~al.}{2017}]{2017ApJ...840...53R}
{Rantala} A.,  {Pihajoki} P.,  {Johansson} P.~H.,  {Naab} T.,  {Lah{\'e}n} N.,
   {Sawala} T.,  2017, \mn@doi [\apj] {10.3847/1538-4357/aa6d65}, \href
  {https://ui.adsabs.harvard.edu/abs/2017ApJ...840...53R} {840, 53}

\bibitem[\protect\citeauthoryear{{Rantala}, {Johansson}, {Naab}, {Thomas}  \&
  {Frigo}}{{Rantala} et~al.}{2018}]{2018ApJ...864..113R}
{Rantala} A.,  {Johansson} P.~H.,  {Naab} T.,  {Thomas} J.,   {Frigo} M.,
  2018, \mn@doi [\apj] {10.3847/1538-4357/aada47}, \href
  {https://ui.adsabs.harvard.edu/abs/2018ApJ...864..113R} {864, 113}

\bibitem[\protect\citeauthoryear{{Rebassa-Mansergas}, {Toonen}, {Korol}  \&
  {Torres}}{{Rebassa-Mansergas} et~al.}{2019}]{reb19}
{Rebassa-Mansergas} A.,  {Toonen} S.,  {Korol} V.,   {Torres} S.,  2019,
  \mn@doi [\mnras] {10.1093/mnras/sty2965}, \href
  {https://ui.adsabs.harvard.edu/abs/2019MNRAS.482.3656R} {482, 3656}

\bibitem[\protect\citeauthoryear{{Regan} \& {Downes}}{{Regan} \&
  {Downes}}{2018}]{2018MNRAS.478.5037R}
{Regan} J.~A.,  {Downes} T.~P.,  2018, \mn@doi [\mnras]
  {10.1093/mnras/sty1289}, \href
  {https://ui.adsabs.harvard.edu/abs/2018MNRAS.478.5037R} {478, 5037}

\bibitem[\protect\citeauthoryear{{Reines}, {Greene}  \& {Geha}}{{Reines}
  et~al.}{2013}]{2013ApJ...775..116R}
{Reines} A.~E.,  {Greene} J.~E.,   {Geha} M.,  2013, \mn@doi [\apj]
  {10.1088/0004-637X/775/2/116}, \href
  {https://ui.adsabs.harvard.edu/abs/2013ApJ...775..116R} {775, 116}

\bibitem[\protect\citeauthoryear{{Reisswig}, {Ott}, {Abdikamalov}, {Haas},
  {M{\"o}sta}  \& {Schnetter}}{{Reisswig} et~al.}{2013}]{2013PhRvL.111o1101R}
{Reisswig} C.,  {Ott} C.~D.,  {Abdikamalov} E.,  {Haas} R.,  {M{\"o}sta} P.,
  {Schnetter} E.,  2013, \mn@doi [\prl] {10.1103/PhysRevLett.111.151101}, \href
  {https://ui.adsabs.harvard.edu/abs/2013PhRvL.111o1101R} {111, 151101}

\bibitem[\protect\citeauthoryear{{Reitze} et~al.,}{{Reitze}
  et~al.}{2019}]{2019arXiv190704833R}
{Reitze} D.,  et~al., 2019, arXiv e-prints, \href
  {https://ui.adsabs.harvard.edu/abs/2019arXiv190704833R} {p. arXiv:1907.04833}

\bibitem[\protect\citeauthoryear{Robson, Cornish, Tamanini  \& Toonen}{Robson
  et~al.}{2018}]{Robson:2018svj}
Robson T.,  Cornish N.~J.,  Tamanini N.,   Toonen S.,  2018, \mn@doi [Phys.
  Rev.] {10.1103/PhysRevD.98.064012}, D98, 064012

\bibitem[\protect\citeauthoryear{{Roedig} \& {Sesana}}{{Roedig} \&
  {Sesana}}{2012}]{2012JPhCS.363a2035R}
{Roedig} C.,  {Sesana} A.,  2012, in Journal of Physics Conference Series. p.
  012035 (\mn@eprint {arXiv} {1111.3742}),
  \mn@doi{10.1088/1742-6596/363/1/012035}

\bibitem[\protect\citeauthoryear{{Roedig}, {Sesana}, {Dotti}, {Cuadra},
  {Amaro-Seoane}  \& {Haardt}}{{Roedig} et~al.}{2012}]{2012A&A...545A.127R}
{Roedig} C.,  {Sesana} A.,  {Dotti} M.,  {Cuadra} J.,  {Amaro-Seoane} P.,
  {Haardt} F.,  2012, \mn@doi [\aap] {10.1051/0004-6361/201219986}, \href
  {https://ui.adsabs.harvard.edu/abs/2012A&A...545A.127R} {545, A127}

\bibitem[\protect\citeauthoryear{{Rossi}, {Lodato}, {Armitage}, {Pringle}  \&
  {King}}{{Rossi} et~al.}{2010}]{2010MNRAS.401.2021R}
{Rossi} E.~M.,  {Lodato} G.,  {Armitage} P.~J.,  {Pringle} J.~E.,   {King}
  A.~R.,  2010, \mn@doi [\mnras] {10.1111/j.1365-2966.2009.15802.x}, \href
  {https://ui.adsabs.harvard.edu/abs/2010MNRAS.401.2021R} {401, 2021}

\bibitem[\protect\citeauthoryear{{Ruiter}, {Belczynski}, {Benacquista},
  {Larson}  \& {Williams}}{{Ruiter} et~al.}{2010}]{rui10}
{Ruiter} A.~J.,  {Belczynski} K.,  {Benacquista} M.,  {Larson} S.~L.,
  {Williams} G.,  2010, \apj, 717, 1006

\bibitem[\protect\citeauthoryear{{Ryu}, {Perna}, {Haiman}, {Ostriker}  \&
  {Stone}}{{Ryu} et~al.}{2018}]{2018MNRAS.473.3410R}
{Ryu} T.,  {Perna} R.,  {Haiman} Z.,  {Ostriker} J.~P.,   {Stone} N.~C.,  2018,
  \mn@doi [\mnras] {10.1093/mnras/stx2524}, \href
  {https://ui.adsabs.harvard.edu/abs/2018MNRAS.473.3410R} {473, 3410}

\bibitem[\protect\citeauthoryear{{Saijo} \& {Hawke}}{{Saijo} \&
  {Hawke}}{2009}]{2009PhRvD..80f4001S}
{Saijo} M.,  {Hawke} I.,  2009, \mn@doi [\prd] {10.1103/PhysRevD.80.064001},
  \href {https://ui.adsabs.harvard.edu/abs/2009PhRvD..80f4001S} {80, 064001}

\bibitem[\protect\citeauthoryear{{Sanders}}{{Sanders}}{2013}]{2013JApA...34...81S}
{Sanders} G.~H.,  2013, \mn@doi [Journal of Astrophysics and Astronomy]
  {10.1007/s12036-013-9169-5}, \href
  {https://ui.adsabs.harvard.edu/abs/2013JApA...34...81S} {34, 81}

\bibitem[\protect\citeauthoryear{{Santamar{\'\i}a} et~al.,}{{Santamar{\'\i}a}
  et~al.}{2010}]{2010PhRvD..82f4016S}
{Santamar{\'\i}a} L.,  et~al., 2010, \mn@doi [\prd]
  {10.1103/PhysRevD.82.064016}, \href
  {https://ui.adsabs.harvard.edu/abs/2010PhRvD..82f4016S} {82, 064016}

\bibitem[\protect\citeauthoryear{{Schneider}, {Ehlers}  \& {Falco}}{{Schneider}
  et~al.}{1992}]{1992grle.book.....S}
{Schneider} P.,  {Ehlers} J.,   {Falco} E.~E.,  1992, {Gravitational Lenses},
  \mn@doi{10.1007/978-3-662-03758-4.
}

\bibitem[\protect\citeauthoryear{{Schnittman} \& {Krolik}}{{Schnittman} \&
  {Krolik}}{2008}]{2008ApJ...684..835S}
{Schnittman} J.~D.,  {Krolik} J.~H.,  2008, \mn@doi [\apj] {10.1086/590363},
  \href {https://ui.adsabs.harvard.edu/abs/2008ApJ...684..835S} {684, 835}

\bibitem[\protect\citeauthoryear{{Schnittman}, {Dal Canton}, {Camp}, {Tsang}
  \& {Kelly}}{{Schnittman} et~al.}{2018}]{2018ApJ...853..123S}
{Schnittman} J.~D.,  {Dal Canton} T.,  {Camp} J.,  {Tsang} D.,   {Kelly} B.~J.,
   2018, \mn@doi [\apj] {10.3847/1538-4357/aaa08b}, \href
  {https://ui.adsabs.harvard.edu/abs/2018ApJ...853..123S} {853, 123}

\bibitem[\protect\citeauthoryear{{Schutz}}{{Schutz}}{1986}]{1986Natur.323..310S}
{Schutz} B.~F.,  1986, \mn@doi [\nat] {10.1038/323310a0}, \href
  {https://ui.adsabs.harvard.edu/abs/1986Natur.323..310S} {323, 310}

\bibitem[\protect\citeauthoryear{Schwaller}{Schwaller}{2015}]{PhysRevLett.115.181101}
Schwaller P.,  2015, \mn@doi [Phys. Rev. Lett.]
  {10.1103/PhysRevLett.115.181101}, 115, 181101

\bibitem[\protect\citeauthoryear{Schwarz \& Stuke}{Schwarz \&
  Stuke}{2009}]{Schwarz:2009ii}
Schwarz D.~J.,  Stuke M.,  2009, \mn@doi [JCAP] {10.1088/1475-7516/2009/11/025,
  10.1088/1475-7516/2010/10/E01}, 0911, 025

\bibitem[\protect\citeauthoryear{{Sereno}, {Sesana}, {Bleuler}, {Jetzer},
  {Volonteri}  \& {Begelman}}{{Sereno} et~al.}{2010}]{2010PhRvL.105y1101S}
{Sereno} M.,  {Sesana} A.,  {Bleuler} A.,  {Jetzer} P.,  {Volonteri} M.,
  {Begelman} M.~C.,  2010, \mn@doi [\prl] {10.1103/PhysRevLett.105.251101},
  \href {https://ui.adsabs.harvard.edu/abs/2010PhRvL.105y1101S} {105, 251101}

\bibitem[\protect\citeauthoryear{{Sereno}, {Jetzer}, {Sesana}  \&
  {Volonteri}}{{Sereno} et~al.}{2011}]{2011MNRAS.415.2773S}
{Sereno} M.,  {Jetzer} P.,  {Sesana} A.,   {Volonteri} M.,  2011, \mn@doi
  [\mnras] {10.1111/j.1365-2966.2011.18895.x}, \href
  {https://ui.adsabs.harvard.edu/abs/2011MNRAS.415.2773S} {415, 2773}

\bibitem[\protect\citeauthoryear{{Sesana} \& {Khan}}{{Sesana} \&
  {Khan}}{2015}]{2015MNRAS.454L..66S}
{Sesana} A.,  {Khan} F.~M.,  2015, \mn@doi [\mnras] {10.1093/mnrasl/slv131},
  \href {https://ui.adsabs.harvard.edu/abs/2015MNRAS.454L..66S} {454, L66}

\bibitem[\protect\citeauthoryear{{Sesana} \& {Vecchio}}{{Sesana} \&
  {Vecchio}}{2010}]{2010PhRvD..81j4008S}
{Sesana} A.,  {Vecchio} A.,  2010, \mn@doi [\prd] {10.1103/PhysRevD.81.104008},
  \href {https://ui.adsabs.harvard.edu/abs/2010PhRvD..81j4008S} {81, 104008}

\bibitem[\protect\citeauthoryear{{Sesana}, {Gair}, {Berti}  \&
  {Volonteri}}{{Sesana} et~al.}{2011}]{2011PhRvD..83d4036S}
{Sesana} A.,  {Gair} J.,  {Berti} E.,   {Volonteri} M.,  2011, \mn@doi [\prd]
  {10.1103/PhysRevD.83.044036}, \href
  {https://ui.adsabs.harvard.edu/abs/2011PhRvD..83d4036S} {83, 044036}

\bibitem[\protect\citeauthoryear{{Seto}}{{Seto}}{2004}]{2004PhRvD..69b2002S}
{Seto} N.,  2004, \mn@doi [\prd] {10.1103/PhysRevD.69.022002}, \href
  {https://ui.adsabs.harvard.edu/abs/2004PhRvD..69b2002S} {69, 022002}

\bibitem[\protect\citeauthoryear{{Shao} et~al.,}{{Shao}
  et~al.}{2015}]{2015aska.confE..42S}
{Shao} L.,  et~al., 2015, in Advancing Astrophysics with the Square Kilometre
  Array (AASKA14). p.~42 (\mn@eprint {arXiv} {1501.00058})

\bibitem[\protect\citeauthoryear{{Shao}, {Sennett}, {Buonanno}, {Kramer}  \&
  {Wex}}{{Shao} et~al.}{2017}]{2017PhRvX...7d1025S}
{Shao} L.,  {Sennett} N.,  {Buonanno} A.,  {Kramer} M.,   {Wex} N.,  2017,
  \mn@doi [Physical Review X] {10.1103/PhysRevX.7.041025}, \href
  {https://ui.adsabs.harvard.edu/abs/2017PhRvX...7d1025S} {7, 041025}

\bibitem[\protect\citeauthoryear{{Shi}, {Krolik}, {Lubow}  \& {Hawley}}{{Shi}
  et~al.}{2012}]{2012ApJ...749..118S}
{Shi} J.-M.,  {Krolik} J.~H.,  {Lubow} S.~H.,   {Hawley} J.~F.,  2012, \mn@doi
  [\apj] {10.1088/0004-637X/749/2/118}, \href
  {https://ui.adsabs.harvard.edu/abs/2012ApJ...749..118S} {749, 118}

\bibitem[\protect\citeauthoryear{Silva, Sakstein, Gualtieri, Sotiriou  \&
  Berti}{Silva et~al.}{2018}]{Silva:2017uqg}
Silva H.~O.,  Sakstein J.,  Gualtieri L.,  Sotiriou T.~P.,   Berti E.,  2018,
  \mn@doi [Phys. Rev. Lett.] {10.1103/PhysRevLett.120.131104}, 120, 131104

\bibitem[\protect\citeauthoryear{{Smartt} et~al.,}{{Smartt}
  et~al.}{2017}]{2017Natur.551...75S}
{Smartt} S.~J.,  et~al., 2017, \mn@doi [\nat] {10.1038/nature24303}, \href
  {https://ui.adsabs.harvard.edu/abs/2017Natur.551...75S} {551, 75}

\bibitem[\protect\citeauthoryear{Soares-Santos et~al.}{Soares-Santos
  et~al.}{2019}]{Soares-Santos:2019irc}
Soares-Santos M.,  et~al., 2019, \mn@doi [Astrophys. J.]
  {10.3847/2041-8213/ab14f1}, 876, L7

\bibitem[\protect\citeauthoryear{{Soltan}}{{Soltan}}{1982}]{1982MNRAS.200..115S}
{Soltan} A.,  1982, \mn@doi [\mnras] {10.1093/mnras/200.1.115}, \href
  {https://ui.adsabs.harvard.edu/abs/1982MNRAS.200..115S} {200, 115}

\bibitem[\protect\citeauthoryear{{Souza Lima}, {Mayer}, {Capelo}  \&
  {Bellovary}}{{Souza Lima} et~al.}{2017}]{2017ApJ...838...13S}
{Souza Lima} R.,  {Mayer} L.,  {Capelo} P.~R.,   {Bellovary} J.~M.,  2017,
  \mn@doi [\apj] {10.3847/1538-4357/aa5d19}, \href
  {https://ui.adsabs.harvard.edu/abs/2017ApJ...838...13S} {838, 13}

\bibitem[\protect\citeauthoryear{{Spallicci}}{{Spallicci}}{2013}]{2013ApJ...764..187S}
{Spallicci} A. D.~A.~M.,  2013, \mn@doi [\apj] {10.1088/0004-637X/764/2/187},
  \href {https://ui.adsabs.harvard.edu/abs/2013ApJ...764..187S} {764, 187}

\bibitem[\protect\citeauthoryear{{Spergel} et~al.,}{{Spergel}
  et~al.}{2015}]{2015arXiv150303757S}
{Spergel} D.,  et~al., 2015, arXiv e-prints, \href
  {https://ui.adsabs.harvard.edu/abs/2015arXiv150303757S} {p. arXiv:1503.03757}

\bibitem[\protect\citeauthoryear{Stephanov}{Stephanov}{2006}]{Stephanov:2007fk}
Stephanov M.~A.,  2006, \mn@doi [PoS] {10.22323/1.032.0024}, LAT2006, 024

\bibitem[\protect\citeauthoryear{{Strominger} \& {Zhiboedov}}{{Strominger} \&
  {Zhiboedov}}{2014}]{2014arXiv1411.5745S}
{Strominger} A.,  {Zhiboedov} A.,  2014, arXiv e-prints, \href
  {https://ui.adsabs.harvard.edu/abs/2014arXiv1411.5745S} {p. arXiv:1411.5745}

\bibitem[\protect\citeauthoryear{{Takahashi} \& {Nakamura}}{{Takahashi} \&
  {Nakamura}}{2003}]{2003ApJ...595.1039T}
{Takahashi} R.,  {Nakamura} T.,  2003, \mn@doi [\apj] {10.1086/377430}, \href
  {https://ui.adsabs.harvard.edu/abs/2003ApJ...595.1039T} {595, 1039}

\bibitem[\protect\citeauthoryear{{Takeda}, {Nishizawa}, {Michimura}, {Nagano},
  {Komori}, {Ando}  \& {Hayama}}{{Takeda} et~al.}{2018}]{2018PhRvD..98b2008T}
{Takeda} H.,  {Nishizawa} A.,  {Michimura} Y.,  {Nagano} K.,  {Komori} K.,
  {Ando} M.,   {Hayama} K.,  2018, \mn@doi [\prd] {10.1103/PhysRevD.98.022008},
  \href {https://ui.adsabs.harvard.edu/abs/2018PhRvD..98b2008T} {98, 022008}

\bibitem[\protect\citeauthoryear{{Tamanini} \& {Danielski}}{{Tamanini} \&
  {Danielski}}{2019}]{2019NatAs.tmp..381T}
{Tamanini} N.,  {Danielski} C.,  2019, \mn@doi [Nature Astronomy]
  {10.1038/s41550-019-0807-y}, \href
  {https://ui.adsabs.harvard.edu/abs/2019NatAs.tmp..381T} {}

\bibitem[\protect\citeauthoryear{{Tamanini}, {Caprini}, {Barausse}, {Sesana},
  {Klein}  \& {Petiteau}}{{Tamanini} et~al.}{2016}]{2016JCAP...04..002T}
{Tamanini} N.,  {Caprini} C.,  {Barausse} E.,  {Sesana} A.,  {Klein} A.,
  {Petiteau} A.,  2016, \mn@doi [\jcap] {10.1088/1475-7516/2016/04/002}, \href
  {https://ui.adsabs.harvard.edu/abs/2016JCAP...04..002T} {2016, 002}

\bibitem[\protect\citeauthoryear{Tamanini, Klein, Bonvin, Barausse  \&
  Caprini}{Tamanini et~al.}{2019}]{Tamanini:2019usx}
Tamanini N.,  Klein A.,  Bonvin C.,  Barausse E.,   Caprini C.,  2019

\bibitem[\protect\citeauthoryear{{Tamburello}, {Capelo}, {Mayer}, {Bellovary}
  \& {Wadsley}}{{Tamburello} et~al.}{2017}]{2017MNRAS.464.2952T}
{Tamburello} V.,  {Capelo} P.~R.,  {Mayer} L.,  {Bellovary} J.~M.,   {Wadsley}
  J.~W.,  2017, \mn@doi [\mnras] {10.1093/mnras/stw2561}, \href
  {https://ui.adsabs.harvard.edu/abs/2017MNRAS.464.2952T} {464, 2952}

\bibitem[\protect\citeauthoryear{{Tamfal}, {Capelo}, {Kazantzidis}, {Mayer},
  {Potter}, {Stadel}  \& {Widrow}}{{Tamfal} et~al.}{2018}]{2018ApJ...864L..19T}
{Tamfal} T.,  {Capelo} P.~R.,  {Kazantzidis} S.,  {Mayer} L.,  {Potter} D.,
  {Stadel} J.,   {Widrow} L.~M.,  2018, \mn@doi [\apjl]
  {10.3847/2041-8213/aada4b}, \href
  {https://ui.adsabs.harvard.edu/abs/2018ApJ...864L..19T} {864, L19}

\bibitem[\protect\citeauthoryear{{Tang}, {MacFadyen}  \& {Haiman}}{{Tang}
  et~al.}{2017}]{2017MNRAS.469.4258T}
{Tang} Y.,  {MacFadyen} A.,   {Haiman} Z.,  2017, \mn@doi [\mnras]
  {10.1093/mnras/stx1130}, \href
  {https://ui.adsabs.harvard.edu/abs/2017MNRAS.469.4258T} {469, 4258}

\bibitem[\protect\citeauthoryear{{Tang}, {Haiman}  \& {MacFadyen}}{{Tang}
  et~al.}{2018}]{2018MNRAS.476.2249T}
{Tang} Y.,  {Haiman} Z.,   {MacFadyen} A.,  2018, \mn@doi [\mnras]
  {10.1093/mnras/sty423}, \href
  {https://ui.adsabs.harvard.edu/abs/2018MNRAS.476.2249T} {476, 2249}

\bibitem[\protect\citeauthoryear{{Tanvir} et~al.,}{{Tanvir}
  et~al.}{2017}]{2017ApJ...848L..27T}
{Tanvir} N.~R.,  et~al., 2017, \mn@doi [\apjl] {10.3847/2041-8213/aa90b6},
  \href {https://ui.adsabs.harvard.edu/abs/2017ApJ...848L..27T} {848, L27}

\bibitem[\protect\citeauthoryear{{Tauris}}{{Tauris}}{2018}]{2018PhRvL.121m1105T}
{Tauris} T.~M.,  2018, \mn@doi [\prl] {10.1103/PhysRevLett.121.131105}, \href
  {https://ui.adsabs.harvard.edu/abs/2018PhRvL.121m1105T} {121, 131105}

\bibitem[\protect\citeauthoryear{{Taylor} \& {Weisberg}}{{Taylor} \&
  {Weisberg}}{1982}]{tay82}
{Taylor} J.~H.,  {Weisberg} J.~M.,  1982, \mn@doi [\apj] {10.1086/159690},
  \href {https://ui.adsabs.harvard.edu/abs/1982ApJ...253..908T} {253, 908}

\bibitem[\protect\citeauthoryear{{The COrE Collaboration} et~al.,}{{The COrE
  Collaboration} et~al.}{2011}]{2011arXiv1102.2181T}
{The COrE Collaboration} et~al., 2011, arXiv e-prints, \href
  {https://ui.adsabs.harvard.edu/abs/2011arXiv1102.2181T} {p. arXiv:1102.2181}

\bibitem[\protect\citeauthoryear{{Thorne}}{{Thorne}}{1992}]{1992PhRvD..45..520T}
{Thorne} K.~S.,  1992, \mn@doi [\prd] {10.1103/PhysRevD.45.520}, \href
  {https://ui.adsabs.harvard.edu/abs/1992PhRvD..45..520T} {45, 520}

\bibitem[\protect\citeauthoryear{{Tinto} \& {Dhurandhar}}{{Tinto} \&
  {Dhurandhar}}{2005}]{2005LRR.....8....4T}
{Tinto} M.,  {Dhurandhar} S.~V.,  2005, \mn@doi [Living Reviews in Relativity]
  {10.12942/lrr-2005-4}, \href
  {https://ui.adsabs.harvard.edu/abs/2005LRR.....8....4T} {8, 4}

\bibitem[\protect\citeauthoryear{{Toonen} \& {Nelemans}}{{Toonen} \&
  {Nelemans}}{2013}]{too13}
{Toonen} S.,  {Nelemans} G.,  2013, \mn@doi [\aap]
  {10.1051/0004-6361/201321753}, \href
  {https://ui.adsabs.harvard.edu/abs/2013A&A...557A..87T} {557, A87}

\bibitem[\protect\citeauthoryear{{Toonen}, {Nelemans}  \& {Portegies
  Zwart}}{{Toonen} et~al.}{2012}]{too12}
{Toonen} S.,  {Nelemans} G.,   {Portegies Zwart} S.,  2012, \aap, 546, A70

\bibitem[\protect\citeauthoryear{{Tremaine}, {Ostriker}  \&
  {Spitzer}}{{Tremaine} et~al.}{1975}]{1975ApJ...196..407T}
{Tremaine} S.~D.,  {Ostriker} J.~P.,   {Spitzer} Jr. L.,  1975, \mn@doi [\apj]
  {10.1086/153422}, \href
  {https://ui.adsabs.harvard.edu/abs/1975ApJ...196..407T} {196, 407}

\bibitem[\protect\citeauthoryear{Tr\"{o}bs, Barke, Theeg, Kracht, Heinzel  \&
  Danzmann}{Tr\"{o}bs et~al.}{2010}]{Trobs:10}
Tr\"{o}bs M.,  Barke S.,  Theeg T.,  Kracht D.,  Heinzel G.,   Danzmann K.,
  2010, \mn@doi [Opt. Lett.] {10.1364/OL.35.000435}, 35, 435

\bibitem[\protect\citeauthoryear{Vachaspati \& Vilenkin}{Vachaspati \&
  Vilenkin}{1985}]{vachaspati_gravitational_1985}
Vachaspati T.,  Vilenkin A.,  1985, \mn@doi [Physical Review D]
  {10.1103/PhysRevD.31.3052}, 31, 3052

\bibitem[\protect\citeauthoryear{{Van Wassenhove}, {Capelo}, {Volonteri},
  {Dotti}, {Bellovary}, {Mayer}  \& {Governato}}{{Van Wassenhove}
  et~al.}{2014}]{2014MNRAS.439..474V}
{Van Wassenhove} S.,  {Capelo} P.~R.,  {Volonteri} M.,  {Dotti} M.,
  {Bellovary} J.~M.,  {Mayer} L.,   {Governato} F.,  2014, \mn@doi [\mnras]
  {10.1093/mnras/stu024}, \href
  {https://ui.adsabs.harvard.edu/abs/2014MNRAS.439..474V} {439, 474}

\bibitem[\protect\citeauthoryear{{Vasiliev}, {Antonini}  \&
  {Merritt}}{{Vasiliev} et~al.}{2015}]{2015ApJ...810...49V}
{Vasiliev} E.,  {Antonini} F.,   {Merritt} D.,  2015, \mn@doi [\apj]
  {10.1088/0004-637X/810/1/49}, \href
  {https://ui.adsabs.harvard.edu/abs/2015ApJ...810...49V} {810, 49}

\bibitem[\protect\citeauthoryear{{Verbiest} et~al.,}{{Verbiest}
  et~al.}{2016}]{2016MNRAS.458.1267V}
{Verbiest} J.~P.~W.,  et~al., 2016, \mn@doi [\mnras] {10.1093/mnras/stw347},
  \href {https://ui.adsabs.harvard.edu/abs/2016MNRAS.458.1267V} {458, 1267}

\bibitem[\protect\citeauthoryear{{Vigna-G{\'o}mez} et~al.,}{{Vigna-G{\'o}mez}
  et~al.}{2018}]{vig18}
{Vigna-G{\'o}mez} A.,  et~al., 2018, \mn@doi [\mnras] {10.1093/mnras/sty2463},
  \href {https://ui.adsabs.harvard.edu/abs/2018MNRAS.481.4009V} {481, 4009}

\bibitem[\protect\citeauthoryear{{Visbal} \& {Haiman}}{{Visbal} \&
  {Haiman}}{2018}]{2018ApJ...865L...9V}
{Visbal} E.,  {Haiman} Z.,  2018, \mn@doi [\apjl] {10.3847/2041-8213/aadf3a},
  \href {https://ui.adsabs.harvard.edu/abs/2018ApJ...865L...9V} {865, L9}

\bibitem[\protect\citeauthoryear{{Volonteri} \& {Natarajan}}{{Volonteri} \&
  {Natarajan}}{2009}]{2009MNRAS.400.1911V}
{Volonteri} M.,  {Natarajan} P.,  2009, \mn@doi [\mnras]
  {10.1111/j.1365-2966.2009.15577.x}, \href
  {https://ui.adsabs.harvard.edu/abs/2009MNRAS.400.1911V} {400, 1911}

\bibitem[\protect\citeauthoryear{{Volonteri} \& {Rees}}{{Volonteri} \&
  {Rees}}{2005}]{2005ApJ...633..624V}
{Volonteri} M.,  {Rees} M.~J.,  2005, \mn@doi [\apj] {10.1086/466521}, \href
  {https://ui.adsabs.harvard.edu/abs/2005ApJ...633..624V} {633, 624}

\bibitem[\protect\citeauthoryear{{Wang} \& {Han}}{{Wang} \&
  {Han}}{2012}]{wan12}
{Wang} B.,  {Han} Z.,  2012, \mn@doi [\nar] {10.1016/j.newar.2012.04.001},
  \href {https://ui.adsabs.harvard.edu/abs/2012NewAR..56..122W} {56, 122}

\bibitem[\protect\citeauthoryear{{Warner}}{{Warner}}{1995}]{war95}
{Warner} B.,  1995, Cambridge Astrophysics Series, \href
  {https://ui.adsabs.harvard.edu/abs/1995CAS....28.....W} {28}

\bibitem[\protect\citeauthoryear{{Will}}{{Will}}{2014}]{2014LRR....17....4W}
{Will} C.~M.,  2014, \mn@doi [Living Reviews in Relativity]
  {10.12942/lrr-2014-4}, \href
  {https://ui.adsabs.harvard.edu/abs/2014LRR....17....4W} {17, 4}

\bibitem[\protect\citeauthoryear{{Will}}{{Will}}{2018}]{2018CQGra..35h5001W}
{Will} C.~M.,  2018, \mn@doi [Classical and Quantum Gravity]
  {10.1088/1361-6382/aab1c6}, \href
  {https://ui.adsabs.harvard.edu/abs/2018CQGra..35h5001W} {35, 085001}

\bibitem[\protect\citeauthoryear{Wong, Baibhav  \& Berti}{Wong
  et~al.}{2019}]{Wong:2019hsq}
Wong K. W.~K.,  Baibhav V.,   Berti E.,  2019

\bibitem[\protect\citeauthoryear{{Woods} et~al.,}{{Woods}
  et~al.}{2018}]{2018arXiv181012310W}
{Woods} T.~E.,  et~al., 2018, arXiv e-prints, \href
  {https://ui.adsabs.harvard.edu/abs/2018arXiv181012310W} {p. arXiv:1810.12310}

\bibitem[\protect\citeauthoryear{{Wu} et~al.,}{{Wu}
  et~al.}{2015}]{2015Natur.518..512W}
{Wu} X.-B.,  et~al., 2015, \mn@doi [\nat] {10.1038/nature14241}, \href
  {https://ui.adsabs.harvard.edu/abs/2015Natur.518..512W} {518, 512}

\bibitem[\protect\citeauthoryear{{Yagi}, {Stein}  \& {Yunes}}{{Yagi}
  et~al.}{2016}]{2016PhRvD..93b4010Y}
{Yagi} K.,  {Stein} L.~C.,   {Yunes} N.,  2016, \mn@doi [\prd]
  {10.1103/PhysRevD.93.024010}, \href
  {https://ui.adsabs.harvard.edu/abs/2016PhRvD..93b4010Y} {93, 024010}

\bibitem[\protect\citeauthoryear{{Yu} \& {Tremaine}}{{Yu} \&
  {Tremaine}}{2002}]{2002MNRAS.335..965Y}
{Yu} Q.,  {Tremaine} S.,  2002, \mn@doi [\mnras]
  {10.1046/j.1365-8711.2002.05532.x}, \href
  {https://ui.adsabs.harvard.edu/abs/2002MNRAS.335..965Y} {335, 965}

\bibitem[\protect\citeauthoryear{{Zel'dovich} \& {Polnarev}}{{Zel'dovich} \&
  {Polnarev}}{1974}]{1974SvA....18...17Z}
{Zel'dovich} Y.~B.,  {Polnarev} A.~G.,  1974, Soviet Astronomy, \href
  {https://ui.adsabs.harvard.edu/abs/1974SvA....18...17Z} {18, 17}

\bibitem[\protect\citeauthoryear{{de Rham} \& {Melville}}{{de Rham} \&
  {Melville}}{2018}]{2018PhRvL.121v1101D}
{de Rham} C.,  {Melville} S.,  2018, \mn@doi [\prl]
  {10.1103/PhysRevLett.121.221101}, \href
  {https://ui.adsabs.harvard.edu/abs/2018PhRvL.121v1101D} {121, 221101}

\bibitem[\protect\citeauthoryear{{de Rham}, {Deskins}, {Tolley}  \& {Zhou}}{{de
  Rham} et~al.}{2017}]{2017RvMP...89b5004D}
{de Rham} C.,  {Deskins} J.~T.,  {Tolley} A.~J.,   {Zhou} S.-Y.,  2017, \mn@doi
  [Reviews of Modern Physics] {10.1103/RevModPhys.89.025004}, \href
  {https://ui.adsabs.harvard.edu/abs/2017RvMP...89b5004D} {89, 025004}

\bibitem[\protect\citeauthoryear{{Nishizawa}, {Sesana}, {Berti}  \&
  {Klein}}{wp3}{}]{wp3}
The missing link in gravitational-wave astronomy: Discoveries waiting in the
  decihertz range, White paper submission for the Voyage 2050

\makeatother
\end{thebibliography}
}

\end{document}